\documentclass[journal]{IEEEtran}

\IEEEoverridecommandlockouts

\usepackage{soul,xcolor}
\usepackage{bm}

\usepackage[normalem]{ulem}
\usepackage{multirow,enumitem}
\usepackage{algorithm}
\usepackage{algpseudocode}
\usepackage{verbatim}
\usepackage[latin1,utf8]{inputenc}
\usepackage{amsmath}
\usepackage{amsfonts}
\usepackage{amssymb}
\usepackage{mathrsfs}
\usepackage{booktabs}
\usepackage{url}
\usepackage{ifthen}
\usepackage{xspace}
\usepackage{dsfont}
\usepackage{bbm}
\usepackage{cite}
\usepackage{epsfig}
\usepackage{epstopdf}
\usepackage{array}
\usepackage{breqn}
\usepackage{esint}
\usepackage{amsthm}
\usepackage{setspace}
\usepackage{lipsum}
\theoremstyle{plain}

\theoremstyle{definition}

\newtheorem{remark}{Remark}
\theoremstyle{remark}

\usepackage{footnote}
\theoremstyle{plain}

\usepackage{fancybox,framed}

\usepackage{arydshln}

\usepackage{bbold}
\usepackage{todonotes}

\usepackage{stmaryrd}
\usepackage{algorithmicx}
\algnewcommand\INPUT{\item[\textbf{Input:}]}%
\algnewcommand\OUTPUT{\item[\textbf{Output:}]}%
\usepackage{lipsum}
\usepackage{mathtools}
\usepackage{cuted}
\usepackage{dblfloatfix}
\DeclareFontFamily{U}{mathx}{}
\DeclareFontShape{U}{mathx}{m}{n}{<-> mathx10}{}
\DeclareSymbolFont{mathx}{U}{mathx}{m}{n}
\DeclareMathAccent{\widecheck}{0}{mathx}{"71}

\usepackage{graphicx}
\usepackage{epstopdf}
\usepackage{afterpage}
\newcommand{\mathleft}{\@fleqntrue\@mathmargin0pt}
\newcommand{\mathcenter}{\@fleqnfalse}
\newcommand{\tcadd}[1]{{{#1}}}
\newcommand{\tcminor}[1]{{{#1}}}

\newcommand{\RN}[1]{%
	\textup{\uppercase\expandafter{\romannumeral#1}}%
}

\usepackage{subcaption}
\usepackage{graphicx}

\ifCLASSINFOpdf

\else

\fi

\begin{document}
	\include{graphics}
	\include{amsmath}
	
	\setulcolor{red}
	\setul{red}{2pt}
	\setstcolor{red}
	\title{Joint UAV Placement and Transceiver Design in Multi-User Wireless Relay Networks}
	\author{Tzu-Hsuan~{Chou}, Nicol\`{o} {Michelusi}, David J. {Love}, and James V. {Krogmeier}
		% \nm{coauthors?}
		%		\begin{singlespace}			
		\thanks{This work was supported in part by the National Science Foundation under grants CNS-1642982, CCF-1816013, EEC-1941529 and CNS-2129015. }
		%			A preliminary version of this paper was presented at IEEE Globecom 2021 \cite{chou2021wideband}.}
		\thanks{T.-H. Chou is with Qualcomm, Inc., San Diego, CA, USA; emails: tzuhchou@qti.qualcomm.com}%
		\thanks{N. Michelusi is with the School of Electrical, Computer and Energy Engineering, Arizona State University, AZ, USA; email: nicolo.michelusi@asu.edu}%
		\thanks{D. J. Love and J. V. Krogmeier are with the School of Electrical and Computer Engineering, Purdue University, West Lafayette, IN, USA; emails: \{djlove, jvk\}@purdue.edu}% 
		
		%	\end{singlespace} 
		% \vspace{-7mm}
	}

	\maketitle
	
	%	\nm{We will submit this to the JSAC issue on transceiver design, but there is no mention of trans. design in the introduction nor abstract!}
	
	%	\nm{Please pay close attention to grammatical and notational errors. There are too many of them. They can be quite annoying (to me and prospective reviewers) and reveal a lack of attention to detail.}
	\begin{abstract}
		In this paper, a novel approach is proposed to improve the minimum signal-to-interference-plus-noise-ratio (SINR) among users in \tcadd{non-orthogonal multi-user wireless relay networks}, by optimizing the placement of unmanned aerial vehicle (UAV) relays, relay beamforming, and receive combining.
		The design is separated into two problems: beamforming-aware UAV placement optimization and \tcadd{transceiver design for minimum SINR maximization}. 
		A significant challenge in beamforming-aware UAV placement optimization is the lack of instantaneous channel state information (CSI) prior to deploying UAV relays, making it difficult to derive \tcadd{the beamforming SINR in non-orthogonal multi-user transmission. }
		{To address this issue, an approximation of the expected beamforming SINR is derived using the narrow beam property of a massive MIMO base station.}
%		the narrow beam property of multiple antenna systems. 
		Based on this, a UAV placement algorithm is proposed to provide UAV positions that improve the minimum expected beamforming SINR among users, using a difference-of-convex framework.
		% Subsequently, with estimated CSI after deploying the UAV relays to the optimized positions, a joint relay beamforming and receive combining (JRBC) algorithm is proposed to optimize the transceiver to improve the minimum beamforming SINR among users, using a block-coordinate descent approach.
  %       \nm{
  %       Subsequently, after deploying the UAV relays to the optimized positions, and with estimated CSI available,
  %       a joint relay beamforming and receive combining (JRBC) algorithm is proposed to optimize the transceiver to improve the minimum beamforming SINR among users, using a block-coordinate descent approach. 
  %       }
        Subsequently, after deploying the UAV relays to the optimized positions, and with estimated CSI available, a joint relay beamforming and receive combining (JRBC) algorithm is proposed to optimize the transceiver to improve the minimum beamforming SINR among users, using a block-coordinate descent approach.
		Numerical results show that the UAV placement algorithm combined with the JRBC algorithm provides a $4.6\text{ dB}$ SINR improvement over state-of-the-art schemes.
	\end{abstract}
	
	%\begin{IEEEkeywords}
	%	Multi-user relaying networks, spatially controlled relay beamforming, cooperative relay beamforming, placement and movement optimization.
	%\end{IEEEkeywords}
	% \vspace{-2mm}
	\section{Introduction}
    \tcadd{Distributed cooperative relay beamforming, where multiple relays collaboratively transmit a signal with coordinated adjustment of phases and amplitudes to enhance received signal quality at the destination, has been extensively studied in wireless applications to augment network coverage, boost spectral efficiency, and improve link reliability  \cite{havary2008distributed,havary2009optimal, zheng2009collaborative, li2010cooperative}.}
	These approaches prove instrumental when direct channels between sources and destinations suffer from severe fading, resulting in poor communication quality.
	Due to limited space and cost constraints in infrastructure, deploying fixed relays is not always feasible. 
    Furthermore, fixed relay positions prevent adaptive placement based on the locations of ground users, limiting their ability to fully exploit spatial diversity and enhance transmission performance.
    % \nm{Furthermore,  fixed relay positions prevent adaptive placement based on the locations of ground users, limiting their ability to fully exploit spatial diversity and enhance transmission performance}. 	
	Recently, unmanned aerial vehicle (UAV) technologies have gained significant attention for applications in wireless communication \cite{brinton2025key, 10615846, zeng2019accessing}, including coverage extension, communication rate enhancement, data harvesting, and resource allocation. 
	Thanks to their 3D mobility, UAVs can swiftly provide flexible service to enhance the quality of service (QoS) for on-demand user equipment.

	%	Therefore, this advantage motivates us to incorporate this property in the multi-user wireless relay network.
	
	The use of UAVs equipped with relays to enhance the wireless communication quality has been investigated in \cite{kalogerias2018spatially, evmorfos2022reinforcement, liu2019comp, xie2020common}. 
	The work \cite{kalogerias2018spatially} addresses the enhancement of signal-to-interference-plus-noise-ratio (SINR) by optimally controlling relay placements for a \emph{single} source-destination pair.
	The work \cite{evmorfos2022reinforcement} proposes a joint relay beamforming and motion control for a spatio-temporally varying channel in single-user wireless relay networks, using reinforcement learning to guide relay motion.
	The works \cite{kalogerias2018spatially, evmorfos2022reinforcement} demonstrate that the joint design of relay placement and distributed relay beamforming improves communication performance in single-user mobile relay networks.
	However, applying these techniques to multi-user networks necessitates orthogonal transmission methods, such as time division multiple access (TDMA), which are not efficient in utilizing wireless resources.
	This motivates us to jointly design the UAV placement and relay beamforming in multi-user wireless relay network scenarios with non-orthogonal transmissions.

    In multi-user wireless networks, employing non-orthogonal multi-user transmissions improves network efficiency but induces inter-user interference, which drastically degrades communication performance.
	Recently, the work \cite{liu2019comp} proposes a wireless network architecture, coordinated multipoint (CoMP) in the sky, which leverages both interference mitigation and the high mobility of UAVs. 
    \tcadd{Additionally, the work \cite{xie2020common} investigates the joint optimization of UAV trajectory and resource allocation to maximize the common (minimum) uplink throughput in a two-user UAV-enabled interference channel scenario, considering the downlink wireless power transfer efficiency.}
    % \tcadd{The work \cite{xie2020common} addresses the optimization of UAV trajectory and resource allocation to maximize common (minimum) throughput for two ground users' cases for UAV-enabled interference channels.}
	Although the works \cite{liu2019comp,xie2020common} are applicable to non-orthogonal transmissions in multi-user relay networks, their UAV placements are designed based on the first-hop channel (i.e., the channel from users to UAV relays) while they assume the second-hop channel (i.e., the channel from UAV relays to base station) is a perfect link.
	In practical scenarios, the second-hop channels are typically wireless links, making it essential to account for them in UAV placement design.
    % which needs to be considered in the UAV placement design.

    The primary challenge in optimizing UAV placement to enhance beamforming SINR in multi-user relay networks is the lack of instantaneous channel state information (CSI) prior to deploying UAVs to new positions. 
    The work \cite{liu2019comp} addresses this issue by optimizing UAV placement based on a lower bound of the ergodic user rate, which uses only statistical CSI information (distance-dependent path loss). 
    % \nm{, which uses only statistical CSI information (distance-dependent path loss)}. 
    However, it assumes the received signals at relays are forwarded to a central processor for joint decoding, neglecting the impact of the second-hop channel. To the best of our knowledge, UAV placement optimization that considers maximizing the minimum beamforming SINR among users while accounting for two-hop channels in multi-user wireless networks remains unaddressed. This gap is the focus of our work.

	% The main issue of UAV placement optimization for \tcadd{enhancing beamforming SINR in multi-user relay networks} is the unavailability of the \textit{instantaneous} channel state information (CSI) before UAV deployment to new positions. 
	% The work \cite{liu2019comp} addressed this issue by considering a lower bound of the ergodic user rate as the objective function in UAV placement optimization.
	% However, the work \cite{liu2019comp} forwards the received signal at relays to a central processor for joint decoding, in which the impact of the second-hop channel is not considered.
 %    To our knowledge, the UAV placement optimization for maximizing the minimum beamforming SINR among users, considering two-hop channels in multi-user wireless networks, has not yet been addressed. This is the focus of this work.
	% To our knowledge, the optimization problem of beamforming-aware UAV placement, considering two-hop channels in multi-user wireless networks, has not yet been addressed. This is the focus of this work.
	%	To the best of our knowledge, the beamforming-aware UAV placement optimization problem considering the two-hop channels in multi-user wireless networks has not been addressed, which is the focus of our work.
	%	only considers the firs-hop channel and applies the ZF equlizer across the distributed UAV relays at the central
	%	assume the , which ignores the second-hop channel

    In this work, we jointly optimize UAV placement, relay beamforming, and receive combining to maximize the minimum SINR among users. 
    The design of relay beamforming and receive combining in the minimum SINR maximization problem requires estimated CSI, particularly the phase information of the channels. 
    However, as CSI is unavailable before UAVs are deployed to new positions, it becomes challenging to optimize UAV placement to improve the minimum SINR among users. 
    To address this, we derive a closed-form approximation of the expected beamforming SINR using the narrow beam property of large multi-antenna systems, enabling a tractable problem formulation for UAV placement optimization. 
    Therefore, we formulate two optimization problems: \emph{beamforming-aware UAV placement optimization} and \emph{transceiver design for minimum SINR maximization}.

	% In our work, we jointly optimize UAV placement, relay beamforming, and receive combining to maximize the minimum SINR among users.
	% The transceiver design of the relay beamforming and receive combining in the minimum SINR maximization problem requires the estimated CSI, especially the channels' phases.
	% However, the CSI is not available before we deploy the UAVs to new positions, which makes it difficult to optimize the UAV placement to improve the minimum SINR among users.
	% To address the issue, we derive a closed-form approximation of the expected beamforming SINR using a narrow beam property of large multi-antenna systems, \tcadd{which facilitates the problem formulation in UAV placement optimization.}
	% Therefore, we formulate two optimization problems: \emph{beamforming-aware UAV placement optimization} and \emph{transceiver design for minimum SINR maximization}.

    For beamforming-aware UAV placement optimization, our goal is to determine UAV positions that maximize the minimum expected beamforming SINR among users in the absence of instantaneous CSI at new positions.
    % , \hl{leveraging the expected SINR}\nm{this sentence is unclear to me} in the absence of instantaneous CSI at new positions. 
    % \nm{We show that this problem can be case as a difference-of-convex (DC) optimization, and propose convex relaxation techniques to solve it.}    
    % We propose a UAV placement algorithm that employs difference-of-convex (DC) functions and convex relaxation techniques to solve this problem. 
    We show that this problem can be formulated as a difference-of-convex (DC) optimization, and propose convex relaxation techniques to solve it.
    Once UAV relays are deployed to the optimized positions, instantaneous CSI can be estimated through pilot-based channel training (e.g., \cite{ma2011pilot}) and used for subsequent transceiver design aimed at minimum SINR maximization. 
    For the transceiver design, we introduce a joint relay beamforming and receive combining (JRBC) algorithm based on the block-coordinate descent (BCD) approach, further improving the minimum SINR among users. To our knowledge, this is the first work in jointly optimizing UAV placement, relay beamforming, and receive combining to enhance the minimum beamforming SINR among users.
    
	% For beamforming-aware UAV placement optimization, we aim to find UAV positions that maximize the minimum of the \emph{expected} beamforming SINR among users.
 %    \tcadd{The expected beamforming SINR is considered for UAV placement optimization due to the lack of instantaneous CSI at new positions}.
	% We propose a UAV placement algorithm to solve this problem using a difference-of-convex (DC) functions with convex relaxation techniques. 
	% Once UAV relays are deployed to the optimized positions, the CSI can be estimated using pilot-based channel training \tcadd{(e.g., \cite{ma2011pilot})} and utilized in subsequent transceiver design for minimum SINR maximization. 
	% For this transceiver design, we propose a joint relay beamforming and combining (JRBC) algorithm, employing the block-coordinate descent (BCD) approach, to further enhance the minimum SINR among users.
	% To our knowledge, this is the first work in optimizing UAV placement, relay beamforming, and receive combining, aiming to improve the minimum beamforming SINR among users.

    \noindent\textbf{Related Works}

	\textit{Wireless relay networks with fixed-positioned relay(s)} have been investigated in \cite{sanguinetti2012tutorial, phan2012beamforming, cheng2012joint, rashid2013relay,phan2013iterative, ruan2019distributed,che2014joint, dimas2019cooperative}. 
    In such works, the relays are stationary at fixed positions, so the large-scale fading effect of the channel remains constant.
    The work \cite{sanguinetti2012tutorial} provided an overview of fundamental results and implementation challenges in designing single-user amplify-and-forward (AF) MIMO relay systems.
    Several works \cite{phan2012beamforming, cheng2012joint, rashid2013relay, phan2013iterative} have utilized the DC framework to design AF relay beamforming for multi-user relay networks, each targeting specific design objectives.
    The work \cite{phan2012beamforming} designed relay beamforming by minimizing relay power under SINR constraints.
    The work \cite{cheng2012joint} jointly optimized relay beamforming and source transmitted power to meet QoS requirements for each source-destination (S-D) pair.
    The work \cite{rashid2013relay} designed the relay beamforming to maximize the minimum throughput among S-D pairs under relay power constraints. 
    The work \cite{phan2013iterative} addressed precoding design in MIMO relays for various objectives, including total relay power minimization, individual power maximin optimization, and SINR maximin optimization.
	To mitigate the impact of channel estimation errors, the work \cite{ruan2019distributed} developed a robust distributed relay beamforming using low-rank and cross-correlation properties.
    For wireless networks with fixed-positioned relays, relay assignment has also been investigated and recognized as an effective technique for AF relay protocol \cite{che2014joint, dimas2019cooperative}.
    The work \cite{che2014joint} addressed a joint optimization problem of relay beamforming and relay assignment to maximize the minimum SINR across users, integrating the DC framework with binary constraints.
    The work \cite{dimas2019cooperative} designed relay beamforming with predictive relay selection by exploiting CSI correlation in single-user multi-relay networks.
    In contrast to \cite{sanguinetti2012tutorial, phan2012beamforming, cheng2012joint, rashid2013relay,phan2013iterative, ruan2019distributed,che2014joint, dimas2019cooperative}, we propose a joint design of relay beamforming and receive combining to maximize the minimum SINR among multiple users, which is an issue that, to the best of our knowledge, has not been addressed in multi-user wireless relay networks.    
    Note that a related work \cite{behbahani2008optimizations} focuses on transceiver design that optimizes MIMO relay beamforming and receive combining to maximize the total SINR of multiple data streams in single-user single-relay networks. However, it does not consider the multi-user scenarios or distributed relays.

    \textit{Wireless relay networks with mobile relay(s)} has been extensively investigated in \cite{kalogerias2018spatially, evmorfos2022deep, evmorfos2022reinforcement, hanna2019uav, hanna2021uav, kang20203d, zhang20213d, ding20203d, huang2021joint, mahmood2023joint, liu2019comp, xie2020common, dinh2019joint, gao20203d, khuwaja2019optimum, wang2022resource}. 
    The mobility of UAV relays has been proposed as an effective approach to construct distributed relay networks, enhancing communication performance.
    % The UAV relay mobility has been proposed as an effective approach to constructing distributed relay networks for enhancing communication performance.
    In \cite{kalogerias2018spatially}, the SINR enhancement was addressed by optimizing relay placements for a \emph{single} source-destination pair.    
    % The work \cite{kalogerias2018spatially} addressed the enhancement SINR by optimally controlling relay placements for a \emph{single} source-destination pair.
    The work \cite{evmorfos2022deep, evmorfos2022reinforcement}
    proposed joint relay beamforming and motion control for single-user wireless relay networks with spatio-temporally varying channels, leveraging reinforcement learning to guide relay motion.
    % proposes a joint relay beamforming and motion control for a spatio-temporally varying channel in single-user wireless relay networks, using reinforcement learning to guide relay motion.
    The work \cite{hanna2019uav} optimized the placement of UAV relays to enhance link capacity by adjusting the distance of the UAV relays to the source and destination.
    In \cite{hanna2021uav}, UAV positions were designed to maximize the MIMO capacity in communication with a multi-antenna ground station. 
    However, these studies focused solely on scenarios involving a single user, whereas our work addresses multi-user scenarios.   
    
    % \nm{break line here and start a new paragraph}     
    % The work \cite{hanna2021uav} designed the positions of a UAV swarm, communicating with a multi-antenna ground station, to attain the maximum MIMO capacity.    
    % However, these works developed the placement of UAV relays only serving a single user, while our work focuses on multi-user scenarios.    
    The work \cite{kang20203d} jointly designed 3D placement and bandwidth allocation, power loading to maximize the minimum achievable expected rates among users, integrating the advantages of the BCD approach and stochastic optimization methods.
    The work \cite{zhang20213d} investigated the system assisted by multiple UAV-mounted BS to improve the coverage by optimizing the UAV positions, UE scheduling, and bandwidth allocation.
    The work \cite{ding20203d} investigated the UAV trajectory design and bandwidth allocation problem to improve the UAVs' energy consumption and the fairness of throughput among users.
    The work \cite{huang2021joint} deployed the UAVs as flying remote radio heads to serve ground users, and optimized the UAV placement to maximize the minimum rate of users.
    The work \cite{mahmood2023joint} jointly optimized the UAV placement, subcarrier allocation and power control to maximize the downlink sum-rate in OFDMA systems with the aid of multiple UAVs.
    These works \cite{kang20203d, zhang20213d, ding20203d, huang2021joint, mahmood2023joint} primarily investigated UAV placement design to enhance multi-user communication in scenarios with orthogonal transmissions, where users are separated across time or frequency resources. 
    However, orthogonal transmissions inherently limit system spectral efficiency. 
    % However, the orthogonal transmission inherently limits\nm{[without THE and plural] orthogonal  transmissions inherently limit...} system spectral efficiency. 
    
    % \nm{start new paragraph}
    % investigated the UAV placement design to improve the multi-user communication with orthogonal transmissions, in which the multi-user transmissions needs to be separated by time or frequency resource, degrading the system spectral efficiency.
    % The multi-user wireless relay network with non-orthogonal transmission were investigated in 
    % The works \cite{liu2019comp, xie2020common, dinh2019joint, gao20203d, khuwaja2019optimum, wang2022resource} addressed 
    The UAV placement problem for multi-user wireless relay networks with non-orthogonal transmissions has been investigated in \cite{liu2019comp, xie2020common, dinh2019joint, gao20203d, khuwaja2019optimum, wang2022resource}. 
    The work \cite{liu2019comp} solved the UAV placement problem to maximize the minimum rate among users.
    In \cite{xie2020common}, a UAV-enabled two-user wireless-powered communication network was investigated, proposing a solution to maximize the minimum throughput of the two users by jointly optimizing the UAV trajectory and downlink/uplink resource allocation.
    The work in \cite{dinh2019joint} proposed methods to enhance QoS by jointly optimizing transmit beamforming, UAV location, and content placement.    
    % \nm{content? Not sure what this means} placement. 
    The content placement is an optimization problem to allocate content data among multiple UAVs, considering their limited storage capacity.
    % to allocate the contents (data) among multiple UAVs due to their limited storage space.
    In \cite{gao20203d}, a decentralized control strategy for UAV placement was developed, utilizing only local information to maximize channel capacity.
    The work in \cite{khuwaja2019optimum} proposed a UAV deployment strategy to maximize coverage area while ensuring ground users’ SNR performance, taking co-channel interference into account.
    The work \cite{wang2022resource} optimized power allocation, UAV service zone scheduling, and user scheduling to improve spectral efficiency, using deep reinforcement learning.
    However, these works were based on a one-hop channel model, assuming the second-hop channel as a perfect link.
    In practical scenarios, the second-hop channels in UAV relay networks are typically wireless, making it essential to account for their impact.
    In contrast, our work considers two-hop channels in multi-user non-orthogonal wireless relay networks. 
    We aim to maximize the minimum SINR among users by jointly optimizing UAV placement, relay beamforming, and receive combining.

	\noindent\textbf{Contributions}	
    
	% In our work, we seek to optimize UAV positions that maximize the minimum beamforming SINR among users, incorporating both relay beamforming and receive combining. 
	% The instantaneous CSI is crucial for deriving the beamforming SINR, but it is unavailable before UAV deployment to new positions.
	% To address this issue, we find an approximation of the expected beamforming SINR using the narrow beam property of multiple antenna systems.
	% Based on this approximation, we propose an algorithm to deploy UAV relays to enhance the minimum expected beamforming SINR among users. 
	% Once the estimated CSI is available after UAV deployment, we propose a JRBC algorithm to improve the minimum SINR among users by designing the relay beamforming and receive combining.\nm{this feels repetitive, since you are repeating the same concepts in the summary below. I would merge this paragraph into the summary below.}

    % \nm{before the summary, I would briefly highlith the open challenges that have not been addressed in prior work, as evident from your literature review above.}
    From our literature review, we observe that, in multi-user non-orthogonal wireless relay networks, the joint design of the relay beamforming and receive combining has not been addressed.
    Additionally, exploiting the mobility of UAVs to enhance communication performance presents a challenge due to the lack of instantaneous CSI, especially when considering two-hop channels. 
    Optimizing UAV relay placement with optimized relay beamforming and receive combining strategies remains an unsolved problem.
    % Additionally, to exploiting the mobility of UAVs for enhancing communication performance, the lack of instantaneous CSI presents a challenge in optimizing the UAV relay placement, with optimized relay beamforming and receive combining strategies, especially when considering two-hop channels. This problem remains unsolved. 
    % Our contributions are as follows:
    % the optimized deployment of UAV relays for enhancing the multi-user non-orthogonal tranmimission, taking into account the two-hop channels is even challenging.
	In summary, the contributions of this work are as follows:
	\begin{itemize}
		\item 
		We propose a joint design of UAV placement, relay beamforming, and receive combining to enhance the minimum beamforming SINR among users. 
        % Due to the lack of instantaneous CSI before UAV deployment, 
        We formulate this into two optimization problems:
		\textbf{beamforming-aware UAV placement optimization} and \textbf{transceiver design for minimum SINR maximization}.
		\item Beamforming-aware UAV placement optimization finds the positions for multiple UAVs to maximize the minimum expected beamforming SINR among users. 
        Due to the lack of instantaneous CSI, we derive an approximation of the expected beamforming SINR using the narrow beam property of large multi-antenna systems.
		We propose a UAV placement algorithm to solve the problem based on this approximation in a DC framework with convex relaxation techniques.
		Numerical results show that the optimized UAV position achieves a $4.6\text{ dB}$ improvement in beamforming SINR over the state-of-the-art UAV placement method \cite{liu2019comp}, assuming the transceivers are optimized by the JRBC algorithm.
		%		\nm{is this 6dB improvement just due to the better positioning? That what it looks like from the way you describe it. Or is it combined with the min SINR maxim (JRBC) explained in the next item?}
		\item Transceiver design for minimum SINR maximization optimizes the relay beamforming and receive combining to improve the minimum SINR among users, based on the estimated CSI. We propose the JRBC algorithm to improve the minimum SINR among users using a BCD approach. 
		Numerical results demonstrate that JRBC yields a $4.5\text{ dB}$ improvement in beamforming SINR compared to the state-of-the-art \cite{behbahani2008optimizations}, assuming the UAV positions are optimized by our UAV placement algorithm.			
		%		\nm{just JRBC alone, or combined with better positioning? Also, if positioning gives 6dB gain and JRBC gives 4.5 gain, does it mean that overall we gain 6+4.5=10.5 dB over the state of art?}
		%		\end{itemize}
		%		\item Numerical results show that ...
	\end{itemize}

	The paper is organized as follows.
	Section \ref{sec_system_model} introduces the system model. 
	Section \ref{sec_prob_formulation} formulates the optimization problems.
	Section \ref{sec_UAV_placement_optimization} proposes the UAV placement algorithm, \tcadd{followed by the complexity analyses}.
    Section \ref{subsec_joint_opt_relay_bf_and_rx_combining} proposes the joint relay beamforming and receive combining algorithm, \tcadd{followed by the complexity analyses}.
	%	max-min algorithm, which jointly optimize relay beamforming and receive combining based on estimated CSI.
	%Section \ref{sec_UAV_movement_alg} develops two UAV movement algorithms, followed by the complexity analyses in Section \ref{sec_complexity}.
	%discusses the computational complexity
	Section \ref{sec_numerical_results} shows the numerical results, and Section \ref{sec_conclusion} concludes the paper.
	
	{\bf Notation:}	Bold lowercase letters $\mathbf{x}$ and bold uppercase letters $\mathbf{X}$ denote vectors and matrices, respectively;
	$\mathbf{X}^\top$, $\mathbf{X}^H$, $\mathbf{X}^{-1}$ represent the transpose, conjugate transpose, and inverse of $\mathbf{X}$, respectively;
	% $\mathbf{X}^\top$, $\mathbf{X}^H$, $\mathbf{X}^{+}$, $\textrm{vec}(\mathbf{X})$, $\textrm{det}(\mathbf{X})$ represent the transpose, conjugate transpose, Moore-Penrose pseudo-inverse, vectorization, and determinant of $\mathbf{X}$, respectively;
	$[\mathbf{X}]_{m,:}$, $[\mathbf{X}]_{:,n}$, $[\mathbf{X}]_{m,n}$ denote the $m$-th row, $n$-th column, and $(m,n)$-th element of $\mathbf{X}$, respectively;
	%$\text{diag}(\mathbf{ X})$ is the vector constructed by the diagonal elements of $\mathbf{X}$; 
	$\text{Diag}(\mathbf{x})$ is the diagonal matrix with the elements of $\mathbf{x}$ as its diagonal; 
	$\text{Diag}(\mathbf{X})$ is the diagonal matrix having the diagonal elements of $\mathbf{X}$ as its diagonal; $\text{Re}(x)$ is the real part of $x$;
	$\odot$ denotes the Hadamard product. 
    We denote $i=\sqrt{-1}$.
	%	\nm{not consistent! You also use j instead of i, see for example Eq 40. But j is used to denote user index. Please check carefully and fix.}
	
	%$[\mathbf{X}]_\Gamma$ (respectively, $[\mathbf{x}]_\Gamma$) is the submatrix with columns of $\mathbf{X}$ (the subvector with elements of $\mathbf{x}$) associated with the indices set $\Gamma$; 
	%$(\mathbf{X})_{\mathbf{n}}$ is the $\mathbf{n}$-th element of $\mathbf{X}$;
	%$\lvert\Psi\rvert$ is the cardinality of the set $\Psi$;	
	%%	the Kronecker product is denoted as $\otimes$;
	%$\otimes$ denotes the Kronecker product;
	%$\mathbf{I}_M$ is an $M \times M$ identity matrix;
	%$\mathcal{F}\{\cdot\}$ is the continuous Fourier transform.
	%$\textrm{bdiag}(\mathbf{X}_1,\dots,\mathbf{X}_N)$ is a block diagonal matrix having the matrices $\mathbf{X}_1,\dots,\mathbf{X}_N$ on its diagonal.
	
	% \vspace{-1em}
	
	\begin{figure}[t]		\centering %\setcounter{figure}{0}
		\includegraphics[width=1.0\linewidth]{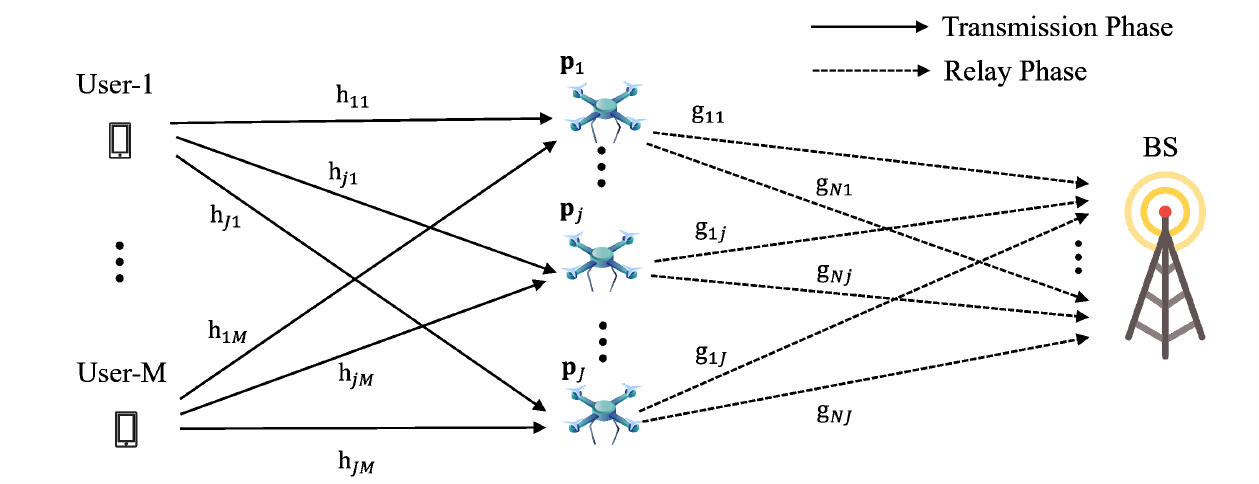}
		\caption{
			The multi-user wireless communication aided by multiple single-antenna UAV relays. 
            % \nm{instead of phase 1 or 2, can you use transmission phase and relay phase?}
		}
		\label{fig:two_hop_channel_block}
		% \vspace{-3mm}
	\end{figure}

	% \vspace{-2mm}
	\section{System Model}\label{sec_system_model}
	%\nt{djl:XXXX 1) When using two indices, I would try to use a comma between them.
	%	2) I would prefer to use H or G as the channel realization.  This is the customary variable.  It is a really bad idea to use f as a channel.  f is universally used for frequency.
	%	3) I would use $i$ or $j $ for $\sqrt{-1}$.
	%	4) You do not really intrdouce the concept of a multiuser linear receiver.  You skip straight to the definition of $\mathbf{B}$.  I would take a paragraph to explain it.
	%	5) I am VERY confused with why the K factors $K_1$ and $K_2$ are defined.  The exponential function is very confusing.  You cite [10], but this is not a common way to define the K factor.  I think you will need a figure and much more explanation.
	%	6) The discussion of LOS and NLOS in the small-scale fading discussion is confusing.  
	%	XXXX }
	
	We consider an uplink multi-user (MU) wireless network, depicted in Fig. \ref{fig:two_hop_channel_block}, where $M$ single-antenna users are serviced by a base station (BS) employing $N$ receive antennas with $N\geq M$.	
	The algorithms in this work are also applicable to downlink multi-user wireless networks, with the roles of the first-hop and second-hop channels reversed.	
	%	The significant path loss is caused by long propagation distances or the blockages between users and BSs, so we assume no available direct link in our considered scenario.
	%Due to the significant path loss caused by long propagation distances\nm{not necessarily long distances.. might be due to other conditions such as blockages}, we consider a negligible\nm{not sure what a negligible direct link is..} direct link between the users and the BS. 
	%	The wireless communication is assisted by multiple single-antenna UAV relays, which follows the AF relay protocol involving two distinct phases.
	The network deployment involves $J$ UAVs, each equipped with a single-antenna relay, to support the uplink communications. 
	We denote $\mathcal{M}=\{1,\dots,M\}$, $\mathcal{N}=\{1,\dots,N\}$, and $\mathcal{J}=\{1,\dots,J\}$ as the set of users, the set of BS antennas, and the set of UAVs, respectively.
    We denote the coordinates of UAV $j$ as $\mathbf{p}_j\in\mathbb{R}^{3}$, the coordinate of user $m$ as $\mathbf{u}_m\in\mathbb{R}^{3}$, and the coordinate of the $n$-th BS antenna as $\mathbf{q}_n\in\mathbb{R}^{3}$.
	% The relay\nm{Each UAV relay} 
    Each UAV relay follows the amplify-and-forward (AF) relay protocol involving two distinct phases, so the entire procedure can be viewed as a two-hop uplink communication. %, depicted in Fig. \ref{fig:two_hop_channel_block}.
	%\nm{out of order.. this should go earlier. please organize the section in a more logical way}
	In the first phase (i.e., the transmission phase in Fig. \ref{fig:two_hop_channel_block}), $M$ users simultaneously transmit their respective signals to $J$ UAV relays. 
	In the second phase (i.e., the relay phase in Fig. \ref{fig:two_hop_channel_block}), the $J$ UAV relays forward the signal to the BS, where the process involves multiplying the incoming signal with a relay beamforming weight, following the AF strategy. 
	The BS receives the signal using $N$ BS antennas. 
    All users transmit the signal operating over the same time and frequency channel resources, thus causing interference with each other's received signal.   
 %    \nm{I would make this a remark...}
 %    \tcadd{Note that this work focuses on the design in underloaded multi-user (MU) scenarios, where the number of users $M$ does not exceed the number of UAV relays $J$ or the number of BS antennas $N$, i.e., $M \leq J$ and $M \leq N$. 
	% In this manner, the receive combining technique can be applied to improve the SINR.
 %    In overloaded MU scenarios (i.e., $M > J$ or $M > N$), significant inter-user interference leads to poor SINR performance, so the user scheduling is necessary to divide users into multiple groups, serving each group independently in either the time domain or frequency domain. 
 %    The joint design of the UAV placement algorithm and user scheduling is an interesting future research direction, but it is beyond the scope of this work.} 
    % \nm{...and I would add a remark regarding the time delay here. Or a high level comment, pointing to a more detailed remark you will make later on}

    \begin{remark}
    \label{overloaded}
        Our work focuses on designing systems for underloaded MU scenarios, where the number of users $M$ does not exceed the number of UAV relays $J$ and the number of BS antennas $N$, i.e., $M \leq J$ and $M \leq N$. 
        In overloaded MU scenarios, i.e., $M > J$ or $M > N$, significant inter-user interference degrades SINR performance, so user scheduling is essential to divide users into multiple groups, with each group served independently in the time or frequency domain.
        \tcminor{% \nm{why are you separating this remark? This should be included in remark 1.}
        User scheduling in overloaded MU scenarios has been investigated in \cite{wu2018joint, wang2022resource,gao2024irs}, addressing different objectives in relay-aided communications.
        User scheduling can be modeled as a constraint in the optimization problem by utilizing binary variables to denote users within a group, served independently across distinct time slots. 
        Incorporating these binary constraints into the optimization problem leads to a mixed-integer programming problem, that is challenging to solve.
        % \nm{rephrase to avoid any confusion: Incorporating these binary constraints into the optimization problem leads to a mixed-integer programming problem, that is challenging to solve.}
        % The optimization problem incorporates this binary constraint, leading to a mixed-integer programming problem that is challenging to solve.
        The joint optimization of UAV placement, transceiver design, and user scheduling in overloaded systems is an interesting future research direction but lies beyond the scope of this work.}
    \end{remark}

	\subsection{Signal Model}
	%\nt{Describe one-way two-hop systems. (2 phases)}
	
	%\nt{Signal model of the first phase}
	Given the transmitted signal of user $m$ as $x_m\in\mathbb{C}$, the received signal at UAV $j$ is denoted as
	\begin{equation*}
%	\vspace{-1mm}
	{r}_{j}=\sum_{m=1}^{M}h_{jm}x_m + \nu_{j},
%	\vspace{-1mm}	
	\end{equation*}
	%\nm{separate line for eq}${r}_{j}=\sum_{m=1}^{M}h_{jm}x_m + \nu_{j}$,
	where $h_{jm}$ is the first-hop channel from user $m$ to UAV $j$.
	The additive noise at UAV $j$ is an independent and identically distributed (i.i.d.) complex Gaussian random variable, denoted as $\nu_j\!\sim\!\mathcal{CN}(0,\sigma_{\nu}^2)$.
	%	The additive noise at UAV $j$ is denoted as $\nu_j$, which is an independent and identically distributed (i.i.d.) complex Gaussian variable with zero mean and variance $\sigma_{\nu}^2$. %, i.e., $v_j\sim\mathcal{CN}(0,\sigma_v^2)$.
	User $m$ transmits the signal with a fixed transmit power, i.e., $\mathbb{E}[\lvert x_m\rvert^2]\!=\!P_t$. %, i.e., $\mathbb{E}[\lvert x_m\rvert^2]=P_t$.
	The transmitted signal of each user is uncorrelated with those of other users.
%	Each user's transmitted signal is uncorrelated with the other user's transmitted signal.
%	Note that the UAVs are assumed to be time aligned such that the signals all arrive at the BS at the same time instant.	
%	\tcadd{We assume the signals of different users arrive at UAVs at the same time. Note that this assumption can be realized on the channel per tone in the orthogonal frequency division multiplexing system with sufficient cyclic prefix.}
	%the unit transmit power of each user, $\mathbb{E}[\lvert s_m\rvert^2]=1$, $m\in\mathcal{M}$.
	%each user transmits the signal with unit transmit power, $\mathbb{E}[\lvert s_m\rvert^2]=1$, $m\in\mathcal{M}$.    
	By stacking $r_j$ at all UAVs, we have
%	 across a column vector $\mathbf r$, we have 
	%\nm{the notation x might confuse readers (tpyicalluy used for tx signals) use r instead?}
	\begin{equation*}
	\mathbf{r} = \mathbf{H} \mathbf{x} + \boldsymbol{\nu},
	\end{equation*}
	where $\mathbf{x}\!=\![{x}_{1},\dots,{x}_{M}]^{\top}$, $\mathbf{r}=[{r}_{1},\dots,{r}_{J}]^{\top}$, $\boldsymbol{\nu}\!=\![\nu_{1},\dots,\nu_{J}]^{\top}$, and $[\mathbf{H}]_{j,m}=h_{jm}$.
	%Superscripts $(\cdot)^\top$ is the transpose of a vector or a matrix.
%	The first-hop channel matrix $\mathbf{H}\in\mathbb{C}^{J\times M}$ is constructed by $[\mathbf{H}]_{j,m}=h_{jm}$.
	%	 $\mathbf{H}=[\mathbf{h}_1,\dots,\mathbf{h}_M]\in\mathbb{C}^{J\times M}$ is constructed by $[\mathbf{H}]_{j,m}=h_{jm}$.

\begin{figure}[t]		\centering %\setcounter{figure}{0}
		\includegraphics[width=1.0\linewidth]{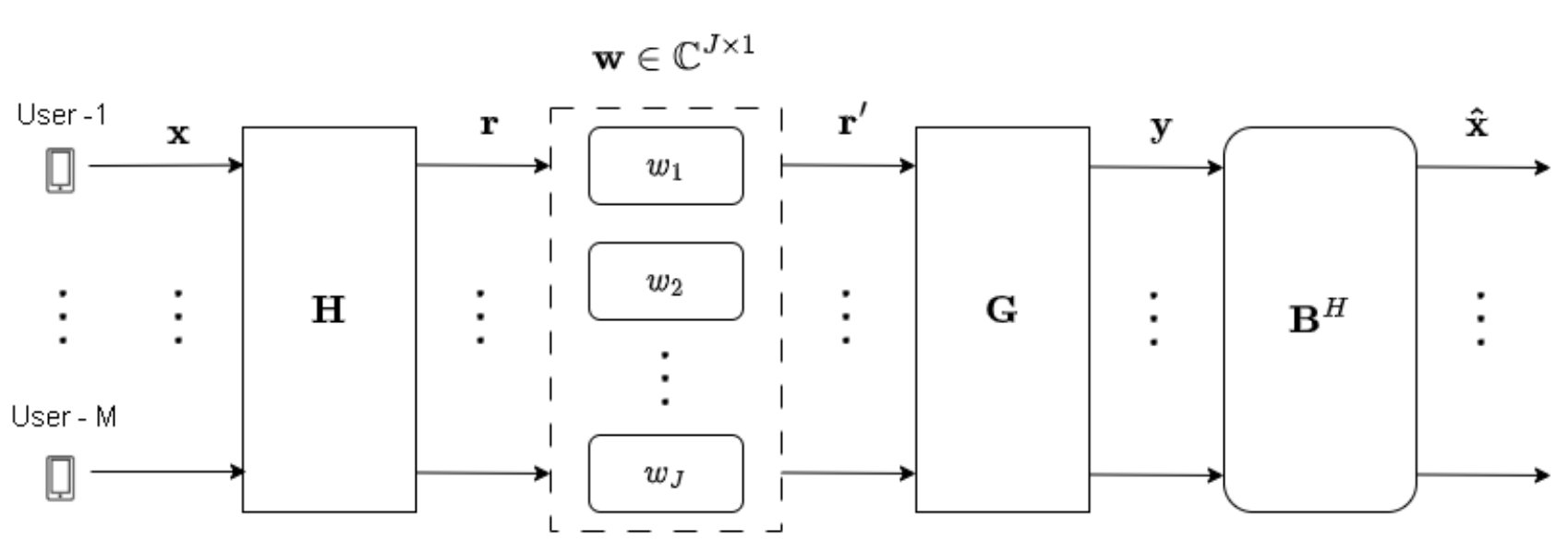}
		\caption{
			Block diagram of the multi-user relay networks.
			% , with the first-hop channel $\mathbf{H}$, second-hop channel $\mathbf{G}$, relay beamforming $\mathbf{w}$, and receive combining $\mathbf{B}$.
		}
		\label{fig:block_diagram}
		% \vspace{-2.5mm}
	\end{figure}

	Next, the UAV relays forward the signal by applying a complex weight at each UAV before transmitting it through the second-hop channel.
	We assume that no direct link between the users and the BS is available due to a significant path loss caused by long propagation distances or blockages between the users and the BS.
	Assuming the beamforming vector across relays is $\mathbf{w}=[w_1,\dots,w_J]^\top\in\mathbb{C}^{J\times 1}$, the received signal vector at the BS is %denoted as
	\begin{align*}
	\mathbf{y}
	=\mathbf{G}\text{Diag}(\mathbf{w})\mathbf{r} + \boldsymbol{\zeta}
	= \mathbf{G}\text{Diag}(\mathbf{w})\mathbf{H}\mathbf{x} + \mathbf{G}\text{Diag}(\mathbf{w})\boldsymbol{\nu} + \boldsymbol{\zeta},
	%	\label{eq_rx_signal}
	\end{align*}
	% \nm{What happens to the direct link? Are you assuming it is blocked? Please explain}\tc{In the beginning of Section II, we already stated that no direct link is available due to a significant path loss caused by long propagation loss or blockages}
	where $\boldsymbol{\zeta}\sim\mathcal{CN}(0,\sigma_{\zeta}^2\mathbf{I})$ is an $ N\times 1$ noise vector at the BS and $\mathbf{G}\in\mathbb{C}^{N\times J}$ denotes the second-hop channel matrix, constructed by $[\mathbf{G}]_{n,j}=g_{nj}$.
	%	 $\mathbf{G}=[\mathbf{g}_1,\dots,\mathbf{g}_J]\in\mathbb{C}^{N\times J}$ denotes the second-hop channel matrix.	
	%	For the estimate of the transmitted symbol $\mathbf{x}$, 
%	\tcadd{
%		We assume that the signals of UAVs all arrive at the BS at the same time, which can be realized by adjusting the transmit signals of UAVs to make them time-aligned in the second phase.}
%		UAVs are assumed to be time-aligned such that the signals all arrive at the BS at the same time instant.}
	To enhance the SINR, we employ a multi-user linear receiver $\mathbf{B}=[\mathbf{b}_1,\dots,\mathbf{b}_M] \in\mathbb{C}^{N\times M}$ to reconstruct the transmit signal $\mathbf x$. 
	Thus, the combined signal associated with user $m$ is given by
	\begin{equation}\label{eq_signal_model}
	{\hat x}_m = \mathbf{b}_m^H\mathbf{y}= \mathbf{b}_m^H\mathbf{G}\text{Diag}(\mathbf{w})\mathbf{H}\mathbf{x} + \mathbf{b}_m^H\mathbf{G}\text{Diag}(\mathbf{w})\boldsymbol{\nu} + \mathbf{b}_m^H\boldsymbol{\zeta}.
	\end{equation}
    % {\nm{In both phases,} 
    In both phases, we assume simultaneous signal arrival at the BS and at the relays, justified by considering the first-hop and second-hop channels as a subcarrier in orthogonal frequency division multiplexing (OFDM) with sufficient cyclic prefix (CP) (see \textbf{Remark \ref{ofdm}}).	
    The overall architecture is depicted in Fig. \ref{fig:block_diagram}.

    \tcadd{
    \begin{remark}
    \label{ofdm}
	In practice, signals may reach the UAVs at different time instants due to different propagation delays or synchronization errors, while we assume the signals arrive simultaneously at the UAVs.
	This assumption is justified by considering the first-hop channel as a subcarrier in OFDM with sufficient CP, so the time delay of the channel is manifested as a phase shift in the frequency domain.
    For the second-hop channel, we also assume that signals from different UAVs arrive at the BS simultaneously, for the same reason. 
    The second-hop channels from different UAVs to the BS may exhibit different delays in the time domain, which are manifested as the sum of phase-shifted signals in OFDM systems.
    \end{remark}}
	
	The receiving SINR of user $m$ is given by
	\begin{align}%\nonumber
	\label{eq_SINR}
	\text{SINR}_{m}(\mathbf{w},\mathbf{b}_m,\mathbf{H},\mathbf{G})
	=\frac{P_{S,m}}{P_{I,m}+P_{N,m}},	
	\end{align}
	%\nm{SINRm is only a function of $\mathbf b_m$ not of the full $\mathbf B$!}
	where $P_{S,m}, P_{I,m}, P_{N,m}$ are the desired signal power, the interference power, and the effective noise power for ${\hat{x}}_m$, respectively, derived as follows.
	%	derived in \eqref{eq_desired_signal_power}, \eqref{eq_interference_signal_power}, \eqref{eq_noise_power}, respectively.
	The desired signal power for ${\hat{x}}_m$ (i.e., the signal power from user $m$) is given by
	% from the ${m}$-th user to the ${m}$-th destination is given by
	\begin{align*}
	% \nonumber
	P_{S,m}\!=\!
	\mathbb{E}\Bigl[\left\lvert\mathbf{b}_m^H\mathbf{G}\text{Diag}(\mathbf{w})\mathbf{h}_m x_m\right\rvert^2\Bigr]\!=\!P_t\left\lvert\mathbf{b}_m^H\mathbf{G}\text{Diag}(\mathbf{w})\mathbf{h}_m \right\rvert^2\!,
	% \label{eq_desired_signal_power}
	\end{align*}
	where $\mathbf{h}_m = [\mathbf{H}]_{:,m}$ is the first-hop channel from user $m$ to the UAV relays.
	%\nm{$\mathbf h_m$ undefined. Define it first and then H.}
	%	The desired signal power can also be expressed as
	%	\begin{equation}
	%		P_{S,m}=\mathbf{w}^H\mathbf{R}_{m}\mathbf{w},
	%	\end{equation} 
	%	where $\mathbf{R}_{m} = P_t\left((\mathbf{b}_m^H\mathbf{G})\odot \mathbf{h}_m^\top\right)^H\left((\mathbf{b}_m^H\mathbf{G})\odot \mathbf{h}_m^\top\right)$.
	The interference power for ${\hat{x}}_m$ (i.e., the signal power received from the users other than user $m$) is given by
	\begin{align*}
	\nonumber
	P_{I,{m}} &= \mathbb{E}\Bigl[\Big\lvert\mathbf{b}_m^H\mathbf{G}\text{Diag}(\mathbf{w}) \sum_{k\in\mathcal{M}\setminus \{m\}}\mathbf{h}_k x_k \Big\rvert^2\Bigr]\\
	&=\sum_{k\in\mathcal{M}\setminus \{m\}}P_t\left\lvert \mathbf{b}_m^H\mathbf{G} \text{Diag}(\mathbf{w}) \mathbf{h}_k\right\rvert^2. 
	%	=\mathbf{w}^H\mathbf{Q}_{m}\mathbf{w}
	% \label{eq_interference_signal_power}
	\end{align*}
	The effective noise power for ${\hat{x}}_m$ is given by
	\begin{align*}
	\nonumber
	P_{N,{m}} &= \mathbb{E}\left[\left\lvert\mathbf{b}_m^H\mathbf{G}\text{Diag}(\mathbf{w})\boldsymbol{\nu} + \mathbf{b}_m^H\boldsymbol{\zeta}\right\rvert^2\right]\\
	&={\sigma_{\nu}^2}\mathbf{b}_m^H\mathbf{G}\text{Diag}(\lvert\mathbf{w}\rvert^2)\mathbf{G}^H\mathbf{b}_m	+ {\sigma_{\zeta}^2}\lVert\mathbf{b}_m\rVert_2^2.
	%\nonumber
	%&=\sigma_v^2\sum_{j=1}^{J}\lvert [\mathbf{H}]_{m,j}w_j\rvert^2+\sigma_n^2[\mathbf{B}]_{:,m}^H[\mathbf{B}]_{:,m}\\
	%	&=\mathbf{w}^H\mathbf{D}_{{m}}\mathbf{w} + {\sigma}_n^2\lVert\mathbf{b}_m\rVert^2,
	% \label{eq_noise_power}
	\end{align*}			
	In addition, the transmitted signal power at UAV $j$ can be derived as $P_{r,j}=\lvert w_j\rvert^2(P_t\sum_{m\in\mathcal{M}}\lvert h_{jm}\rvert^2+\sigma_{\nu}^2)$, which is subject to the power budget $P_{r}$ per UAV relay.

	\subsection{Channel Model}
	%\nm{why are you showing time dependence here? You where not showing time dependence before. Be consistent. Remove time dependence?}
	% For the first-hop channel $\mathbf{H}\in\mathbb{C}^{J\times M}$, we denote\nm{you already defined it} the channel coefficient \cite{you20193d,kang20203d} between user $m$ and UAV $j$ as 
 %    \nm{We model the first-hop channel $h_{jm}$  between user $m$ and UAV $j$ as \cite{you20193d,kang20203d}}
    % \sst{For the first-hop channel $\mathbf{H}\in\mathbb{C}^{J\times M}$,} 
    We model the first-hop channel coefficient $h_{jm}$ between user $m$ and UAV $j$ as \cite{you20193d,kang20203d}
	\begin{align}\label{first_hop_channel_Rician}
	[\mathbf{H}]_{j,m} = h_{jm}=\beta_{jm} \bar{h}_{jm},
	\end{align}
	%\begin{align}\label{first_hop_channel_Rician}
	%f_{jm}(t)=\sqrt{\beta_{1,jm}(t)} f_{jm}^{S}(t),
	%\end{align}
	where $\beta_{jm}$
	%	\nm{I think the superscript makes the notation cumbersome. Why not use just beta here and gamma (or smt else) in place of $\beta^g$ later?} 
	is the large-scale channel coefficient accounting for the path loss in the first-hop channel between user $m$ and UAV $j$ and $\bar{h}_{jm}$ represents the small-scale fading channel component.
	% which follows Friis' transmission formula.
	% The large-scale channel coefficient of the first-hop channel can be modeled as 
 %    \nm{We model the large-scale channel coefficient of the first-hop channel  as }
    We model the large-scale channel coefficient of the first-hop channel as
	\begin{equation}\label{eq_large_scale_channel_gain_first_hop}
	\beta_{jm}=
	\beta_0 \lVert \mathbf{p}_j - \mathbf{u}_m \rVert_2^{-\ell/2},
	\end{equation}
	%\begin{equation}\label{eq_large_scale_channel_gain_first_hop}
	%	\beta_{1,jm}(t) = \beta_0 \lVert \mathbf{p}_j(t) - \mathbf{u}_m(t) \rVert_2^{-\ell},
	%%	d_{1,jm}^{-\ell}(t),
	%\end{equation}
	where $\ell$ is the path loss exponent, $\beta_0$ is the channel gain at a reference distance of $1$ m, and $\lVert \mathbf{p}_j - \mathbf{u}_m \rVert_2$ is the Euclidean distance between user $m$ and UAV $j$.
	%Considering the high altitude of the UAV relays compared to its surroundings, it is reasonable to assume that \hl{the small-scale fading predominantly occurs in the LOS state}\nm{what does this mean?} with a high probability \cite{al2014optimal}.
	%It is reasonable to assume that the UAV swarms has a higher altitude than its surroundings, so the small-scale fading is in the LOS state with high probability \cite{al2014optimal}.
	% To model the small-scale fading channel component in the LOS scenario, we employ the Rician fading model \cite{kang20203d,keshavamurthy2023maestro}, expressed as
    We model the small-scale channel coefficient as Rician fading \cite{kang20203d,keshavamurthy2023maestro}, so that
	\begin{equation}\label{eq_small_scale_fading_ch_first_hop}
	\bar{h}_{jm} = \sqrt{\frac{K_r}{K_r+1}}e^{i\theta_{jm}}+\sqrt{\frac{1}{K_r+1}}{h}^{\prime}_{jm},
	\end{equation}
	where $K_r\geq 0$ denotes the Rician factor and $\theta_{jm}$ denotes the phase shift of the LOS component.
	%	\nm{define ${h}^{\prime}_{jm}$ and K}
	The second term ${h}^{\prime}_{jm}\sim\mathcal{CN}(0,1)$ represents the random scattering channel component.
	Given that the wavelength is typically of a small scale, even a minor variation in the position of the UAV or the user can result in a significant change in the channel phase shift.
	To characterize the impact of such positional variations, the channel's phase shift is modeled as $\theta_{jm}\sim\mathcal{U}(0,2\pi)$ \cite{liu2019comp}.

	For the second-hop channel $\mathbf{G}\in\mathbb{C}^{N\times J}$, 
    % \nm{we assume a LOS dominated channel [REF], hence}
    we assume a LOS dominated channel \cite{yan2019comprehensive}, hence we model the channel coefficient between UAV $j$ and the $N$ BS antennas as
	\begin{equation} \label{second_channel_ULA_form}
	[\mathbf{G}]_{:,j} = c_j \mathbf{v}_j\in\mathbb{C}^{N\times 1},
	\end{equation}
	where $c_j$ is the channel gain between UAV $j$ and the BS antenna array, and $\mathbf{v}_j\in\mathbb{C}^{N\times 1}$ is the unit-norm array response vector.
	%	\nm{can you please provide math expression? Are you assuming a LOS dominated channel for the second hop? Please clarify/justify accordingly.}
	%	\tc{I first derive the }
 %    \nm{this next sentence is a bit ambiguous/confusing. Remove and simply define the array response vector?}
	% \hl{Even if the channel between the UAV and the BS antenna has a random phase, the channel phases from the same UAV to different BS antennas can be correlated. }
	We assume the BS uses a uniform linear array (ULA) with half-wavelength antenna spacing.
	Hence, the array response vector \cite{chou2023compressed} is defined as 
	\begin{equation}
	\mathbf{v}_j=\frac{1}{\sqrt{N}}\left[1, e^{-i\pi\sin(\vartheta_j)}, \dots, e^{-i(N-1)\pi\sin(\vartheta_j)}\right]^\top,
	\end{equation}	
	where $\vartheta_j$ is the angular direction of the LOS path between UAV $j$ and the BS.
	Note that, with a suitable definition of array response vector, our approach is applicable to any antenna array design, not just ULAs.
	The channel gain between UAV $j$ and the BS antenna array is defined as $c_j = \sqrt{N} g_{j}$.
	The baseband channel coefficient $g_{j}$ is modeled as	
	\begin{align} \label{eq_channel_gain_second_hop}
	g_{j}=\gamma_{j} \bar{g}_{j},
	\end{align}
	where $\gamma_{j}$ is the large-scale channel coefficient accounting for the path loss of the second-hop channel between UAV $j$ and the BS antenna array, and $\bar{g}_{j}$ is the small-scale fading channel component.
	Similar to the first-hop channel, the components $\gamma_{j}$ and $\bar{g}_{j}$ are constructed as in \eqref{eq_large_scale_channel_gain_first_hop} and \eqref{eq_small_scale_fading_ch_first_hop}, respectively, by substituting $(\mathbf{p}_j,\mathbf{q}_1)$ in place of $(\mathbf{u}_m,\mathbf{p}_j)$.

	\begin{remark}
		To facilitate analysis, we assume that the users and the BS are stationary. However, they may move in real-world scenarios. Our approach remains applicable as long as their positions change at a slower timescale compared to the time required to acquire the positions of the users and BS, solve the UAV placement optimization problem, and reposition the UAVs in the system.
        % \hl{acquire CSI}\nm{do you need CSI to reposition the uav? I thought you didn't}, 
        
        % \nm{and reposition the UAVs in the system}.
		%		to solve the optimization problem and acquire CSI.
		
		%		\hl{Our approach remains applicable in such scenarios if the positions of the users and the BS can be updated for our proposed UAV placement algorithm.	}\nm{not clear. Please rephrase. I would say: the approach is applicable as long as these positions change at a slower time scale than the time needed to solve the optimization problem and acquire CSI. Can you please provide some rule of thumb for this? What is considered a significant positional change that requires recomputing UAV relay positions? going at x km per hour, how long would it take to move such distance?}
		%		our work is applicable for the scenarios with moving users and BS
		%		For ease of analysis, the users and the BS are assumed to be stationary. In real scenarios, the movement of users and BS is permitted, as long as their positions are updated by the feedback information. \nm{not very clear. Please rephrase.}
	\end{remark}
	
	% \vspace{-5mm}
	\section{Problem Formulation}\label{sec_prob_formulation}
	Our goal is to enhance the minimum beamforming SINR among users in a multi-user network supported by UAV relays, achieved by optimizing the UAV positions $\mathbf{p}_{j},\ j\in\mathcal{J}$, the relay beamforming $\mathbf{w}$, and the receive combining $\mathbf{b}_m,\ m\in\mathcal{M}$.
    A key challenge in optimizing UAV positions lies in the absence of instantaneous CSI prior to UAV deployment, making it difficult to determine the beamforming SINR.
    Hence, we optimize UAV positions based on the expected beamforming SINR, which takes expectation over channel statistics, in which the objective function is the worst expected beamforming SINR among users accounting for the optimized relay beamforming and receive combining.
    % \nm{you also need to clarify that we account for the optimization of the beamforming}.
    Accordingly, we divide our work into two optimization problems, as follows.
    
	% It is challenging to optimize the UAV positions due to the lack of instantaneous CSI before the UAV deployment, making it difficult to determine the beamforming SINR.  
	% Hence, we optimize the UAV positions to improve its expected beamforming SINR, which takes expectation over channel statistics.
	% In this manner, we divide our work into two optimization problems, as follows.

	First, we formulate a \tcadd{\emph{beamforming-aware UAV placement optimization}} which optimizes the UAV positions that maximize the minimum of the expected beamforming SINR among users, described in Section \ref{UAV_movement_opt_formulation}.
	Next, once the UAVs are deployed to their designated positions, we optimize the relay beamforming and receive combining based on the estimated CSI.
	The optimization problem is the \tcadd{\emph{transceiver design for minimum SINR maximization}}, detailed in Section \ref{SINR_maxmin_formulation}.
	
	% \vspace{-3mm}
	\subsection{\tcadd{Beamforming-aware UAV Placement Optimization}}\label{UAV_movement_opt_formulation}
	%\nt{UAV Movement to attain the QoS maximization, based on the worst case.}
	We focus on designing the UAV placement in a multi-user relay network. %, i.e., \text{UAV placement optimization}.
	The goal is to optimize the UAVs' positions to maximize the minimum \emph{expected} beamforming SINR among users.
    % \hl{place the UAVs at the positions}\add{optimize the UAVs' positions} to maximize the minimum \emph{expected} beamforming SINR among users.
	%	Since the receive combining does not have much impact on the scheduling of the UAV positions, we consider the receive combining as an identity matrix to simplify the optimization problem, i.e., the receive combining $\mathbf{B}= \mathbf{I}_M$ so that the effective second hop channel matrix $\mathbf{G}$.
	However, the instantaneous CSI cannot be acquired before the UAVs are deployed to the designated positions.
	%	, \hl{so the small-scale fading CSI is not available at this stage}\nm{repetitive: you just said it on the same line}.\nm{these explanations are not quite clear. I'd say:} 
	%	\add{
 %    \nm{To account for the lack of CSI, we define}
	% For the problem formulation, we define 
	To account for the lack of CSI, we define two functions $\bm{\mathcal{W}}$ and $\bm{\mathcal{B}}_m$, which map the channels $\mathbf H$ and $\mathbf G$, collectively denoted by $\boldsymbol{\varphi}=\{\mathbf{H}, \mathbf{G}\}$,	
	to choices of relay weights $\mathbf w=\bm{\mathcal{W}}(\boldsymbol{\varphi})$ and receive combining $\mathbf{b}_m=\bm{\mathcal{B}}_m(\boldsymbol{\varphi}),\forall m=1,\dots,M$, respectively. 
	With these definitions, we define the expected SINR for user $m$ as
	$\mathbb E_{\boldsymbol{\varphi}}[\text{SINR}_{m}(\bm{\mathcal{W}}(\boldsymbol{\varphi}),\bm{\mathcal{B}}_m(\boldsymbol{\varphi}),\boldsymbol{\varphi})],$
	where $\text{SINR}_{m}$ is given as in \eqref{eq_SINR}, and the expectation is taken with respect to the small-scale fading statistics. 
    This expected SINR is a function of the relay positions $\mathcal{P}$ as well as the relay weights and receive combining maps ($\bm{\mathcal{W}}$, $\bm{\mathcal{B}}_m$), all of which need to be jointly optimized. 
    This approach enables optimization of the expected SINR without requiring knowledge of \emph{instantaneous} CSI, relying instead on the fading statistics necessary to evaluate the expectation.
	%	}

	%	We know that the channel conditions of the first-hop and second-hop channels are dependent on the UAV relay's position (as in \eqref{eq_large_scale_channel_gain_first_hop} and \eqref{eq_large_scale_channel_gain_second_hop}).
	%	However, the UAV positions $\mathcal{P}$ would simultaneously vary the first-hop channel and second-hop channel, which makes this a difficult optimization problem.
	
	%	\nm{not clear. You are maximizing over positions the "optimized" SINR, optimized over beamforming.}
	In this work, we particularly aim at maximizing the minimum expected beamforming SINR among users, where the maximization is with respect to the UAV positions and the maps $\bm{\mathcal{W}}$ and $\bm{\mathcal{B}}_m$. This optimization problem is expressed as
	%	To address the above issue, we formulate an optimization problem to maximize the objective function over UAV relay positions $\mathcal{P}$, expressed as 
	%	\add{expressed as}\sst{depicted as below:}
	%	 the minimum of ``expected SINR'' among users, depicted below:
	%	\nm{please use the notation $\bm{\mathcal{W}}$ for consistency!}
	\begin{align}\label{opt_uav_placement}
	&\max_{\mathcal{P}}\left\{\max_{\bm{\mathcal{W}},\bm{\mathcal{B}}_m}
	\min_m\mathbb{E}_{\boldsymbol{\varphi}}\left[ \text{SINR}_{m}(\bm{\mathcal{W}}(\boldsymbol{\varphi}),\bm{\mathcal{B}}_m(\boldsymbol{\varphi}),\boldsymbol{\varphi})\right]\right\},\\
	\label{constraint_UAV_relay_pwr_V1}
	&\text{s.t. }
	\lvert {\bm{\mathcal{W}}_j(\boldsymbol{\varphi})}\rvert^2\left(P_t\sum_{m\in\mathcal{M}}\lvert h_{jm}\rvert^2+\sigma_{\nu}^2\right)\leq  {P_{r}},\ \forall j\in \mathcal{J}, {\ \forall\boldsymbol{\varphi}};\\
	\label{constraint_UAV_relay_distance_V1}
	&\hspace{1.2em}\lVert\mathbf{p}_j - \mathbf{p}_{j^\prime}\rVert_2\geq\epsilon_p,\ j\neq j^\prime, \ \forall j,j^\prime \in \mathcal{J},
	\end{align}
	%	\nm{fix below based on my rephrasing}
	where $\mathcal{P}=\{\mathbf{p}_{1},\dots,\mathbf{p}_{J}\}$ is the set of $J$ UAV positions
 %    \sst{. 
	% %	We denote the relay beamforming vector and receive combining as
	% %	\begin{align}
	% %	\boldsymbol{\mathcal{w}} &= \boldsymbol{\mathcal{w}}(\mathbf{H},\mathbf{G}),\\
	% %	\bm{\mathcal{B}}=\bm{\mathcal{B}}(\mathbf{H}, \mathbf{G})&=\left[\boldsymbol{b}_1(\mathbf{H},\mathbf{G}),\dots,\boldsymbol{b}_M(\mathbf{H},\mathbf{G})\right],
	% %	\end{align}
	% %	respectively, which are functions of {the CSI} $(\mathbf{H}, \mathbf{G})$ instead of fixed-valued variables since the CSI is unknown at the time of UAV deployment here.
	% The objective function is the minimum of the expected beamforming SINR among users, in which the beamforming SINR is maximized over $(\bm{\mathcal{W}},\bm{\mathcal{B}}_m)$ }\nm{Repetitions of what you just said before eq 9}
    % $(\boldsymbol{\mathcal{w}}\nm{notation},\bm{\mathcal{B}}_m)$ 
    and the expectation is taken with respect to the first-hop and second-hop channels.
	%	, denoted by $\boldsymbol{\varphi}=\{\mathbf{H}, \mathbf{G}\}$.	
	%	The objective function is the expected SINR of user-$m$,
	%	\nm{inaccurate. Explain in words what the objective is.}
	%	which takes expectation with respect to the first-hop and second-hop channels, denoted by $\boldsymbol{\varphi}=\{\mathbf{H}, \mathbf{G}\}$.
	%	the small-scale fading effect of the channel $\mathbf{H}$ and $\mathbf{G}$, denoted by $\boldsymbol{\varphi}=\{\mathbf{H}, \mathbf{G}\}$
	\tcadd{The constraint \eqref{constraint_UAV_relay_pwr_V1} is the power constraint on the transmission of each UAV relay, where $P_{r}$ is the power budget per UAV relay which needs to be enforced for every possible realization of the first-hop channel.
    The constraint \eqref{constraint_UAV_relay_distance_V1} models collision avoidance, where $\epsilon_p$ denotes the minimum distance between UAVs.
    This problem will be addressed by the UAV placement algorithm in Section \ref{sec_UAV_placement_optimization}}.

	\subsection{\tcadd{Transceiver Design for Minimum SINR Maximization}}
	% \subsection{Joint Relay Beamforming and combining Optimization}
	\label{SINR_maxmin_formulation}
	With UAVs deployed to the given positions, e.g., the positions optimized in \eqref{opt_uav_placement}, the instantaneous channel can be estimated using pilot-based channel estimation techniques, such as the ones developed in \cite{ma2011pilot}.
    \tcadd{This work assumes perfect CSI for designing relay beamforming and receive combining. 
    Although channel estimation errors may lead to a degradation of beamforming SINR, this issue can be mitigated by employing a sufficiently long pilot sequence at the cost of increased training overhead.}
    % Here, we assume the perfect estimated CSI for designing the relay beamforming and receive combining.
    % Although it would lead to degradation in beamforming SINR due to channel estimation errors, this can be mitigated by using a sufficiently long pilot sequence with the cost of a larger training overhead.
	%	\nm{such as the ones developed in ...}.
	%	Note that numerous cascaded channel estimation methods for relay-assisted scenarios have been explored in prior studies; however, they fall outside the scope of this current work.
	%When the UAVs are deployed to the given positions and being stationary, the CSI can be estimated using pilot-based channel estimation techniques.
	Given the estimation of $\mathbf{H}$ and $\mathbf{G}$, we proceed with the design of the relay beamforming $\mathbf{w}$ and receive combining $\mathbf{B}$ to improve the SINR, given by
	%	We seek to maximize ``the minimum SINR among users'' by designing the relay beamforming $\mathbf{w}$ and receive combining $\mathbf{B}$, which is a max-min optimization problem, formulated as
	% known as \textbf{SINR Maximization Optimization}, formulated as
	% By having access to the small-scale CSI, we can then proceed with the design of the relay beamforming $\mathbf{w}$ and receive combining $\mathbf{B}$ to improve the QoS of communication systems, i.e., \textbf{SINR Maximization Optimization}, formulated as
	\begin{align}\label{opt_max_min_modified}
	&\max_{\mathbf{w}, \mathbf{B}}\min_m\ \text{SINR}_{m}(\mathbf{w},\mathbf{b}_m,\mathbf{H},\mathbf{G}),\\
	&\text{s.t. }\lvert w_j\rvert^2\left(P_t\sum_{m\in\mathcal{M}}\lvert h_{jm}\rvert^2+\sigma_{\nu}^2\right)\leq P_{r},\ j\in \mathcal{J}.
	\label{constraint_relay_pwr_v1}
	\end{align}
	% where $\mathbf{B}=[\mathbf{b}_1,\dots,\mathbf{b}_M]$ denotes the multi-user linear receiver.
	\tcadd{The objective function is the minimum of the beamforming SINR among users, and \eqref{constraint_relay_pwr_v1} is the relay power constraint.
    This problem will be solved by the joint relay beamforming and receive combining algorithm in Section \ref{subsec_joint_opt_B_w}.}
	
	Note that the relay beamforming design without a receive combining based on instantaneous CSI, i.e., the problem \eqref{opt_max_min_modified} with $\mathbf{B}=\mathbf{I}_M$ and $N=M$, has been addressed in \cite{phan2013iterative,che2014joint}.
	%	\nm{do these work address the lack of CSI? Based on how you write it, it seems so..}\tc{These previous works proposed the relay beamforming design based on the perfect CSI. This subsection is about the relay beamforming and receive combining design. The lack of CSI issue only happens at beamforming-aware UAV placement optimization.}
	Our work extends these works by formulating
    % \add{these works by formulating}\sst{ to formulate} 
    an optimization problem that jointly adjusts the relay beamforming and receive combining, which will be addressed in Section \ref{subsec_joint_opt_relay_bf_and_rx_combining}.
	We will numerically compare our proposed algorithm with these works \cite{phan2013iterative,che2014joint} and show the superior beamforming performance of our work.
	%	\nm{I don't think this novelty has been properly highlighted in the intro!}
	
	%	\nm{please highlight the differences between 12 and 15! I.e., expected SINR in 12 vs actual instantaneous SINR in 15.}
    
	The primary distinction between \eqref{opt_uav_placement} and \eqref{opt_max_min_modified} lies in the availability of instantaneous CSI.
	\tcadd{In UAV placement optimization, we lack the instantaneous CSI of the two-hop channels $(\mathbf{H}, \mathbf{G})$ before deploying UAVs to designated positions.
    Hence, the relay beamforming and receive combining strategies must be designed as functions that map any possible CSI realization to corresponding beamforming and combining weight choices. 
    The objective of the UAV placement optimization problem is to adjust UAV positions to enhance the minimum of the expected beamforming SINR among users based on the statistical CSI, as in \eqref{opt_uav_placement}.	
    After UAV deployment, when estimated CSI becomes available, we can further optimize the transceiver $(\mathbf{w},\mathbf{B})$ to improve the minimum SINR among users, as in \eqref{opt_max_min_modified}.
    In this case, since we seek to optimize the relay beamforming and receive combining for a specific channel realization, we formulate the optimization problem with the decision variables $(\mathbf{w},\mathbf{B})$ instead of their function expressions.}
    
    % At this stage, we have the estimation of instantaneous CSI $(\mathbf{H}, \mathbf{G})$, we could formulate the optimization problem using the decision variables $(\mathbf{w},\mathbf{b}_m)$.
    
    % Note that, for the UAV placement optimization, since we consider the expected beamforming SINR as the objective function, we formulate the optimization problem based on the functions $(\bm{\mathcal{W}},\bm{\mathcal{B}}_m)$ so that we could have the relay beamforming and receive combining for each channel realization.
    % For the transceiver design, as we have the estimation of instantaneous CSI $(\mathbf{H}, \mathbf{G})$, we could formulate the optimization problem using the decision variables $(\mathbf{w},\mathbf{b}_m)$.
    
    % The UAV placement optimization is developed based on the function $(\bm{\mathcal{W}},\bm{\mathcal{B}}_m)$ because its objective function is the expected beamforming SINR
    % \nm{question: why cannot one simply use the functions W and B to do this? You need to clarify this..is it because of the use of approximations.}
	
	%	after UAV are placed at the target positions, the channels $(\mathbf{H}, \mathbf{G})$ can be estimated via pilot-based channel estimation, thus we can consider the instantaneous SINR as the objective function.

		% \vspace{-1em}
	\section{UAV Placement Algorithm}\label{sec_UAV_placement_optimization}
	We propose a UAV placement algorithm for the optimization problem \eqref{opt_uav_placement} to maximize the minimum expected beamforming SINR among users.
	Since the instantaneous channels are unavailable before the UAV deployment, we can only rely on the expected beamforming SINR, with optimized relay beamforming and receive combining strategies, as the objective function to optimize the UAV positions.
    % \nm{, with optimized relay and receive combining strategies,} as the objective function to optimize the UAV positions.
    However, it is difficult to express the expected beamforming SINR in closed form, with optimized relay beamforming and receive combining strategies.
    To address this issue, we derive a lower bound on the optimized expected beamforming SINR, which serves as a surrogate objective function for position optimization, as in Section \ref{subesection_UAV_placement_reformulation}.
    % \nm{what issue? I think the issue is the difficulty of expressing in closed form the optimization over relayand cominingh strategies. please clarify}, 
    % \nm{we derive a lower (is it lower?) bound to the optimized expected beamforing SINR, serving as a surrogate objective functions for position optimization, as in...}
    % we derive the expected beamforming SINR and introduce a substitute form to serve as a new objective function, as in Section \ref{subesection_UAV_placement_reformulation}.
	% In Section \ref{subesection_UAV_placement_reformulation},  we first compute the expected beamforming SINR and introduce a substitute form to serve as a new objective function. 
    Then, we formulate a new optimization problem as in \eqref{opt_prob_to_be_solved}. 
    In Section \ref{susection_DC_approach}, we propose a UAV placement algorithm that employs a DC framework with convex relaxations to solve the optimization problem \eqref{opt_prob_to_be_solved}.

    % \nt{Add a paragraph to detail the procedure of solving the UAV placement optimization problem}
    
	% \vspace{-4mm}
	
	\subsection{Reformulation of UAV Placement Optimization}
	\label{subesection_UAV_placement_reformulation}
	To tackle the complicated objective function in \eqref{opt_uav_placement}, we address the following subproblem, expressed for a given set of UAVs' positions, as
    % \sst{with the given UAV relay positions, given by}\add{, expressed for a given set of UAVs' positions as}
	\begin{equation}\label{eq_max_min_exp_sinr_opt}
	\max_{\bm{\mathcal{W}},\bm{\mathcal{B}}_m}
	\min_m\mathbb{E}_{\boldsymbol{\varphi}}\left[ \text{SINR}_{m}(\bm{\mathcal{W}}(\boldsymbol{\varphi}),\bm{\mathcal{B}}_m(\boldsymbol{\varphi}),\boldsymbol{\varphi})\right], \text{ s.t. }\eqref{constraint_UAV_relay_pwr_V1}.
	\end{equation}
	%Letting $\mathbf{w} = f_w(\mathbf H, \mathbf G)$ and $\mathbf{B} = f_{B}(\mathbf H, \mathbf G)=[f_{b_1}(\mathbf H, \mathbf G),\dots,f_{b_M}(\mathbf H, \mathbf G)]$, we have
	%\begin{equation}
	%	\max_{f_w, f_B}	\min_m\mathbb{E}_{\boldsymbol{\varphi}}\left[ \text{SINR}_{m}(f_w(\mathbf H, \mathbf G),f_{b_m}(\mathbf H, \mathbf G),\mathbf{H},\mathbf{G})\right]
	%\end{equation}
	%For ease of exposition, we use $\mathbf{w}$ and $\mathbf{B}$ as in \eqref{eq_max_min_exp_sinr_opt} for the following derivation, while 
	%	It is worth noting that the variables $\bm{\mathcal{W}}$ and $\bm{\mathcal{B}}_m$ are functions of $\boldsymbol{\varphi}=\{\mathbf{H},\mathbf{G}\}$.
	%We would like to show that
	%\begin{align}%\nonumber
	%	&\max_{f_w, f_B}	\min_m\mathbb{E}_{\boldsymbol{\varphi}}\left[ \text{SINR}_{m}(f_w(\mathbf H, \mathbf G),f_{b_m}(\mathbf H, \mathbf G),\mathbf{H},\mathbf{G})\right] \\
	%	\equiv 	\min_m\mathbb{E}_{\boldsymbol{\varphi}}\left[ \max_{f_w}\max_{f_B}\text{SINR}_{m}(f_w(\mathbf H, \mathbf G),f_{b_m}(\mathbf H, \mathbf G),\mathbf{H},\mathbf{G})\right]
	%\end{align}
	%	Since the linear receiver $\mathbf{b}_m=[\mathcal{B}]_{:,m}$ is designed for each user respectively, \eqref{eq_max_min_exp_sinr_opt} can be reformulated as
	%	\nm{is it equivalent? How did you move the max inside the expectation? Please explain}
	Since $\bm{\mathcal{B}}_m$ can be designed for each user $m$ separately and $\bm{\mathcal{B}}_m$ of a certain user does not affect the SINR of another user,
	%	\nm{and the bm of a certain user does not affect the SINR of another user,}, 
	we can switch the order between the optimizations of $\max_{\bm{\mathcal{B}}_m}$ and $\min_m$.	
	Thus, we have the equivalent form of \eqref{eq_max_min_exp_sinr_opt} as
	\begin{equation}\label{eq_opt_new_V1b}
	\max_{\bm{\mathcal{W}}}	\min_m\max_{\bm{\mathcal{B}}_m}\mathbb{E}_{\boldsymbol{\varphi}}\left[ \text{SINR}_{m}(\bm{\mathcal{W}}(\boldsymbol{\varphi}),\bm{\mathcal{B}}_m(\boldsymbol{\varphi}),\boldsymbol{\varphi})\right], \text{ s.t. }\eqref{constraint_UAV_relay_pwr_V1}.
	\end{equation}
	$\bm{\mathcal{B}}_m$ is the function which maps each channel realization $\boldsymbol{\varphi}$ to the combining vector $\mathbf{b}_m$, i.e., $\mathbf{b}_m=\bm{\mathcal{B}}_m(\boldsymbol{\varphi})$.
    \tcadd{For each channel realization $\boldsymbol{\varphi}$, the function $\bm{\mathcal{W}}$ maps to the relay beamforming vector $\mathbf{w}=\bm{\mathcal{W}}(\boldsymbol{\varphi})$ with the constraint \eqref{constraint_UAV_relay_pwr_V1}. The largest beamforming SINR of user $m$ can be obtained for every possible channel realization by the maximization over $\mathbf{b}_m=\bm{\mathcal{B}}_m(\boldsymbol{\varphi})$.}
    % \st{Since the constraint \eqref{constraint_UAV_relay_pwr_V1} needs to be enforced for each channel realization, the largest beamforming SINR of user $m$ is obtained by the maximization over $\mathbf{b}_m=\bm{\mathcal{B}}_m(\boldsymbol{\varphi})$ for every possible channel realization.}	
	Thus, we move $\mathbf{b}_m$ into the expectation in \eqref{eq_opt_new_V1b}, so that the optimization problem can be expressed as
    % \sst{given by }\add{so that the optimization problem can be expressed as}
	\begin{equation}\label{eq_opt_new_V1c}
	\max_{\bm{\mathcal{W}}}	\min_m\mathbb{E}_{\boldsymbol{\varphi}}\left[\max_{\mathbf{b}_m} \text{SINR}_{m}(\bm{\mathcal{W}}(\boldsymbol{\varphi}),\mathbf{b}_m,\boldsymbol{\varphi})\right], \text{ s.t. }\eqref{constraint_UAV_relay_pwr_V1}.
	\end{equation}
	%	The receive combining $\mathbf{b}_m$ can now be designed for each channel realization, thus the receive combining $\mathbf{b}_m$  is viewed as a complex-valued vector.
	For a given channel realization $\boldsymbol{\varphi}$ and $\mathbf w=\bm{\mathcal{W}}(\boldsymbol{\varphi})$,  we first solve the problem within the expected function expressed as
	\begin{equation}\label{eq_subproblem_receive_combining}
	\max_{\mathbf{b}_m}\text{SINR}_{m}(\mathbf{w},\mathbf{b}_m,\boldsymbol{\varphi}). %\nm{check punctuation of equations}
	\end{equation}
%	Given the SINR form in \eqref{eq_SINR}, we have the SINR of user $m$ as	
	%	To solve this subproblem in the expected function, we first derive the SINR of user-$m$ as
	%	\nm{please refer to equations shown earlier to explain Eq 20}
	The SINR of user $m$ in \eqref{eq_SINR} can be reformulated as
	\begin{align}\nonumber
	&\text{SINR}_m(\mathbf{w},\mathbf{b}_m,\boldsymbol{\varphi})=\\
	&\frac{P_t\mathbf{b}_m^H \left[\mathbf{a}_m\mathbf{a}_m^H\right]\mathbf{b}_m}
	{\mathbf{b}_m^H\left[P_t\sum_{k\in\mathcal{M}\setminus \{m\}}\mathbf{a}_k \mathbf{a}_k^H + \sigma_{\nu}^2\mathbf{G}\text{Diag}(\lvert\mathbf{w}\rvert^2)\mathbf{G}^H+\sigma_{\zeta}^2\mathbf{I}\right]\mathbf{b}_m},
	\label{SINRm_V2}
	\end{align}
	where $\mathbf{a}_m = \mathbf{G}\text{Diag}(\mathbf w)\mathbf{h}_m$ and ${|\mathbf{w}|^2} = [{|w_1|^2},\dots,{|w_J|^2}]^\top$.
	The maximization of \eqref{SINRm_V2} is a Rayleigh quotient problem \cite{golub2013matrix} for $\mathbf{b}_m$, whose maximizer has a closed-form solution given by
	\begin{equation}\label{eq_combining_opt}
	\mathbf{b}_m^*=\Bigg[P_t\sum_{k\in\mathcal{M}\setminus \{m\}}\mathbf{a}_k \mathbf{a}_k^H + \sigma_{\nu}^2\mathbf{G}\text{Diag}(\lvert\mathbf{w}\rvert^2)\mathbf{G}^H+\sigma_{\zeta}^2\mathbf{I}\Bigg]^{-1}\mathbf{a}_m.
	\end{equation}
	%	\add{Replacing this solution into REF leads to the maximized SINR expression give by}
	Substituting this maximizer $\mathbf{b}_m^*$ into \eqref{SINRm_V2} leads to the maximized SINR form, which is the solution of \eqref{eq_subproblem_receive_combining}, given by
	\begin{align}\nonumber
	&{\text{ SINR}}_m^{\prime}(\mathbf{w},\boldsymbol{\varphi})
	=\max_{\mathbf{b}_m} \text{ SINR}_m(\mathbf{w}, \mathbf{b}_m,\boldsymbol{\varphi})\\ %=\text{ SINR}_m(\mathbf{w}, \mathbf{b}_m^*,\boldsymbol{\varphi})\\
	\label{maxSINR_over_B}
	&=\mathbf{a}_m^H\left[\!\sum_{k\in\mathcal{M}\setminus \{m\}}\!\mathbf{a}_k \mathbf{a}_k^H \!+\! \frac{\sigma_{\nu}^2}{P_t}\mathbf{G}\text{Diag}(\lvert\mathbf{w}\rvert^2)\mathbf{G}^H\!+\!\frac{\sigma_{\zeta}^2}{P_t}\mathbf{I}\right]^{-1}\mathbf{a}_m.
	\end{align}
	Given ${\text{SINR}}_m^{\prime}(\mathbf{w},\boldsymbol{\varphi})$ maximized over $\mathbf{b}_m$, we have a new form of the optimization problem \eqref{eq_opt_new_V1c}, given by
	\begin{equation}\label{eq_opt_new_V1d}
	\max_{\bm{\mathcal{W}}}	\min_m\mathbb{E}_{\boldsymbol{\varphi}}\left[{\text{ SINR}}_m^{\prime}(\bm{\mathcal{W}}(\boldsymbol{\varphi}),\boldsymbol{\varphi})\right], \text{ s.t. }\eqref{constraint_UAV_relay_pwr_V1}.
	\end{equation}
    % \nm{What about this? I think it follows a more logical flow and provides some key missing steps (I was having a hard time understanding/remembering what we did while re-reading the paper after a long time)}

% \vspace{-1mm}
    However, the solution to this problem is intractable in closed form due to the intricate dependence between the relay weights and the SINR values. To achieve a tractable solution suitable for position optimization, we now introduce approximations that enable a closed-form expression.
    
	Specifically, assuming a significantly larger number of BS antennas compared to UAV relays, i.e., when $N \gg J$, we achieve ample angular resolution with the BS antenna array and ensure adequate spatial separation among UAVs, leading to $\mathbf{v}_j^H\mathbf{v}_{j^\prime}=0$ for $j\neq j^\prime$.
	% \begin{align}\label{V_orthogonal_property}
	% \mathbf{v}_j^H\mathbf{v}_{j^\prime} = 
	% \begin{cases}
	% 1 & \text{if $j= j^\prime$} \\
	% 0 & \text{if $j\neq j^\prime$},
	% \end{cases}
	% \end{align}
	% for $j,j^\prime \in\mathcal{J}$.	
	Such orthogonality property holds under the constraint \eqref{constraint_UAV_relay_distance_V1} if $\epsilon_p$ is greater than the beamwidth, which is inversely proportional to the number of BS antennas.

    \begin{figure*} [hb!]%[!ht]%[!htb]
		%	\begin{strip}[b]
		%		\noindent\rule{\textwidth}{0.75pt}	
		\hrulefill	
        % \vspace{-1mm}
		\begin{equation} \tag{38}\label{eq_lower_bound}
		\Gamma_m(\mathcal{P})
		=\sum_{j\in\mathcal{J}}\frac{\beta_0^2 \lVert \mathbf{p}_j - \mathbf{u}_m \rVert_2^{-\ell}}{\sum_{k\in\mathcal{M}\setminus\{m\}}\beta_0^2 \lVert \mathbf{p}_j - \mathbf{u}_k \rVert_2^{-\ell}+\frac{\sigma_{\nu}^2}{P_t}
			+\frac{\sigma_{\nu}^2\sigma_{\zeta}^2}{P_tP_r}\frac{1}{N\beta_0^2 \lVert \mathbf{p}_j - \mathbf{q}_1\rVert_2^{-\ell}}+\frac{\sigma_{\zeta}^2}{P_r}\frac{\sum_{k\in\mathcal{M}} \lVert \mathbf{p}_j - \mathbf{u}_k \rVert_2^{-\ell}}{N\lVert \mathbf{p}_j - \mathbf{q}_1\rVert_2^{-\ell}}}.
            % \nm{gamma also depends on position!}
		\end{equation}
		%		\noindent\rule{\textwidth}{0.75pt}	
	\end{figure*}

    % \vspace{-2mm}
To derive the SINR expression under such approximation, we first express the second-hop channel in \eqref{second_channel_ULA_form} as $\mathbf{G} =\mathbf V\cdot\text{Diag}(\mathbf{c})$,
	where we have defined $\mathbf V=[\mathbf{v}_1,\mathbf{v}_2,\dots,\mathbf{v}_J]$ and $\mathbf{c}=[c_1, c_2,\dots,c_J]^\top$,
and substitute it	 into ${\text{ SINR}}_m^{\prime}(\mathbf{w},\boldsymbol{\varphi})$ (as in \eqref{maxSINR_over_B}), yielding
	% To solve \eqref{eq_opt_new_V1db}, we substitute the second-hop channel  $\mathbf{G}$ (as in \eqref{second_channel_ULA_form}) into ${\text{ SINR}}_m^{\prime}(\mathbf{w},\boldsymbol{\varphi})$ (as in \eqref{maxSINR_over_B}), leading to a new SINR form, given by
	\begin{align}\nonumber
	&{\text{ SINR}}_m^{\prime\prime}(\mathbf{w},\mathbf{H},\mathbf{c}, \{\mathbf{v}_j\})=\\
	&\mathbf{a}_m^H\left[\sum_{k\in\mathcal{M}\setminus \{m\}}\!\mathbf{a}_k \mathbf{a}_k^H \!+\! \frac{\sigma_{\nu}^2}{P_t}\sum_{j\in\mathcal{J}}|w_j|^2|c_j|^2\mathbf{v}_j\mathbf{v}_j^H+\frac{\sigma_{\zeta}^2}{P_t}\mathbf{I}\right]^{-1}\!\mathbf{a}_m,
	\label{maxSINR_over_B2}
	%	\nonumber
	\end{align}	
	where we remind that $\mathbf{a}_m = \mathbf V\text{Diag}(\mathbf{c}\odot\mathbf w)\mathbf{h}_m$. Then, substituting the expression of $\mathbf{a}_m$, we can rewrite $\text{SINR}_m^{\prime\prime}$ as
		\begin{align}\nonumber
	&{\text{ SINR}}_m^{\prime\prime}(\mathbf{w},\mathbf{H},\mathbf{c}, \{\mathbf{v}_j\})=\mathbf{h}_m^H
	\text{Diag}(\mathbf{c}\odot\mathbf w)^H\mathbf V^H\\
	&	
	 \quad\times\left[\mathbf V\mathbf Q\mathbf V^H+\frac{\sigma_{\zeta}^2}{P_t}\mathbf{I}\right]^{-1}\mathbf V\text{Diag}(\mathbf{c}\odot\mathbf w)\mathbf{h}_m,
	\label{maxSINR_over_B2_V2}
	%	\nonumber
	\end{align}
 %    \nm{consider moving $\mathbf{h}_m^H
	% \text{Diag}(\mathbf{c}\odot\mathbf w)^H\mathbf V^H$ to the first line and adding $\times$}
	where we have defined %\nm{make sure Q has not been used}
	\begin{align}\nonumber
\mathbf Q=	
&\sum_{k\in\mathcal{M}\setminus \{m\}}
	\text{Diag}(\mathbf{c}\odot\mathbf w)\mathbf{h}_k\mathbf{h}_k^H
	\text{Diag}(\mathbf{c}\odot\mathbf w)^H\\
 &+ \frac{\sigma_{\nu}^2}{P_t}
 \text{Diag}\left(|\mathbf{c}|^2\odot|\mathbf w|^2\right).
	\end{align}
	% with $|\mathbf{u}|^2 = [|u_1|^2, \dots,|u_J|^2]^\top$.
	Then, by leveraging the orthogonality property of $\mathbf{v}_j$, we obtain
	$$
	\mathbf V^H
	\left[\mathbf V\mathbf Q\mathbf V^H+\frac{\sigma_{\zeta}^2}{P_t}\mathbf{I}\right]^{-1}\mathbf V=
	\left[\mathbf Q+\frac{\sigma_{\zeta}^2}{P_t}\mathbf{I}\right]^{-1}.
	$$
	Substituting in \eqref{maxSINR_over_B2_V2} and simplifying, we then obtain
	\begin{align}	\widetilde{\text{SINR}}_m(\mathbf{w},\mathbf{H},\mathbf{c})=\hspace{0em}\mathbf{{h}}_m^H\mathbf{S}_m(|\mathbf{w}|^2)^{-1}\mathbf{{h}}_m,
	\label{eq_SINR_after_ULA}
	\end{align}	
	where  
    % \nm{index m was missing from S! Also, i think it is better to express S as function of $|\mathbf{w}|^2$}
	\begin{equation*}%\label{eq_P_matrix}
	\mathbf{S}_m(|\mathbf{w}|^2)\!=\!\!\!\sum_{k\in\mathcal{M}\setminus \{m\}}\!\!\!\mathbf{{h}}_k\mathbf{{h}}_k^H + \frac{\sigma_{\nu}^2}{P_t}\mathbf{I}+\frac{\sigma_{\zeta}^2}{P_t}\text{Diag}(|\mathbf{c}|^2\odot |\mathbf{w}|^2)^{-1}.
		% \vspace{-1mm}
	\end{equation*}
	%	\begin{align}\nonumber
	%	\tilde{\text{SINR}}_m&(\mathbf{w},\mathbf{H},\mathbf{c})=\hspace{0em}\mathbf{{h}}_m^H\Big[\sum_{k\in\mathcal{M}\setminus \{m\}}\mathbf{{h}}_k\mathbf{{h}}_k^H + \frac{\sigma_{\nu}^2}{P_t}\mathbf{I}\\
	%	&+\frac{\sigma_{\zeta}^2}{P_t}\text{Diag}(|\mathbf{c}|^2)^{-1}\text{Diag}(|\mathbf{w}|^2)^{-1}\Big]^{-1}\mathbf{{h}}_m,
	%	\label{eq_SINR_after_ULA}
	%	\end{align}	
	% \nm{$$\tilde{\text{SINR}}_m(\mathbf{w},\mathbf{H},\mathbf{c})=\mathbf{h}_m^H\mathbf{P}^{-1}\mathbf{h}_m$$
	% where we have defined 
	% $$\mathbf{P}\!=\!\sum_{k\in\mathcal{M}\setminus \{m\}}\mathbf{{h}}_k\mathbf{{h}}_k^H \!+\! \frac{\sigma_{\nu}^2}{P_t}\mathbf{I}\!+\!\frac{\sigma_{\zeta}^2}{P_t}\text{Diag}(|\mathbf{c}|^2)^{-1}\text{Diag}(|\mathbf{w}|^2)^{-1}$$}
	%	where $|\mathbf{c}|^2 = [|c_1|^2, \dots,|c_J|^2]^\top$.
	Thanks to the orthogonality of $\{\mathbf{v}_j\}$, $\widetilde{\text{SINR}}_m(\mathbf{w},\mathbf{H},\mathbf{c})$ is not dependent on each second-hop channel direction $\mathbf{v}_j$.
	Notably, each SINR value depends on $\mathbf w$ through the relay weight magnitudes, but not their phases, a fact that we will use next to solve the optimization problem \eqref{eq_opt_new_V1d}, reformulated as
	\begin{equation}\label{eq_opt_new_V1e}
	\max_{\bm{\mathcal{W}}}	\min_m\mathbb{E}_{\boldsymbol{\varphi}}\left[
	\widetilde{\text{SINR}}_m(\bm{\mathcal{W}}(\boldsymbol{\varphi}),\mathbf{H},\mathbf{c})\right], \text{ s.t. }\eqref{constraint_UAV_relay_pwr_V1}.
	\end{equation}
	To solve this problem, we now show that, for a given channel realization $(\mathbf{H},\mathbf{c})$, each SINR value $\widetilde{\text{SINR}}_m$ is maximized by relay weights $\mathbf w^*$ satisfying the power constraint \eqref{constraint_UAV_relay_pwr_V1} with equality, i.e.,
	$$
	\widetilde{\text{SINR}}_m(\bm{\mathcal{W}}(\boldsymbol{\varphi}),\mathbf{H},\mathbf{c})
	\leq
	\widetilde{\text{SINR}}_m(\mathbf w^*,\mathbf{H},\mathbf{c}),\ \forall m\in\mathcal{M}
	$$
	with
	\begin{equation}\label{relay_bf_amplitude}
		% \vspace{-1mm}
	w_j^*=\left(\frac{P_r}{P_t\sum_{k\in\mathcal{M}}|h_{jk}|^2 + \sigma_{\nu}^2}\right)^{1/2}.    
		% \vspace{-1mm}
	\end{equation}
	To show this property, consider $\mathbf S_m^*\triangleq \mathbf S_m(|\mathbf{w}^*|^2)$, evaluated under
	such $\mathbf w^*$.
	It is straightforward to see that $\mathbf S_m(|\mathbf{w}|^2)\succeq \mathbf S_m^*\succ 0$, for any $|\mathbf{w}|^2$ satisfying the constraint \eqref{constraint_UAV_relay_pwr_V1}. 
%	From Lemma \ref{lemma_PQ_inv} in Appendix \ref{appendix_of_lemma_PQ_inv}, 
	{From Corollary 7.7.4 in \cite{horn2012matrix}, it follows that
	$0\prec\mathbf S_m(|\mathbf{w}|^2)^{-1}\preceq \mathbf S_m^{*-1}$. }
	Using properties of semidefinite positive matrices ($\mathbf A\succeq \mathbf B$ if and only if $\mathbf z^H\mathbf A\mathbf z\geq \mathbf z^H\mathbf B\mathbf z,\forall \mathbf z$), we conclude that
	\begin{align*}  
		% \vspace{-1mm}
	&\widetilde{\text{SINR}}_m(\mathbf{w},\mathbf{H},\mathbf{c})=\mathbf{{h}}_m^H\mathbf{S}_m(|\mathbf{w}|^2)^{-1}\mathbf{{h}}_m\\
	&\leq
	\mathbf{{h}}_m^H\mathbf{S}_m^{*-1}\mathbf{{h}}_m
	=\widetilde{\text{SINR}}_m(\mathbf{w}^*,\mathbf{H},\mathbf{c}).  	
    % \vspace{-1mm}
	\end{align*}
	Since such choice of $\mathbf w$ simultaneously maximizes the SINR values across all users $m$, for any channel realization, it follows that $\mathbf{w}^*$ is optimal for the problem \eqref{eq_opt_new_V1e}.
    % \eqref{constraint_UAV_relay_pwr_V1}\nm{I dont think this is the right ref. you mean eq 26?}.
	Plugging $\mathbf{w}^*$ into the SINR expression, we obtain
	% \nm{use $\triangleq$ when defining for the first time}
	\begin{align}\nonumber
	&\widehat{\text{SINR}}_m(\mathbf{H},\mathbf{c})\triangleq\widetilde{\text{ SINR}}_m(\mathbf{w}^*,\mathbf{H},\mathbf{c})\\
	\nonumber 
    &\hspace{1em}=\mathbf{{h}}_m^H\Bigg[\sum_{k\in\mathcal{M}\setminus \{m\}}\mathbf{{h}}_k\mathbf{{h}}_k^H + \frac{\sigma_{\nu}^2}{P_t}\mathbf{I}+\frac{\sigma_{\zeta}^2\sigma_{\nu}^2}{P_tP_r}\text{Diag}\left(\mathbf{|c|}^2\right)^{-1}\\
	&\hspace{4em} + \frac{\sigma_{\zeta}^2}{P_r}\text{Diag}\left(\mathbf{|c|}^2\right)^{-1}\text{Diag}\left(\mathbf{H}\mathbf{H}^H\right)\Bigg]^{-1}\mathbf{{h}}_m.
		% \vspace{-1mm}
	\label{eq_SINR_maximized_by_w}
	\end{align}
    
    We are thus left with evaluating
	\begin{equation}\label{eq_opt_new_V3}
	\min_m\mathbb{E}_{\boldsymbol{\varphi}}\left[\widehat{\text{SINR}}_m(\mathbf{H},\mathbf{c})\right].
	\end{equation}
    % \vspace{-3mm}
	For the first-hop channel $\mathbf{H}=[\mathbf{h}_1,\dots, \mathbf{h}_M]$, we decompose the channel of user $m$ as
	\begin{equation}\label{eq_first_hop_channel_decomp}
	\mathbf{h}_m=\text{Diag}\left(e^{i{\boldsymbol{\psi}_m}}\right)\mathbf{f}_m, 
	\end{equation}	
	where we have defined the channel gain vector $\mathbf{f}_m = \lvert\mathbf{h}_m\rvert\in\mathbb{R}^{J\times 1}$, and the channel phase vector $\boldsymbol{\psi}_m=\measuredangle\mathbf{h}_m\in[0,2\pi]^{J\times 1}$.	
    Substituting \eqref{eq_first_hop_channel_decomp} into $\mathbb{E}_{\boldsymbol{\varphi}}\left[\widehat{\text{SINR}}_m(\mathbf{H},\mathbf{c})\right]$, we explicitly express the SINR with its dependence on the magnitude $\mathbf{f}_m$ and phase $\boldsymbol{\psi}_m$ of the first-hop channel, given by
    \begin{align}\label{eq_new_SINR_decompose}
    \mathbb{E}_{\boldsymbol{\varphi}}\left[\widehat{\text{SINR}}_m(\mathbf{H},\mathbf{c})\right]
	=\mathbb{E}_{\boldsymbol{\varphi}}\left[\mathbf{f}_m^\top \mathbf{Y}_m^{-1}\mathbf{f}_m\right],	
	\end{align}
	% Substituting \eqref{eq_first_hop_channel_decomp} into $\mathbb{E}_{\boldsymbol{\varphi}}\left[\widehat{\text{SINR}}_m(\mathbf{H},\mathbf{c})\right]$, we have a new SINR form as \nm{what is $\widecheck{\text{SINR}_{m}}$? Not defined, can create confusion. Rather, I would keep using $\widehat{\text{SINR}}$ and simply say that you explicitly express the dependence on magnitude and phase of channel.}
	%	With \eqref{eq_first_hop_channel_decomp}, we have a new SINR form as 
	% \begin{align}\label{eq_new_SINR_decompose}	\nonumber
	% \mathbb{E}_{\boldsymbol{\varphi}}[\widecheck{\text{SINR}_{m}}(\boldsymbol{\psi}_m,\mathbf{f}_m,\mathbf{c})] 
	% &=\mathbb{E}_{\boldsymbol{\varphi}}\left[\widehat{\text{SINR}}_m(\mathbf{H},\mathbf{c})\right]\\ &
	% =\mathbb{E}_{\boldsymbol{\varphi}}\left[\mathbf{f}_m^\top \mathbf{Y}_m^{-1}\mathbf{f}_m\right],	
	% \end{align}
	where 
	\begin{align}\nonumber
	&\mathbf{Y}_m\!=\!\!\!\!\sum_{k\in\mathcal{M}\setminus \{m\}}\!\!\!\text{Diag}(e^{i(\boldsymbol{\psi}_k-\boldsymbol{\psi}_m)})\mathbf{f}_k\mathbf{f}_k^{\top}\text{Diag}(e^{-i(\boldsymbol{\psi}_k-\boldsymbol{\psi}_m)})+\frac{\sigma_{\nu}^2}{P_t}\mathbf{I}\\
	% \nonumber
	&\hspace{1em}+\frac{\sigma_{\nu}^2\sigma_{\zeta}^2}{P_tP_r}\text{Diag}\left(|\mathbf{c}|^2\right)\!^{-1}
	+\frac{\sigma_{\zeta}^2}{P_r}\text{Diag}\left(|\mathbf{c}|^2\right)\!^{-1}\sum_{k\in\mathcal{M}}\text{Diag}\left(\lvert\mathbf{f}_k\rvert^2\right).
    % \vspace{-2mm}
	\label{eq_Y_form}
	\end{align}
	Since a direct calculation of $\mathbb{E}_{\boldsymbol{\varphi}}\left[\mathbf{f}_m^\top \mathbf{Y}_m^{-1}\mathbf{f}_m\right]$ is challenging, we derive a lower bound of $\mathbb{E}_{\boldsymbol{\varphi}}\left[\mathbf{f}_m^\top \mathbf{Y}_m^{-1}\mathbf{f}_m\right]$ to replace the objective function.
	To simplify the derivation, for $\mathbf{f}_m=[{f}_m(1),\dots,{f}_m(J)]^\top$ and $\mathbf{c}=[c_1,\dots,c_J]^\top$, we let
	\begin{align}
	\label{eq_pathloss_1st_hop}
	&{f}_m(j) \approx \beta_{jm}=\beta_0 \lVert \mathbf{p}_j - \mathbf{u}_m \rVert_2^{-\ell/2},\\
	\label{eq_pathloss_2nd_hop}
	&|{c}_j| \approx \sqrt{N}\gamma_{j}=\sqrt{N}\beta_0 \lVert \mathbf{p}_j - \mathbf{q}_1 \rVert_2^{-\ell/2}.
	\end{align}
	This approximation uses the large-scale component as the path gain, which is reasonable in the UAV setting considered in this paper, since the channel is dominated by the LOS path.
	% \sst{ 	We approximate the path gain by its large-scale component to simplify the derivation, which is reasonable given that the channel is dominated by the LOS path.}
	%	\nm{what do these approximations mean? Please provide intuition. The channel is dominated by LOS?}
	Thus, the only random variables are the first-hop channel phase $\boldsymbol{\psi}$, so we have $\mathbb{E}_{\boldsymbol \varphi}[\mathbf{Y}] \approx \mathbb{E}_{\boldsymbol \psi}[\mathbf{Y}]$.
	We observe that \eqref{eq_new_SINR_decompose} is a matrix fractional function \cite{boyd2004convex} of $(\mathbf{f}_m, \mathbf{Y})$ and $\mathbf{Y}\succ 0$, so \eqref{eq_new_SINR_decompose} is convex with respect to $(\mathbf{f}_m, \mathbf{Y})$.
	From Jensen's inequality \cite{boyd2004convex}, we have the lower bound of the expected SINR as
	\begin{align}\nonumber
	\mathbb{E}_{\boldsymbol \psi}\!\left[\widehat{\text{SINR}}_m(\mathbf{H},\mathbf{c})\right]
	\!=\!\mathbb{E}_{\boldsymbol \psi}[\mathbf{f}_m^\top \mathbf{Y}_m^{-1}\mathbf{f}_m]
	\geq \mathbf{f}_m^\top\mathbb{E}_{\boldsymbol \psi}[\mathbf{Y}_m]^{-1}\mathbf{f}_m.
	% \label{SINR_lower_bound}
	% \vspace{-0.5em}
	\end{align}
	For the first term of $\mathbb{E}_{\boldsymbol \psi}[\mathbf{Y}_m]$ (with $\mathbf{Y}_m$ as in \eqref{eq_Y_form}), with independent channel phase $\boldsymbol{\psi}_m$, we derive
	\begin{align}\nonumber
	&\mathbb{E}_{\boldsymbol\psi}\left[\sum_{k\in\mathcal{M}\setminus \{m\}}\text{Diag}(e^{i(\boldsymbol{\psi}_k-\boldsymbol{\psi}_m)})\mathbf{f}_k\mathbf{f}_k^{\top}\text{Diag}(e^{-i(\boldsymbol{\psi}_k-\boldsymbol{\psi}_m)})\right]\\
	&=\sum_{k\in\mathcal{M}\setminus \{m\}}\text{Diag}(|\mathbf{f}_k|^2).
    % \vspace{-2mm}
	\end{align}
	% where $|\mathbf{f}_k|^2 = [|{f}_m(1)|^2,\dots,|{f}_m(J)|^2]^\top$.\nm{$|\mathbf{f}_k|^2$ should have been defined earlier since already used }
	Thus, we have a lower bound of the expected SINR as 
	\begin{align}
	\nonumber
	&\mathbb{E}_{\boldsymbol \psi}[\widehat{\text{SINR}}_m(\mathbf{H},\mathbf{c})]\\
	\nonumber
	&\geq\mathbf{f}_m^\top\Big[\sum_{k\in\mathcal{M}\setminus \{m\}} \text{Diag}(|\mathbf{f}_k|^2)+\frac{\sigma_{\nu}^2}{P_t}\mathbf{I}+\frac{\sigma_{\nu}^2\sigma_{\zeta}^2}{P_tP_r}\text{Diag}\left(|\mathbf{c}|^2\right)^{-1}\\
	\label{eq_expectedSINR_lower_bnd}
	%		\nonumber
	&\hspace{2.5em}+\frac{\sigma_{\zeta}^2}{P_r}\text{Diag}(\lvert\mathbf{c}\rvert^2)^{-1}\sum_{k\in\mathcal{M}}\text{Diag}\left(\lvert\mathbf{f}_k\rvert^2\right) \Big]^{-1}\mathbf{f}_m\\
	&=\sum_{j\in\mathcal{J}}
	\frac{|{f}_m(j)|^2}
	{\sum\limits_{k\in\mathcal{M}\setminus\{m\}}\!\!|{f}_k(j)|^2+\frac{\sigma_{\nu}^2}{P_t}
		+\frac{\sigma_{\nu}^2\sigma_{\zeta}^2}{P_tP_r}\frac{1}{|{c}_j|^2}\!+\!\frac{\sigma_{\zeta}^2}{P_r}\frac{\sum_{k\in\mathcal{M}}|{f}_k(j)|^2}{|{c}_j|^2}}.\label{lower_bnd_SINR_form}
	%\\
	%&={\text{SINR}_{m}^*}(\mathbf{f}_1,\dots,\mathbf{f}_M,\mathbf{c})\label{lower_bnd_SINR_form}
	%		&=\sum_{j\in\mathcal{J}}\frac{(\beta_{jm}^h)^2}{\sum_{k\neq m}(\beta_{jk}^h)^2+\frac{\sigma_{\nu}^2}{P_t}+\frac{\sigma_n^2\sigma_{\nu}^2}{P_tP_r}\frac{1}{(\beta_j^v)^2}+\frac{\sigma_n^2}{P_r}\frac{1}{(\beta_j^v)^2}\sum_{k=1}^M(\beta_{jk}^h)^2},
    % \vspace{-1mm}
	\end{align}
	%	By substituting the assumptions \eqref{eq_pathloss_1st_hop} and \eqref{eq_pathloss_2nd_hop} into \eqref{lower_bnd_SINR_form}, we have the lower bound of the expected SINR as $\Gamma_m(\mathcal{P})$ in \eqref{eq_lower_bound}.
	%\begin{align}\nonumber
	%&\Gamma_m(\mathcal{P})\\
	%%		&=\sum_{j\in\mathcal{J}}
	%%		\frac{|\mathbf{f}_m(j)|^2}
	%%		{\sum_{k\in\mathcal{M}\setminus\{m\}}|\mathbf{f}_m(j)|^2+\frac{\sigma_{\nu}^2}{P_t}
	%%		+\frac{\sigma_n^2\sigma_{\nu}^2}{P_tP_r}\frac{1}{|\mathbf{c}(j)|^2}+\frac{\sigma_n^2}{P_r}\frac{\sum_{k\in\mathcal{M}}|\mathbf{f}_m(j)|^2}{|\mathbf{c}(j)|^2}}\\
	%&=\sum_{j\in\mathcal{J}}\frac{\left(\beta_{jm}^{(h)}\right)^2}{\sum_{k\in\mathcal{M}\setminus\{m\}}\left(\beta_{jk}^{(h)}\right)^2+\frac{\sigma_{\nu}^2}{P_t}
	%	+\frac{\sigma_{\nu}^2\sigma_n^2}{P_tP_r}\frac{1}{N(\beta_{1j}^{(g)})^2}+\frac{\sigma_n^2}{P_r}\frac{\sum_{k\in\mathcal{M}}(\beta_{jk}^{(h)})^2}{N(\beta_{1j}^{(g)})^2}}.
	%%		+\frac{\sigma_n^2}{P_r}\frac{1}{(\beta_{1j}^{(v)})^2}\sum_{k\in\mathcal{M}}(\beta_{jk}^{(h)})^2}
	%\label{eq_lower_bound}
	%\end{align}	
	% \nm{make a better connection to the next step. "at this point, we found the solution of ... now, we consider the position optmita. "}
	%	We reformulate the UAV placement optimization problem as
    % \vspace{-2mm}
    \tcadd{At this point, we found a lower bound of the expected beamforming SINR of user $m$ as in \eqref{lower_bnd_SINR_form}, with the optimized relay beamforming and receive combining. 
	By substituting the channel approximations 
    % \nm{something wrong with eq labels... also label 39 appears twice for two different eqs.}
    \eqref{eq_pathloss_1st_hop} and \eqref{eq_pathloss_2nd_hop} into \eqref{lower_bnd_SINR_form}, we obtain the function $\Gamma_m(\mathcal{P})$ in \eqref{eq_lower_bound}, which is a function of UAV positions $\mathcal{P}$, representing the lower bound of the expected beamforming SINR.
	We formulate the UAV placement optimization in \eqref{opt_uav_placement} into a new form as 
	\begin{align}
	%	\nonumber
	\label{opt_prob_to_be_solved} \setcounter{equation}{38}
	&\max_{\mathcal{P}}\min_m\ \Gamma_m(\mathcal{P}),\\ 		
	&\text{s.t. }\lVert\mathbf{p}_j - \mathbf{p}_{j^\prime}\rVert_2\geq\epsilon_p,\ j\neq j^\prime, \ \forall j,j^\prime \in \mathcal{J}.	
	\end{align}
    It is worth noting that the objective function \eqref{opt_prob_to_be_solved} in this optimization problem incorporates the power constraint on the transmission of each UAV relay (i.e., \eqref{constraint_UAV_relay_pwr_V1}) through the optimal choice of relay beamforming weight found in \eqref{relay_bf_amplitude}.}
    
    % It is worth noting that the objective function in this optimization problem \eqref{opt_prob_to_be_solved} incorporates the power constraint on the transmission of each UAV relay (i.e., \eqref{constraint_UAV_relay_pwr_V1}) by the optimal choise of relay beamforming weight we found in \eqref{relay_bf_amplitude}.

% \nm{one of the reviewers was asking about "constraints". I think you need to clarify here and to the revierew that the problem 47 already incorporates power constraint, since it incorporates the optimal choice of weight designs found earler.}

	% \vspace{-3mm}
	\subsection{Difference-of-Convex-Based (DC-Based) Approach}
	\label{susection_DC_approach}
	% \vspace{-1em}
	We aim to solve problem \eqref{opt_prob_to_be_solved} by developing a DC-based approach as follows.
    First, we rewrite the optimization problem as shown in \eqref{sub_obj_func1}. 
    We replace the objective $\Gamma_m(\mathcal{P})$ with $\phi_m(\boldsymbol{\alpha},\mathbf{z})=\sum_{j\in\mathcal{J}}\frac{\alpha_{jm}^2}{z_{jm}}$, and add two inequality constraints \eqref{add_constraint_1} and \eqref{add_constraint_2} to make $\phi_m(\boldsymbol{\alpha},\mathbf{z})$ a lower bound of $\Gamma_m(\mathcal{P})$.
    % \nm{First, we rewrite the optimization problem as shown in [Eq 49]. We replace the objective $\Gamma_m(\mathcal{P})$ with $\phi_m(\boldsymbol{\alpha},\mathbf{z})=\sum_{j\in\mathcal{J}}\frac{\alpha_{jm}^2}{z_{jm}}$, and add two ...}
    % First, we rewrite the optimization problem as shown in \eqref{sub_obj_func1}. 
    % We replace th    
	% We assume an objective function $\phi_m(\boldsymbol{\alpha},\mathbf{z})=\sum_{j\in\mathcal{J}}\frac{\alpha_{jm}^2}{z_{jm}}$ to replace $\Gamma_m(\mathcal{P})$ in \eqref{opt_prob_to_be_solved}.
	% We add two inequality constraints \eqref{add_constraint_1} and \eqref{add_constraint_2} to make $\phi_m(\boldsymbol{\alpha},\mathbf{z})$ a lower bound of $\Gamma_m(\mathcal{P})$.
	The constraint \eqref{add_constraint_1} is introduced to bound the desired signal power using variables $\alpha_{jm}^2$,  $j\in \mathcal{J}$, $m\in\mathcal{M}$. 
	We include the constraint \eqref{add_constraint_2} to restrict the interference and noise by the variable $z_{jm}$,  $j\in \mathcal{J}$, $m\in\mathcal{M}$.
	%	, which is a lower bound of $\Gamma_m(\mathcal{P})$ in \eqref{opt_prob_to_be_solved}
	%	for $\Gamma_m(\mathcal{P})$ in \eqref{opt_prob_to_be_solved}, where 
	% We reformulate \eqref{opt_prob_to_be_solved} into a new optimization problem as
    These steps yield the equivalent optimization problem
	\begin{align}
	\label{sub_obj_func1}
	&\max_{\mathcal{P},\boldsymbol{\alpha},\mathbf{z}}\min_m\ \phi_m(\boldsymbol{\alpha},\mathbf{z})=\sum_{j\in\mathcal{J}}\frac{\alpha_{jm}^2}{z_{jm}},\\
	\label{add_constraint_1}
	&\text{s.t. }\beta_0^2\lVert \mathbf{p}_j - \mathbf{u}_m \rVert_2^{-\ell}\geq \alpha_{jm}^2,\ j\in\mathcal{J},\ m\in\mathcal{M};\\
	\nonumber
	&\hspace{0.8em}\sum_{k\in\mathcal{M}\setminus\{m\}}\beta_0^2\lVert \mathbf{p}_j - \mathbf{u}_k \rVert_2^{-\ell}+\frac{\sigma_{\nu}^2}{P_t}+\frac{\sigma_{\nu}^2\sigma_{\zeta}^2}{NP_tP_r}\frac{1}{\beta_0^2\lVert \mathbf{p}_j - \mathbf{q}_1 \rVert_2^{-\ell}}\\
	\label{add_constraint_2}
	&\hspace{0.8em}+\frac{\sigma_{\zeta}^2}{NP_r}\frac{\sum_{k\in\mathcal{M}}\beta_0^2\lVert \mathbf{p}_j - \mathbf{u}_k \rVert_2^{-\ell}}{\beta_0^2\lVert \mathbf{p}_j - \mathbf{q}_1 \rVert_2^{-\ell}} \leq z_{jm},\ j\in\mathcal{J},\ m\in\mathcal{M};\\
	\label{add_constraint_3}
	&\hspace{0.8em}\lVert\mathbf{p}_j - \mathbf{p}_{j^\prime}\rVert_2\geq\epsilon_p,\ j\neq j^\prime, \ \forall j,j^\prime \in \mathcal{J}.
	\end{align}
	%	We have a new objective function $\phi_m(\boldsymbol{\alpha},\mathbf{z})$, which is a lower bound of $\Gamma_m(\mathcal{P})$ preserved by the added inequality contraints \eqref{add_constraint_1}, \eqref{add_constraint_2}.
	%	The inequality constraint \eqref{add_constraint_1} is introduced to bound the desired signal power using variables $\alpha_{jm}^2$,  $j\in \mathcal{J}$, $m\in\mathcal{M}$. 
	%	Also, we include the inequality constraint \eqref{add_constraint_2} to restrict the interference and noise by the variable $z_{jm}$,  $j\in \mathcal{J}$, $m\in\mathcal{M}$.
	Note that $\frac{\alpha_{jm}^2}{z_{jm}}$ is a convex function in $(\alpha_{jm},z_{jm})\in\mathbb{R}\times\mathbb{R_+}$,
	%\nm{in which domain? I believe z must be positive}
	which can be verified by the positive semi-definiteness of its Hessian matrix.
	The function $\phi_m(\boldsymbol{\alpha},\mathbf{z})$ is a linear combination of convex functions, which is also convex.
	Letting ${\eta}_1(\boldsymbol{\alpha},\mathbf{z}) = \max_m \sum_{k\in\mathcal{M}\setminus\{m\}}{\phi}_k(\boldsymbol{\alpha},\mathbf{z})$ and ${\eta}_2(\boldsymbol{\alpha},\mathbf{z}) = \sum_{k\in\mathcal{M}}{\phi}_k(\boldsymbol{\alpha},\mathbf{z})$, we rephrase the optimization problem as
	\begin{align}\label{optimization_UAV_movement_with_large_scale_DCP}
	&\min_{\mathcal{P},\boldsymbol{\alpha},\mathbf{z}} {\eta}_1(\boldsymbol{\alpha},\mathbf{z})-{\eta}_2(\boldsymbol{\alpha},\mathbf{z}),\\
	\nonumber
	&\text{s.t. }\eqref{add_constraint_1}, \eqref{add_constraint_2},\eqref{add_constraint_3}.
	\end{align}
	Given that ${\eta}_1(\boldsymbol{\alpha},\mathbf{z})$ and ${\eta}_2(\boldsymbol{\alpha},\mathbf{z})$ are convex functions of $(\boldsymbol{\alpha},\mathbf{z})$, we state that the optimization problem \eqref{optimization_UAV_movement_with_large_scale_DCP} can be formulated as a DC optimization framework.
	The DC optimization problem has the objective function as the difference of two convex functions, with convex constraints.
	The DC iterative procedure is developed to locate the optimal solutions, by applying the linear approximation on ${\eta}_2(\boldsymbol{\alpha},\mathbf{z})$ and the non-convex constraints in each iteration.
    % \nm{and linear approx of constraints? You should state it here.}.
	%The objective function of the DC framework is the difference of two convex functions, 
	% the iterative procedure to locate the optimal solutions, by.
	%\nm{explain briefly how DC works before delving in the linear apprx}
	It requires the linear approximation of ${\eta}_2(\boldsymbol{\alpha},\mathbf{z})$  and of \eqref{add_constraint_1}, \eqref{add_constraint_2}, \eqref{add_constraint_3} to convexify the objective function and the constraints, derived next.
	
	\textbf{Linear approximation of $\eta_2(\boldsymbol{\alpha},\mathbf{z})$}: 
	We derive the linear approximation of $\eta_2(\boldsymbol{\alpha},\mathbf{z})$ at the point $(\boldsymbol{\alpha}^{(\iota)},\mathbf{z}^{(\iota)})$, given by
	%We apply a linear approximation to make $f_2(\mathbf{w},\mathbf{z})$ an affine function, given by
	\begin{align}
	\nonumber
	&\tilde{\eta}_2(\boldsymbol{\alpha},\mathbf{z};\boldsymbol{\alpha}^{(\iota)},\mathbf{z}^{(\iota)})= \eta_2(\boldsymbol{\alpha}^{(\iota)},\mathbf{z}^{(\iota)}) \\
	&+	\big<\nabla_{\boldsymbol{\alpha}} \eta_2(\boldsymbol{\alpha}^{(\iota)},\mathbf{z}^{(\iota)}),\boldsymbol{\alpha}-\boldsymbol{\alpha}^{(\iota)}\big>\!+\!	\big<\nabla_{\mathbf{z}} \eta_2(\boldsymbol{\alpha}^{(\iota)},\mathbf{z}^{(\iota)}),\mathbf{z}-\mathbf{z}^{(\iota)}\big>,
	\nonumber
	\end{align}
	where
	\begin{align}
	\nonumber
	&\left<\nabla_{\boldsymbol{\alpha}} \eta_2\left(\boldsymbol{\alpha}^{(\iota)},\mathbf{z}^{(\iota)}\right),\boldsymbol{\alpha}-\boldsymbol{\alpha}^{(\iota)}\right>\\
	&=\sum_{k\in\mathcal{M}}\sum_{j\in\mathcal{J}}\frac{2\alpha_{jk}^{(\iota)}}{z_{jk}^{(\iota)}}\left(\alpha_{jk}-\alpha_{jk}^{(\iota)}\right),\\
	\nonumber
	&\left<\nabla_{\mathbf{z}} \eta_2\left(\boldsymbol{\alpha}^{(\iota)},\mathbf{z}^{(\iota)}\right),\mathbf{z}-\mathbf{z}^{(\iota)}\right>\\
	&=-\sum_{k\in\mathcal{M}}\sum_{j\in\mathcal{J}}\left(\frac{\alpha_{jk}^{(\iota)}}{z_{jk}^{(\iota)}}\right)^2\left(z_{jk}-z_{jk}^{(\iota)}\right).
	\end{align}	
	%	where 
	%	\begin{align*}
	%	%\nonumber
	%	&\left<\nabla_{\mathbf{y}} \eta_2(\mathbf{y}^{(k)},\mathbf{z}^{(k)}),\mathbf{y}-\mathbf{y}^{(k)}\right>=-\sum_{m=1}^{M}\frac{y_m-{y}_m^{(k)}}{\left({y}_m^{(k)}\right)^2\left({z}_m^{(k)}+\sigma_n^2\right)},\\
	%	%\nonumber
	%	&\left<\nabla_{\mathbf{z}} f_2(\mathbf{y}^{(k)},\mathbf{z}^{(k)}),\mathbf{z}-\mathbf{z}^{(k)}\right>=-\sum_{m=1}^{M}\frac{z_m-{z}_m^{(k)}}{{y}_m^{(k)}\left({z}_m^{(k)}+\sigma_n^2\right)^2}.
	%	\end{align*}
	By substituting $\tilde{\eta}_2(\boldsymbol{\alpha},\mathbf{z};\boldsymbol{\alpha}^{(\iota)},\mathbf{z}^{(\iota)})$ for ${\eta}_2(\boldsymbol{\alpha},\mathbf{z};\boldsymbol{\alpha}^{(\iota)},\mathbf{z}^{(\iota)})$, we have the convex approximation of the objective function \eqref{optimization_UAV_movement_with_large_scale_DCP}.
	% is transformed into a convex function.
	
	Next, we need to find the convex relaxation form for \eqref{add_constraint_1}, \eqref{add_constraint_2} and \eqref{add_constraint_3}, which are detailed as follows:
	
	\textbf{Constraint relaxation of \eqref{add_constraint_1}}: We have the equivalent form of \eqref{add_constraint_1} by taking an exponent to $-1/\ell$ as
	%	the inequality constraint as 
	%	\nm{explain how you did this. Which equation did you plug in? You need to provide guidance to the reader...}
	\begin{equation}
	\beta_0^{-2/\ell}\lVert \mathbf{p}_j - \mathbf{u}_m \rVert_2\leq \alpha_{jm}^{-2/\ell},\ j\in\mathcal{J},\ m\in\mathcal{M}.
	\end{equation}
	We observe that the L.H.S. is the norm of $\mathbf{p}_j$, which is convex.
	However,  $\alpha_{jm}^{-2/\ell}$ is also convex when $\alpha_{jm}>0$ and $-2/\ell<0$, which requires the convex relaxation.
	We derive the new inequality constraint by taking the linear approximation of $\alpha_{jm}^{-2/\ell}$ at the point $\alpha_{jm}^{(\iota)}$, given by
	\begin{align}\label{relaxed_add_constraint_1}
	\beta_0^{-\frac{2}{\ell}}\lVert \mathbf{p}_j - \mathbf{u}_m \rVert_2 \!\leq\! \left(\alpha_{jm}^{(\iota)}\right)^{-\frac{2}{\ell}} \!-\!\frac{2}{\ell}\left(\alpha_{jm}^{(\iota)}\right)^{-\frac{2}{\ell}-1}\left(\alpha_{jm}\!-\!\alpha_{jm}^{(\iota)}\right).
	\end{align}
    
    % \nm{be consistent. You use both $\Vert\Vert_2$ ad $\Vert\Vert$ (without subscript 2). Please check throughout.}
	
	%	\textbf{Constraint relaxation of \eqref{add_constraint_1}}: We have the inequality constraint as 
	%	\begin{equation}
	%		\beta_0^2\lVert \mathbf{p}_j - \mathbf{u}_m \rVert_2^{-\ell}\geq \alpha_{jm}^2,\ j\in\mathcal{J},\ m\in\mathcal{M};
	%	\end{equation}
	%	Assuming slack variables $\zeta_{jm}\geq 0$, we have
	%	\begin{equation}\label{slack_var_1}
	%		\lVert \mathbf{p}_j - \mathbf{u}_m \rVert_2\leq \zeta_{jm}^{2/\ell}, (j,m)\in\mathcal{J}\times\mathcal{M}.
	%	\end{equation}
	%	The pathloss exponent $\ell$ is normally in the range of $2$ to $4$, so $\zeta_{jm}^{2/\ell}$ is a concave function.
	%	Thus, \eqref{slack_var_1} is an convex inequality constraints.
	%	Substituting \eqref{slack_var_1} into \eqref{add_constraint_1}, we have $\beta_0^2\zeta_{jm}^{-2} \geq \alpha_{jm}^2$.
	%	Since $\zeta_{jm}^{-2}$ is convex for $\zeta>0$, we derive the linear approximation of $\zeta^{-2}$ at the point $\zeta_{jm}^{(\iota)}$, given by $$(\zeta_{jm}^{(\iota)})^{-2} - 2(\zeta_{jm}^{(\iota)})^{-3}(\zeta_{jm}-\zeta_{jm}^{(\iota)}).$$
	%	In this manner, we relax the constraint \eqref{add_constraint_1} by the following inequality constraints
	%	\begin{align}
	%		&\beta_0^2\left(\zeta_{jm}^{(\iota)})^{-2} - 2(\zeta_{jm}^{(\iota)})^{-3}(\zeta_{jm}-\zeta_{jm}^{(\iota)}\right) \geq \alpha_{jm}^2;\\
	%		&\lVert \mathbf{p}_j - \mathbf{u}_m \rVert_2\leq \zeta_{jm}^{2/\ell}, (j,m)\in\mathcal{J}\times\mathcal{M};\\
	%		&\zeta_{jm}>0;
	%	\end{align}
	
	\textbf{Constraint relaxation of \eqref{add_constraint_2}}: 
    The constraint \eqref{add_constraint_2} can be expressed as
	% By substituting $\beta_{jk}$ in \eqref{eq_large_scale_channel_gain_first_hop} and $\gamma_{j}$ in \eqref{eq_channel_gain_second_hop} into the constraint \eqref{add_constraint_2}, we have
	%	\nm{explain how you did this. Which equation did you plug in? You need to provide guidance to the reader...}
	\begin{align}
	\nonumber
	&\sum_{k\in\mathcal{M}\setminus\{m\}}\beta_0^2\lVert \mathbf{p}_j - \mathbf{u}_k \rVert_2^{-\ell}+\frac{\sigma_{\zeta}^2\sigma_{\nu}^2}{NP_tP_r\beta_0^2}\lVert \mathbf{p}_j - \mathbf{q}_1 \rVert_2^{\ell}\\
	&+\frac{\sigma_{\zeta}^2}{NP_r}\lVert \mathbf{p}_j - \mathbf{q}_1 \rVert_2^{\ell}\sum_{k\in\mathcal{M}}\lVert \mathbf{p}_j - \mathbf{u}_k \rVert_2^{-\ell} \leq z_{jm}-\frac{\sigma_{\nu}^2}{P_t},
	\label{add_constraint_2_V2}
	\end{align}
	for $(j,m)\in\mathcal{J}\times \mathcal{M}$.
	With slack variables 	$[\boldsymbol{\xi}]_{jm}=\xi_{jm}>0$ and $[\boldsymbol{\Omega}]_{j}=\Omega_{j}>0$, we add two inequality constraints
	\begin{align}
	\label{slack_constraint_1}
	\xi_{jm}^{-2/\ell}\leq \lVert\mathbf{p}_j - \mathbf{u}_m\rVert_2,\\
	\label{slack_constraint_2}		
	\lVert\mathbf{p}_j - \mathbf{q}_1\rVert_2\leq\Omega_{j}^{-1/\ell}.
	\end{align}
	Substituting \eqref{slack_constraint_1} and \eqref{slack_constraint_2} into \eqref{add_constraint_2_V2}, we derive
	\begin{align}
	\nonumber
	&\sum_{k\in\mathcal{M}\setminus\{m\}}\beta_0^2\xi_{jk}^{2}+\left(\frac{\sigma_{\zeta}^2\sigma_{\nu}^2}{NP_tP_r\beta_0^2}\right)\frac{1}{\Omega_{j}}
	+\left(\frac{\sigma_{\zeta}^2}{NP_r}\right)\sum_{k\in\mathcal{M}}\frac{\xi_{jk}^{2}}{\Omega_{j}}\\
	& \hspace{3mm}\leq z_{jm}-\frac{\sigma_{\nu}^2}{P_t},\ j\in\mathcal{J},\ m\in\mathcal{M}.
	\label{relaxed_add_constraint_2}
	\end{align}
	It is a convex inequality constraint by verifying that the L.H.S. is a convex function in $(\xi_{jm},\Omega_{j})$.
	However, \eqref{slack_constraint_1} and \eqref{slack_constraint_2} are non-convex constraints, so we replace their R.H.S. by the lower-bound first-order Taylor approximation around points $\mathbf{p}^{(\iota)}$ and $\Omega_j^{(\iota)}$, respectively, as follows
	\begin{align}
	\label{relaxed_slack_constraint_1}
	\xi_{jm}^{-2/\ell}\leq 
	\lVert\mathbf{p}_j^{(\iota)} - \mathbf{u}_m\rVert_2
	+\frac{(\mathbf{p}_j^{(\iota)} - \mathbf{u}_m)^\top}{\lVert\mathbf{p}_j^{(\iota)} - \mathbf{u}_m\rVert_2}\left(\mathbf{p}_j-\mathbf{p}_j^{(\iota)}\right),\\
	\label{relaxed_slack_constraint_2}		
	\lVert\mathbf{p}_j - \mathbf{q}\rVert_2\leq\left(\Omega_{j}^{(\iota)}\right)^{-\frac{1}{\ell}}-\frac{1}{\ell}\left(\Omega_{j}^{(\iota)}\right)^{-\frac{1}{\ell}-1}\left(\Omega_{j}-\Omega_{j}^{(\iota)}\right).
	\end{align}

	\textbf{Constraint relaxation of \eqref{add_constraint_3}}: We derive the new inequality constraint by taking the linear approximation of $\lVert\mathbf{p}_j - \mathbf{p}_{j^\prime}\rVert_2$ at $(\mathbf{p}_j^{(\iota)},\mathbf{p}_{j^\prime}^{(\iota)})$, given by
	\begin{align}\nonumber
	\epsilon_p\leq \lVert\mathbf{p}_j^{(\iota)} - \mathbf{p}_{j^\prime}^{(\iota)}\rVert_2
	&+\frac{(\mathbf{p}_j^{(\iota)} - \mathbf{p}_{j^\prime}^{(\iota)})^\top}{\lVert\mathbf{p}_j^{(\iota)} - \mathbf{p}_{j^\prime}^{(\iota)}\rVert_2}\left(\mathbf{p}_j-\mathbf{p}_j^{(\iota)}\right)\\
	&+\frac{(\mathbf{p}_{j^\prime}^{(\iota)} - \mathbf{p}_{j}^{(\iota)})^\top}{\lVert\mathbf{p}_j^{(\iota)} - \mathbf{p}_{j^\prime}^{(\iota)}\rVert_2}\left(\mathbf{p}_{j^\prime}-\mathbf{p}_{j^\prime}^{(\iota)}\right).
	\label{relaxed_slack_constraint_3}
	\end{align}
	
	With the above relaxations, we are able to formulate a new convex optimization problem at the $\iota$-th iteration as
	\begin{align}\label{optimization_UAV_movement_DCP_covex}
	&\min_{\mathcal{P},\boldsymbol{\alpha},\mathbf{z},\boldsymbol{\xi},\boldsymbol{\Omega}} {\eta}_1(\boldsymbol{\alpha},\mathbf{z})-\tilde{\eta}_2\left(\boldsymbol{\alpha},\mathbf{z};\boldsymbol{\alpha}^{(\iota)},\mathbf{z}^{(\iota)}\right),\\
	\nonumber
	&\text{s.t. }\eqref{relaxed_add_constraint_1}, \eqref{relaxed_add_constraint_2}, \eqref{relaxed_slack_constraint_1}, \eqref{relaxed_slack_constraint_2}, \eqref{relaxed_slack_constraint_3}, \boldsymbol{\xi}>0, \boldsymbol{\Omega}>0.
	\end{align}
	The new constraints \eqref{relaxed_add_constraint_1}, \eqref{relaxed_add_constraint_2}, \eqref{relaxed_slack_constraint_1}, \eqref{relaxed_slack_constraint_2}, \eqref{relaxed_slack_constraint_3}, $\boldsymbol{\xi}>0$, and $\boldsymbol{\Omega}>0$ are the relaxations of the original constraints \eqref{add_constraint_1}, \eqref{add_constraint_2}, and \eqref{add_constraint_3}, so the feasible set of \eqref{optimization_UAV_movement_DCP_covex} is a subset of the feasible set of \eqref{optimization_UAV_movement_with_large_scale_DCP}.
	Thus, the solution of \eqref{optimization_UAV_movement_DCP_covex} is also feasible to  \eqref{optimization_UAV_movement_with_large_scale_DCP}.
	
	%	\nm{can you explain about feasibility? IS the solution to this OP feasible with respect to the original problem? in other words, is the feasible set of this new OP a subset of the feasible set of the original one?}
	
	Since ${\eta}_1$ is convex and $\tilde{\eta}_2$ is affine, \eqref{optimization_UAV_movement_DCP_covex} is a convex optimization problem, which can be solved by the CVX package \cite{cvx}, e.g., an interior-point method, leading to an optimized solution $(\mathcal{P}_*^{(\iota)},\boldsymbol{\alpha}_*^{(\iota)},\mathbf{z}_*^{(\iota)},\boldsymbol{\xi}_*^{(\iota)},\boldsymbol{\Omega}_*^{(\iota)})$ in the $\iota$-th DC iteration (DCI).
	%	 $(\boldsymbol{\alpha}^{(\iota)}_*,\mathbf{z}^{(\iota)}_*)$ in the $\iota$-th iteration of DCIs.
	In the next (i.e., $(\iota+1)$-th) DCI, we consider the optimization problem \eqref{optimization_UAV_movement_DCP_covex} by substituting the objective function with $\eta_1(\boldsymbol{\alpha},\mathbf{z})-\tilde \eta_2(\boldsymbol{\alpha},\mathbf{z};\boldsymbol{\alpha}^{(\iota+1)},\mathbf{z}^{(\iota+1)})$, where $(\boldsymbol{\alpha}^{(\iota+1)},\mathbf{z}^{(\iota+1)})=(\boldsymbol{\alpha}^{(\iota)}_*,\mathbf{z}^{(\iota)}_*)$.
	The solution can be obtained by solving this convex optimization problem, yielding  $(\mathcal{P}_*^{(\iota+1)},\boldsymbol{\alpha}_*^{(\iota+1)},\mathbf{z}_*^{(\iota+1)},\boldsymbol{\xi}_*^{(\iota+1)},\boldsymbol{\Omega}_*^{(\iota+1)})$.
	% $(\mathbf{w}^{(k+1)}_*,\mathbf{z}^{(k+1)}_*)$ as the solution by solving the convex optimization.
	The DC-based algorithm stops when the solution converges, i.e., $\lVert\boldsymbol{\alpha}- \boldsymbol{\alpha}^{(\iota)} \rVert_2 + \lVert\mathbf{z}- \mathbf{z}^{(\iota)} \rVert_2+ \lVert\boldsymbol{\xi}- \boldsymbol{\xi}^{(\iota)} \rVert_F+ \lVert\boldsymbol{\Omega}- \boldsymbol{\Omega}^{(\iota)} \rVert_2 + \sum_{j\in\mathcal{J}}\lVert \mathbf{p}_j - \mathbf{p}_{j}^{(\iota)}\rVert_2<\epsilon$, where $\epsilon>0$ is a given threshold that controls the quality of the optimization.
    % \nm{that controls the quality of the approximation (?)}.
    The UAV placement algorithm is detailed in \textbf{Algorithm \ref{alg_uav_placement}}.
    
    % the solution converges, i.e., $\lVert\boldsymbol{\alpha}- \boldsymbol{\alpha}^{(\iota)} \rVert + \lVert\mathbf{z}- \mathbf{z}^{(\iota)} \rVert+ \lVert\boldsymbol{\xi}- \boldsymbol{\xi}^{(\iota)} \rVert+ \lVert\boldsymbol{\Omega}- \boldsymbol{\Omega}^{(\iota)} \rVert + \sum_{j\in\mathcal{J}}\lVert \mathbf{p}_j - \mathbf{p}_{j}^{(\iota)}\rVert<\epsilon$, where $\epsilon>0$ is a constant threshold.
	% The UAV placement algorithm is detailed in \textbf{Algorithm \ref{alg_uav_placement}}.
	%	.\nm{define convergence,. when the updates to the optimization variables become sufficiently small?}
	
	\begin{algorithm}[t!]
		\caption{\tcadd{UAV Placement Algorithm.}}
		\label{alg_uav_placement}
		\begin{algorithmic}[1]
			% \REQUIRE 
			\INPUT User position $\mathbf{u}_m,\ m\in\mathcal{M}$, BS antenna position $\mathbf{q}_1$
			\OUTPUT UAV positions $\mathcal{P}=\{\mathbf{p}_j,\ j\in\mathcal{J}\}$
			%			\STATE \emph{Optimize $\mathcal{P}$}									
			%			\STATE \emph{Optimize relay beamforming $\mathbf{w}$ with fixed $\mathbf{B}$}
			\State $\iota=0$;
            \Repeat
            % \STATE 132
            % \Until{123}
			% \WHILE {$E> \epsilon$}%} \textbf{and} $k<k_{max}$}
			%			\STATE \emph{Solve }
                % \WHILE{Not convergence of the objective function \eqref{optimization_UAV_movement_DCP_covex}}
			\State Derive $(\mathcal{P},\boldsymbol{\alpha},\mathbf{z},\boldsymbol{\xi},\boldsymbol{\Omega})$ by solving \eqref{optimization_UAV_movement_DCP_covex}, given $(\mathcal{P}^{(\iota)},\boldsymbol{\alpha}^{(\iota)},\mathbf{z}^{(\iota)},\boldsymbol{\xi}^{(\iota)},\boldsymbol{\Omega}^{(\iota)})$;
			% \STATE $E=\sum_{j\in\mathcal{J}}\lVert \mathbf{p}_j - \mathbf{p}_{j}^{(\iota)}\rVert + \lVert\boldsymbol{\alpha}- \boldsymbol{\alpha}^{(\iota)} \rVert + \lVert\mathbf{z}- \mathbf{z}^{(\iota)} \rVert+ \lVert\boldsymbol{\xi}- \boldsymbol{\xi}^{(\iota)} \rVert+ \lVert\boldsymbol{\Omega}- \boldsymbol{\Omega}^{(\iota)} \rVert$;
			% \item[]			
			\Statex \emph{\ \ \ \ Record variables for next iteration}
			\State $\iota := \iota + 1$;
			\State 
			$(\mathcal{P}^{(\iota)},\boldsymbol{\alpha}^{(\iota)},\mathbf{z}^{(\iota)},\boldsymbol{\xi}^{(\iota)},\boldsymbol{\Omega}^{(\iota)})= (\mathcal{P},\boldsymbol{\alpha},\mathbf{z},\boldsymbol{\xi},\boldsymbol{\Omega})$;
			%			$\mathcal{P},\boldsymbol{\alpha},\mathbf{z},\boldsymbol{\xi},\boldsymbol{\Omega}$
			%			$\mathbf{\bar{w}}_{old}=\mathbf{\bar{w}}$; $\mathbf{\bar{z}}_{old}=\mathbf{\bar{z}}$;
			%			\STATE $i_1\coloneqq i_1+1$; \% inner-loop index
			% \ENDWHILE
			\Until{\tcadd{Convergence of the decision variables;}}
			%			\STATE $\mathbf{\hat{w}} = \mathbf{w}_{k}$; $\mathbf{\hat{B}}=\mathbf{B}_{k}$;					
		\end{algorithmic}
		% \vspace{-1mm}
	\end{algorithm}

        \subsection{Complexity Analysis}
        \textbf{Algorithm \ref{alg_uav_placement}} employs a DC-based approach to address a complex optimization problem by breaking it down into convex subproblems, by iteratively replacing a non-convex function with a sequence of convex approximations.
        In each DCI, the subproblem is optimized using a convex solver based on the interior point method, with a complexity of $\mathcal{O}((JM)^{3.5}\log(1/\epsilon))$ \cite{ben2001lectures}, where $\epsilon > 0$ is the solution accuracy, $J$ is the number of UAV relays, and $M$ is the number of users. 
        Given convergence after $K$ DCIs, the total computational complexity of the UAV placement algorithm is $\mathcal{O}(K(JM)^{3.5}\log(1/\epsilon))$.
        For comparison, the state-of-the-art \cite{liu2019comp} has a complexity of $\mathcal{O}(K(JM)^{3.5}\log(1/\epsilon))$,
        % \sst{with the solution accuracy $\epsilon > 0$}\nm{already defined}, 
        where $J$, $M$ and $K$ represent the number of UAV relays, the number of users, and the iteration numbers to converge, respectively.
        We see that our proposed UAV placement algorithm has a comparable computational complexity to the state-of-the-art \cite{liu2019comp}. 
        However, the UAV position optimized by our algorithm achieves a $4.6 \text{ dB}$ improvement in beamforming SINR compared to the UAV position optimized by the state-of-the-art scheme \cite{liu2019comp}, as shown in Fig. \ref{fig:min_sinr_CDF}.
        % over the UAV positions optimized by the state-of-the-art, as shown in Fig. \ref{fig:min_sinr_CDF}. 
        % However, our algorithm can optimize the UAV positions\nm{I am confused: isn't [10] also a placement optimization algorithm? IF not, can you specify what it is optimizing, to avoid misunderstanding?}, resulting in a $4.6 \text{ dB}$ SINR improvement over the state-of-the-art with optimized relay beamforming and receive combining, as shown in Fig. \ref{fig:min_sinr_CDF}.
        % For comparison, the state-of-the-art of the UAV placement algorithm \cite{liu2019comp} has a complexity of $\mathcal{O}(K(JM)^{3.5}\log(1/\epsilon))$ with the solution accuracy $\epsilon > 0$, where $K$ is the iteration number.}
        % \nm{repeated sentence!! Also, can you use K instead of K1?}
        % \nm{please point out that the complexity is the same. /what are. you gaining with your scheme?}

        % \nm{note to self: continue from here}
	
	% \vspace{-1mm}
	% \section{Joint Relay Beamforming and combining optimization} 
	% \section{Minimum SINR Maximization}
        \section{Joint Relay Beamforming and Receive Combining Algorithm}
	\label{subsec_joint_opt_relay_bf_and_rx_combining}
        When UAVs are deployed to the optimized positions, the CSI required for the transceiver design can be estimated via channel training \cite{ma2011pilot}.
        This section focuses on optimizing the relay beamforming and receive combining based on the estimated CSI to further enhance the minimum SINR among users, as in \eqref{opt_max_min_modified}.
	% This section focuses on enhancing the minimum SINR among users by optimizing the relay beamforming and receive combining based on the estimated CSI, as in \eqref{opt_max_min_modified}.
	%	This section focuses on solving the optimization problem in \eqref{opt_max_min_modified} to further improve the minimum SINR among users by designing the relay beamforming $\mathbf{w}$ and receive combining $\mathbf{B}$ based on the instantaneous CSI $(\mathbf{H},\mathbf{G})$. 
	%	Although we have developed a set of solution
	%	We consider the scenario in which the UAV relays are deployed to the designated position, and the estimated CSI is available. 
	% We assume the relays transmit the signal only when they are stationary, so 
	% \nm{swap the two sentences, seems more logical to me: "When UAVs are deployed.... This section focuses...."}
	
	%	This instantaneous CSI contains the knowledge of channel phases that are critical to the design of $(\mathbf{w}, \mathbf{B})$.
	%	We jointly design the relay beamforming vector $\mathbf{w}\in\mathbb{C}^{J\times 1}$ and receive combining $\mathbf{B}\in\mathbb{C}^{N\times M}$ based on the instantaneous channel $(\mathbf{H},\mathbf{G})$.
	
	Note that Section \ref{sec_UAV_placement_optimization} also offers an optimized solution, $\mathbf{w}^*$ in \eqref{relay_bf_amplitude} and $\mathbf{b}_m^*$ in \eqref{eq_combining_opt}, to enhance the beamforming SINR in the UAV placement optimization.
	Due to the lack of CSI (particularly the channel phase) in the UAV placement algorithm, the optimized relay beamforming $\mathbf{w}^*$ is formulated based on an approximation of the beamforming SINR, leveraging the high angular resolution of the BS antenna array.
    This solution for the relay beamforming $\mathbf{w}^*$ in \eqref{relay_bf_amplitude} may not be adequate when CSI becomes available, as it relies on the orthogonality assumptions of the channels to the BS. However, the general system model of the multi-user wireless relay networks might not always exhibit such narrow beam properties, motivating us to develop a new joint design of relay beamforming and receive combining based on the estimated CSI.

    With the estimated CSI, we apply the BCD approach to construct an alternating algorithm, which separates the problem \eqref{opt_max_min_modified} into two subproblems.
	First, we optimize the receive combining $\mathbf{B}$ given a fixed $\mathbf{w}$, and the solution is provided in \eqref{eq_combining_opt}.
	Next, we optimize the relay beamforming $\mathbf{w}$ given a fixed $\mathbf{B}$, detailed in Section \ref{subsec_w_opt_with_fixed_B}. 
	The alternating algorithm is introduced in Section \ref{subsec_joint_opt_B_w}.
    % \nm{remove this next statement? It is a repetition of what you say next in Sec A}
    % \tcadd{It is worth noting that the optimization problem without the receive combining has been addressed in \cite{phan2013iterative,che2014joint}, which is equivalent to our work considering $\mathbf{B}=\mathbf{I}$ and $N=M$.}

	% \vspace{-3mm}
	
	\subsection{Optimize Relay Beamforming $\mathbf{w}$ with Fixed $\mathbf{B}$} \label{subsec_w_opt_with_fixed_B}
	% \vspace{-2mm}	
	We formulate a subproblem to maximize the minimum SINR among users by optimizing the relay beamforming with a fixed receive combining. 
    \tcadd{Note that by considering the equivalent second-hop channel obtained by integrating the original second-hop channel $\mathbf{G}$ with the receive combining $\mathbf{B}$, the methods in \cite{phan2013iterative,che2014joint} are applicable to optimize the $\mathbf{w}$ given $\mathbf{B}$. For completeness, we include this derivations below.}
    
	% It is worth noting that the optimization problem without the receive combining has been addressed in \cite{phan2013iterative,che2014joint}, which is a special case of our framework with $\mathbf{B}=\mathbf{I}$ and $N=M$.
	% Unlike \cite{phan2013iterative,che2014joint}, we provide a more general form of the optimization problem.
    
    % \nm{If this is the only difference, I don't understand how your approach is different from 15 17. If you consider the  equivalent channel obtained by combining the second hop with B,  $\mathbf G\mathbf B$, in the equivalent model: $\mathbf H'=\mathbf G\mathbf B$,  $B'=I$ and $N'=M$. Doesnt this equivalence make 15/17 readily applicable?}

    % \nm{IF the following analysis is just a repetition of the work done in previous papers, There should be a sentence along the lines: "Note that by considering the equiv channel obtained by combining ... the methods in  X and Y are applicable to optimize w given B. For completeness, we include these derivations below."}

	%	\nm{Hasnt this problem already been addressed in the literature?}
	Given a receive combining $\mathbf{B}$ and the SINR form in \eqref{eq_SINR}, the optimization problem can be formulated as 
	\begin{align}\label{opt_max_min_modified_w}
	&\max_{\mathbf{w}}\min_m\ \frac{\mathbf{w}^H\mathbf{R}_{{m}}\mathbf{w}}{\mathbf{w}^H(\mathbf{Q}_{m}+\mathbf{D}_{m})\mathbf{w} + \sigma_{\zeta}^2\lVert\mathbf{b}_m\rVert_2^2},\\
	&\text{s.t. }\lvert w_j\rvert^2\left(P_t\sum_{m\in\mathcal{M}}\lvert h_{jm}\rvert^2+\sigma_{\nu}^2\right)\leq P_{r},\  j\in\mathcal{J},
	%&\hspace{1.8em}\text{SINR}_m\geq \gamma,\ m=1,\dots,M.	
	\end{align}
	where
	\begin{align}
	\mathbf{R}_{m} &= P_t\left((\mathbf{b}_m^H\mathbf{G})\odot \mathbf{h}_m^\top\right)^H\left((\mathbf{b}_m^H\mathbf{G})\odot \mathbf{h}_m^\top\right)\\
	\mathbf{Q}_{{m}}&=P_t\sum_{k\in\mathcal{M}\setminus\{m\}}\Big(\mathbf{b}_m^H\mathbf{G}\odot \mathbf{h}_k^\top\Big)^H
	\Big(\mathbf{b}_m^H\mathbf{G}\odot \mathbf{h}_k^\top\Big)\\
	\mathbf{D}_m&=\sigma_{\nu}^2\text{Diag}(|\mathbf{G}^H \mathbf{b}_m|^2).
	\end{align}
	%	\hl{Assuming a variable vector $\mathbf{z}=[z_1,\dots, z_M]^\top$}\nm{?? Please explain, do you mean "we introduce a slack variable...."}, 
	By introducing the slack variables $\mathbf{z}=[z_1,\dots, z_M]^\top$,  we have the equivalent optimization problem as 
	\begin{align}\label{opt_DCP_modificaiton}
	&\max_{\mathbf{w},\mathbf{z}}\min_m\ \phi_m(\mathbf{w},\mathbf{z})=\frac{\mathbf{w}^H\mathbf{R}_{m}\mathbf{w}}{z_m + \sigma_{\zeta}^2\lVert\mathbf{b}_m\rVert_2^2},\\
	\label{constraint:indep_relay_pwr}
	&\text{s.t. }\lvert w_j\rvert^2(P_t\sum_{m\in\mathcal{M}}\lvert h_{jm}\rvert^2+\sigma_{\nu}^2)\leq P_{r},\  j\in\mathcal{J};\\
	\label{constraint:interference_bound}	
	&\hspace{1.8em}\mathbf{w}^H(\mathbf{Q}_{m}+\mathbf{D}_{m})\mathbf{w}\leq z_m,\ m\in\mathcal{M}.
	\end{align}
	The vector $\mathbf{z}$ is added as the upper bound of the interference power given by inequality constraint \eqref{constraint:interference_bound}.
	The SINR is maximized over $\mathbf{z}$ by suppressing the interference power.
	%	\nm{plesae clarify equivalence.}
	The SINR $\phi_m(\mathbf{w},\mathbf{z})=\frac{\mathbf{w}^H\mathbf{R}_{m}\mathbf{w}}{z_m + \sigma_{\zeta}^2\lVert\mathbf{b}_m\rVert_2^2}=\frac{\lvert(\mathbf{b}_m^H\mathbf{G}\odot \mathbf{h}_m^\top)\mathbf{w}\rvert^2}{z_m+ \sigma_{\zeta}^2\lVert\mathbf{b}_m\rVert_2^2}$ is a convex function in $(\mathbf{w}, \mathbf{z})\in\mathbb{R}\times\mathbb{R_+}$, which can be verified by the positive semi-definiteness of its Hessian matrix.
	Letting $f_1(\mathbf{w},\mathbf{z}) = \max_m \sum_{k\in\mathcal{M}\setminus\{m\}}\phi_k(\mathbf{w},\mathbf{z})$ and $f_2(\mathbf{w},\mathbf{z}) = \sum_{k\in\mathcal{M}}\phi_k(\mathbf{w},\mathbf{z})$,
	%With the properties [XXX], 
	the optimization problem is formulated as 
	\begin{align}\label{opt_problem:beamforming_DCP}
	\min_{\mathbf{w},\mathbf{z}} \left\{f_1(\mathbf{w},\mathbf{z})-f_2(\mathbf{w},\mathbf{z})\right\}, \text{ s.t. }\eqref{constraint:indep_relay_pwr},\eqref{constraint:interference_bound}.
	\end{align}
	%where 
	%$f_1(\mathbf{w},\mathbf{z}) = \max_m \sum_{i\neq m}\phi_i(\mathbf{w},{z}_i)$ and $f_2(\mathbf{w},\mathbf{z}) = \sum_{m=1}^{M}\phi_m(\mathbf{w},{z}_m)$.
	Note that $f_1$ and $f_2$ are convex functions with respect to $(\mathbf{w},\mathbf{z})$.
	The objective function in \eqref{opt_problem:beamforming_DCP} is the difference of two convex functions, which can be addressed by the DC framework \cite{phan2013iterative}.
	%	 \nm{you called it diff of convex before. Isnt it the same thing? Please clarify and be consistent.} \cite{phan2013iterative}.
	To solve such a problem, we derive a linear approximation of $f_2(\mathbf{w},\mathbf{z})$ at the point $(\mathbf{w}^{(\iota)},\mathbf{z}^{(\iota)})$, given by
	%	\nm{you are already using k to denote user index above!}
	%apply a linear approximation to make $f_2(\mathbf{w},\mathbf{z})$ an affine function at the point $(\mathbf{w}^{(k)},\mathbf{z}^{(k)})$, given by
	\begin{align}
	\nonumber
	&\tilde {f}_2(\mathbf{w},\mathbf{z};\mathbf{w}^{(\iota)},\mathbf{z}^{(\iota)}) \\
	\nonumber
	= &f_2(\mathbf{w}^{(\iota)},\mathbf{z}^{(\iota)})+	\left<\nabla_{\mathbf{w}} f_2(\mathbf{w}^{(\iota)},\mathbf{z}^{(\iota)}),\mathbf{w}-\mathbf{w}^{(\iota)}\right>\\
	&+	\left<\nabla_{\mathbf{z}} f_2(\mathbf{w}^{(\iota)},\mathbf{z}^{(\iota)}),\mathbf{z}-\mathbf{z}^{(\iota)}\right>,
	\end{align}
	where 
	\begin{align}
	\nonumber
	&\left<\nabla_{\mathbf{w}} f_2(\mathbf{w}^{(\iota)},\mathbf{z}^{(\iota)}),\mathbf{w}-\mathbf{w}^{(\iota)}\right>\\
	=&2 \text{Re}\left(\sum_{m\in\mathcal{M}}\frac{(\mathbf{w}^{(\iota)})^H\mathbf{R}_m}{z_m^{(\iota)}+\sigma_{\zeta}^2\lVert\mathbf{b}_m\rVert_2^2}(\mathbf{w}-\mathbf{w}^{(\iota)})\right),
	\end{align}
	% \vspace{-1mm}
	\begin{align}\nonumber
	&\left<\nabla_{\mathbf{z}} f_2(\mathbf{w}^{(\iota)},\mathbf{z}^{(\iota)}),\mathbf{z}-\mathbf{z}^{(\iota)}\right>\\
	=&-\sum_{m\in\mathcal{M}}\frac{(\mathbf{w}^{(\iota)})^H\mathbf{R}_m\mathbf{w}^{(\iota)}}{\left(z_m^{(\iota)}+\sigma_{\zeta}^2\lVert\mathbf{b}_m\rVert_2^2\right)^2}(z_m-z_m^{(\iota)}).
	\end{align}
	%where 
	%\begin{align}
	%	&\left<\nabla f_2(\mathbf{\tilde w}^{(k)},\mathbf{\tilde q}^{(k)}),(\mathbf{w}-\mathbf{\tilde w}^{(k)},\mathbf{q}-\mathbf{\tilde q}^{(k)})\right>\\
	%	&=\sum_{m=1}^{M}\left[\frac{2\mathbf{w}^H\mathbf{R}_m}{q_m+\tilde{\sigma}_n^2}(\mathbf{w}-\mathbf{\tilde w}^{(k)})-\frac{\mathbf{w}^H\mathbf{R}_m\mathbf{w}}{(q_m+\tilde{\sigma}_n^2)^2}(q_m-\tilde{q}_m^{(k)})\right]
	%\end{align}
	The DC programming operates as follows. At the $\iota$-th DCI, the optimization problem is transformed into
	\begin{align}\label{opt_problem:DCI}
	% \nonumber
	&\min_{\mathbf{w},\mathbf{z}} \Big\{f_1(\mathbf{w},\mathbf{z})-\tilde f_2(\mathbf{w},\mathbf{z};\mathbf{w}^{(\iota)},\mathbf{z}^{(\iota)})\Big\},\\
	%	&\hspace{2em}-\left<\nabla f_2(\mathbf{w}^{(k)},\mathbf{q}^{(k)}),(\mathbf{w}-\mathbf{w}^{(k)},\mathbf{q}-\mathbf{q}^{(k)})\right>\Big\},\\
	&\text{s.t. }\eqref{constraint:indep_relay_pwr},\eqref{constraint:interference_bound}.	
	\end{align}
	Since $f_1$ is convex and $\tilde{f}_2$ is affine, \eqref{opt_problem:DCI} is a convex optimization, solved by the CVX package \cite{cvx}, 
	% e.g., an interior-point method, 
	leading to a solution $(\mathbf{w}^{(\iota)}_*,\mathbf{z}^{(\iota)}_*)$ in the $\iota$-th DCI.
	In the next (i.e., $(\iota+1)$-th) DCI, we consider the optimization problem \eqref{opt_problem:DCI} by substituting the objective function with $f_1(\mathbf{w},\mathbf{z})-\tilde f_2(\mathbf{w},\mathbf{z};\mathbf{w}^{(\iota+1)},\mathbf{z}^{(\iota+1)})$, where $(\mathbf{w}^{(\iota+1)},\mathbf{z}^{(\iota+1)})=(\mathbf{w}^{(\iota)}_*,\mathbf{z}^{(\iota)}_*)$.
	The solution can be obtained by solving the convex optimization problem, yielding  $(\mathbf{w}^{(\iota+1)}_*,\mathbf{z}^{(\iota+1)}_*)$.
	{The DCI implementation stops when the solution converges, i.e., $\lVert\mathbf{w}_*^{(\iota+1)}- \mathbf{w}_*^{(\iota)} \rVert_2+\lVert\mathbf{z}_*^{(\iota+1)}- \mathbf{z}_*^{(\iota)} \rVert_2\leq \epsilon^\prime$, where $\epsilon^\prime>0$ is a given threshold.}

	\begin{algorithm}[t!]
		\caption{\tcadd{Joint Relay Beamforming and Receive Combining Algorithm (JRBC).}}
		% 	\label{alg_gain_delay_refine}
		\label{alg_joint_relay_combining_design}
		\begin{algorithmic}[1]
			% \REQUIRE 
			\INPUT first-hop channel $\mathbf{H}$, second-hop channel $\mathbf{G}$%measurement $\mathbf{Y}_k$, dictionary matrix $\mathbf{\Theta}_k$, support set $\Tilde{\Xi}$			
			\OUTPUT relay beamforming ${\mathbf{w}}$, receive combining $\mathbf{B}$
			%			\STATE \emph{Initialization}
			%			\STATE $E_{otr}=\infty$; 
			%			\STATE $\mathbf{w}_{ini}\in\mathbb{C}^{J\times 1}$ with amplitude \eqref{relay_bf_amplitude} and random phase;
			%			\item[]
			\Statex \emph{Optimize $\mathbf{w}$ and $\mathbf{B}$ iteratively}
			% \State $E_{otr}=\infty$;
			\State Set $\mathbf{w}$ with amplitude of \eqref{relay_bf_amplitude} and random phase;
			%			\STATE $E_{otr}=\infty$;
			%			\STATE Initiate $\mathbf{w}^{(1)}\in\mathbb{C}^{J\times 1}$ with amplitude \eqref{relay_bf_amplitude} and random phase;
			% \WHILE{$E_{otr}>\epsilon$ }%\textbf{and} $t<t_{max}$}
            \Repeat
			\Statex \emph{\ \ \ \ Optimize receive combining $\mathbf{B}$ with a fixed $\mathbf{w}$}
			\State Derive $\mathbf{b}_m$ by \eqref{eq_combining_opt} with $\mathbf{w}$;
			%		solving \eqref{opt_receiver_design} with $\mathbf{w}_{t+1}$;
			\State $\mathbf{B}(:,m)=\mathbf{b}_m,\ m\in\mathcal{M}$; 
			\item[]
			%		\STATE Derive $\mathbf{K}$ by solving \eqref{opt_max_min_SDP} with $\mathbf{B}_t$;
			\Statex \emph{\ \ \ \ Optimize relay beamforming $\mathbf{w}$ with a fixed $\mathbf{B}$}
			% \State $E_{in}=\infty$;% $i_1=1$;
            % \State Set 
            \State $\iota = 0$;
            \Repeat
			% \WHILE {$E_{in}> \epsilon^\prime$}%} \textbf{and} $k<k_{max}$}
			\State Derive $(\mathbf{\bar{w}}, \mathbf{\bar{z}})$ by solving \eqref{opt_problem:DCI} with $(\mathbf{\bar{w}}^{(\iota)}, \mathbf{\bar{z}}^{(\iota)},\mathbf{B})$;        
            % \State \emph{Record variables for next iteration}
            \Statex \emph{\ \ \ \ \ \ \ \ \ Record variables for next iteration}
            \State $\iota := \iota + 1$;
            \State $(\mathbf{\bar{w}}^{(\iota)}, \mathbf{\bar{z}}^{(\iota)})=(\mathbf{\bar{w}}, \mathbf{\bar{z}})$;
			% \State $\mathbf{\bar{w}}_{old}=\mathbf{\bar{w}}$; $\mathbf{\bar{z}}_{old}=\mathbf{\bar{z}}$;
			%			\STATE $i_1\coloneqq i_1+1$; \% inner-loop index
			% \ENDWHILE
            \Until{\tcadd{Convergence of the decision variables $(\mathbf{\bar{w}}, \mathbf{\bar{z}})$;}}
            \State $\mathbf{w}=\mathbf{\bar{w}}$;
			% \State $\mathbf{w}=\mathbf{\bar{w}}$;
			%		\STATE Derive $\mathbf{w}_{t+1}$ from $\mathbf{K}$;			
			% \item[]
			% \State $E_{otr}=\lVert\mathbf{w}-\mathbf{w}_{old}\rVert + \lVert\mathbf{B}-\mathbf{B}_{old}\rVert_F$
			% \State $\mathbf{w}_{old}=\mathbf{w}$; $\mathbf{B}_{old}=\mathbf{B}$;
			%			\STATE $k\coloneqq k+1$; \% outer-loop index
			% \
            \Until{\tcadd{Convergence of the decision variables;}}
			%			\STATE $\mathbf{\hat{w}} = \mathbf{w}_{k}$; $\mathbf{\hat{B}}=\mathbf{B}_{k}$;					
		\end{algorithmic}
		% \vspace{-1mm}
	\end{algorithm}
	
	% \vspace{-3mm}
	\subsection{Joint Optimization of $\mathbf{B}$ and $\mathbf{w}$}
	\label{subsec_joint_opt_B_w}
	%Given the first-hop and second-hop channel, 
	The joint relay beamforming and receive combining algorithm using a BCD approach is shown in \textbf{Algorithm \ref{alg_joint_relay_combining_design}}.
	% The algorithm to optimize the relay beamforming and receive combining using a BCD approach is detailed in \textbf{Algorithm \ref{alg_joint_relay_combining_design}}.
	First, we optimize the receive combining $\mathbf{B}$ with a fixed $\mathbf{w}$ by \eqref{eq_combining_opt}.
	%	Assuming $g(\mathbf{w},\mathbf{B})=\min_m SINR_m(\mathbf{w},\mathbf{B})$, we have $g(\mathbf{w}_{k},\mathbf{B}_{k+1})\geq g(\mathbf{w}_{k},\mathbf{B}_{k})$, where $\mathbf{B}_{k+1}$ is obtained by solving DCI problem \eqref{opt_problem:DCI}.
	Next, we optimize the relay beamforming $\mathbf{w}$ with a fixed $\mathbf{B}$.
	%	\nm{again, k used for users already}
	Letting $g(\mathbf{w},\mathbf{B})=\min_m \text{SINR}_m(\mathbf{w},\mathbf{B})$, after updating $\mathbf{B}$, we have $g(\mathbf{w}_{k},\mathbf{B}_{k+1})\geq g(\mathbf{w}_{k},\mathbf{B}_{k})$, where $k$ denotes the iteration number.
	Then, we update $\mathbf{w}$ by solving \eqref{opt_problem:DCI}, leading to $g(\mathbf{w}_{k+1},\mathbf{B}_{k+1})\geq g(\mathbf{w}_{k},\mathbf{B}_{k+1})$.
	This shows that the iterative update on $\mathbf{B}$ and $\mathbf{w}$ improves the minimum beamforming SINR among users. 
    The iterative algorithm stops when $\lVert\mathbf{w}_{k+1}-\mathbf{w}_{k}\rVert_2  + \lVert\mathbf{B}_{k+1}-\mathbf{B}_{k}\rVert_F <\epsilon$, where $\epsilon>0$ is a given threshold.

	\tcadd{\subsection{Complexity Analysis}
        The JRBC algorithm utilizes the BCD approach to develop an alternating algorithm, dividing the optimization problem into two subproblems.
        First, we optimize the receive combining with a fixed relay beamforming, with the solution in \eqref{eq_combining_opt}. The complexity of \eqref{eq_combining_opt} is dominated by a matrix inversion operation, which has a complexity of $\mathcal{O}(N^3)$, where $N$ is the number of BS receive antennas.
        Next, we optimize the relay beamforming given a fixed receive combining, using the DC-based approach detailed in Section \ref{subsec_w_opt_with_fixed_B}.
        In each DCI, we solve the subproblem using a convex solver based on the interior point method, with a complexity of $\mathcal{O}((J+M)^{3.5}\log(1/\epsilon))$ \cite{ben2001lectures}, where $\epsilon > 0$ is the solution accuracy. $J$ and $M$ represent the number of UAV relays and the number of users, respectively.
        Given convergence after $K_d$ DCIs, the total computational complexity of the UAV placement algorithm is $\mathcal{O}(K_d(J+M)^{3.5}\log(1/\epsilon))$. 
        Assuming BCD converges after $K_b$ iterations, the total computational complexity of the JRBC algorithm is $\mathcal{O}(K_bN^3 + K_bK_d(J+M)^{3.5}\log(1/\epsilon))$.
        % \nm{what is K3 here?}.
        For comparison, the state-of-the-art of the transceiver design \cite{behbahani2008optimizations} has a complexity of $\mathcal{O}(\min\left(J^2M,JM^2\right)+N^3)$.
        The proposed JRBC algorithm requires higher complexity than the state-of-the-art \cite{behbahani2008optimizations}. However, JRBC achieves a better $\mathbb{E}[\min_m \text{SINR}_m] = 13\text{ dB}$, outperforming the state-of-the-art by $4.5\text{ dB}$ in the simulation setting $(M,J,N)=(4,4,4)$ and $P_t=P_r=30\text{ dBm}$, as shown in Fig. \ref{fig:expected_sinr_vs_Pr}.}
        % \nm{how do you compare your complexity with state of art? What are you trading off?}

	% \vspace{-1mm}
	\section{Numerical Results}\label{sec_numerical_results}
	%	\nm{The structure of these evaluations is not quite clear. Please add subsections to differentiate between different types of evaluations, i.e.: position optimization, JRBC, joint optimization, SoA comparisons, etc.}
	In this section, we evaluate the performance of the UAV placement algorithm and the JRBC algorithm in a multi-user relay network. %\sst{, and the}\add{. The} 
    The numerical parameters are listed in Table \ref{table:simulation_parameters}. 
	% To evaluate the UAV placement algorithm, 
	% Unless otherwise stated, w
    We consider an uplink transmission in a 3-D environment with $M$ users communicating with an $N$-antenna BS, aided by $J$ single-antenna UAV relays.
% \sst{	We consider $(M, J, N) = (4,4,40)$ for evaluations. }
	We assume no direct transmission is available between users and BS.
	The users' positions, denoted as $\mathbf{u}_m=(u_{mx},u_{my},u_{mz})$ for $m\in\mathcal{M}$, are on the ground with $u_{mz} = 0\text{ m}$ and randomly distributed within an area as $(u_{mx},u_{my})\in[0,80]\text{ m}\times[0,80]\text{ m}$. 
	The BS is equipped with an $N\times 1$ ULA vector with antennas spaced by half-wavelength along the y-axis, with the reference antenna in position $\mathbf{q}_1 = (50,40,40)\text{ m}$.
    \tcminor{The BS antenna height of $40\text{ m}$ complies with the deployment requirements for rural macro scenario as defined in 3GPP TR 38.901 \cite{3GPP38901}, which specifies the applicable range of BS antenna heights as $10-150\text{ m}$.
	The $J$ UAV relays fly within an altitude range of $15$ m to $25$ m above ground level, adhering to the UAV altitude limits outlined in 3GPP TS 22.125 \cite{3gpp_ts_22125_v1920}.}
        Although height optimization is not originally included in our UAV placement problem formulation, it can be incorporated without violating the problem's convexity.
	% \hl{This constraint can be included in our UAV placement optimization without violating the convexity.}\nm{please be more specific. do you mean the following? although height optimization was not originally included in the placement problem, .....}
	% The minimum distance between UAVs is set as $\epsilon_p=3$ m.
	% The carrier frequency is set as $2.4$ GHz.
	% For the channel parameters, we consider the path loss exponent $\ell = 2.3$, the channel gain $\beta_0 = -35$ dB at the reference distance of $1$ m, and the Rician factor $K_r=8$ \cite{liu2019comp}.
	% Each user's transmit power and each UAV relay's power budget are set to $P_t=P_r=30$ dBm. 
	% The noise power at the relays and the BS antennas are set as $\sigma_{\nu}^2=\sigma_{n}^2 = -95$ dBm.
	% The above parameter settings are assumed unless they are otherwise specified. 
	%	\nm{I don't think you defined the values of J M N}
	%	We assume the above parameter setting unless we otherwise specified. %in the experiments.
	
    \begin{table}[t]%\small
		\centering	
		\caption{Common Simulation Parameters}
		% \vspace{-1em}
		\label{table:simulation_parameters}		
		\resizebox{0.45\textwidth}{!}
		{		
			\begin{tabular}{|r   | c|l|} % centered columns (4 columns)
				%			\hline
				\hline %inserts double horizontal lines
				%			&&&&&&\\
				Parameter & Symbol & Value \\ [0.5ex] % inserts table
                % $J$ & $N_u$ & $\mathbb{E}[\min_m \text{SINR}_m]$\\ [0.5ex]
				%heading
				\hline % inserts single horizontal line
				Carrier frequency & $f_c$  & $2.4\text{ GHz}$ \\
                Pathloss exponent & $\ell$  & $2.3$ \\ 			          
                Rician factor & $K_r$  & $8$ \\
                Reference channel gain & $\beta_0$  & $-35\text{ dB}$ \\  
                Noise power at UAV, BS& $\sigma_\nu^2,\sigma_n^2$  & $-95\text{ dBm}$ \\
                Number of users & $M$  & $4$ \\
                Number of UAV relays & $J$  & $4$ \\
                Number of BS antennas & $N$  & $40$ \\
                User transmit power & $P_t$  & $30\text{ dBm}$  \\
                UAV relay power budget & $P_r$  & $30\text{ dBm}$ \\
                Minimum UAV distance & $\epsilon_p$  & $3\text{ m}$ \\ 	
				\hline %inserts single line
			\end{tabular}						
		}	
		
\end{table}

	% \begin{figure}[t]
	% 	\centering
	% 	\begin{subfigure}[t]{0.5\textwidth}
	% 		\centering
	% 		\includegraphics[width=0.8\textwidth]{Journal_Fig1_DCI_trajectory_V1.eps}\\
	% 		(a)
	% 		%		\caption{}
	% 		%		\label{fig:beam_forming_pattern} 
	% 	\end{subfigure}
	% 	\hfill
	% 	\begin{subfigure}[t]{0.5\textwidth}
	% 		\centering
	% 		\includegraphics[width=0.8\textwidth]{Journal_Fig1_DCI_trajectory_birdview_V1.eps}\\
	% 		(b)
	% 		%					\caption{}
	% 		%		\label{fig:expected_beam_forming_pattern}
	% 	\end{subfigure}	
	% 	\caption[UAV trajectory]{(a) 3D plot figure (b) Birds eye view figure. The dash line is the trajectory of each UAV along the DCIs.
	% 		%			\nm{I find Fig a not much informative. Remove to save space? Or perhaps Fig 3 and 4 can be made smaller to fit one line.}
	% 	}
	% 	\label{fig:uav_trajectory}
	% \end{figure}

        \begin{figure}[t]		\centering %\setcounter{figure}{0}
		\includegraphics[scale=0.55]{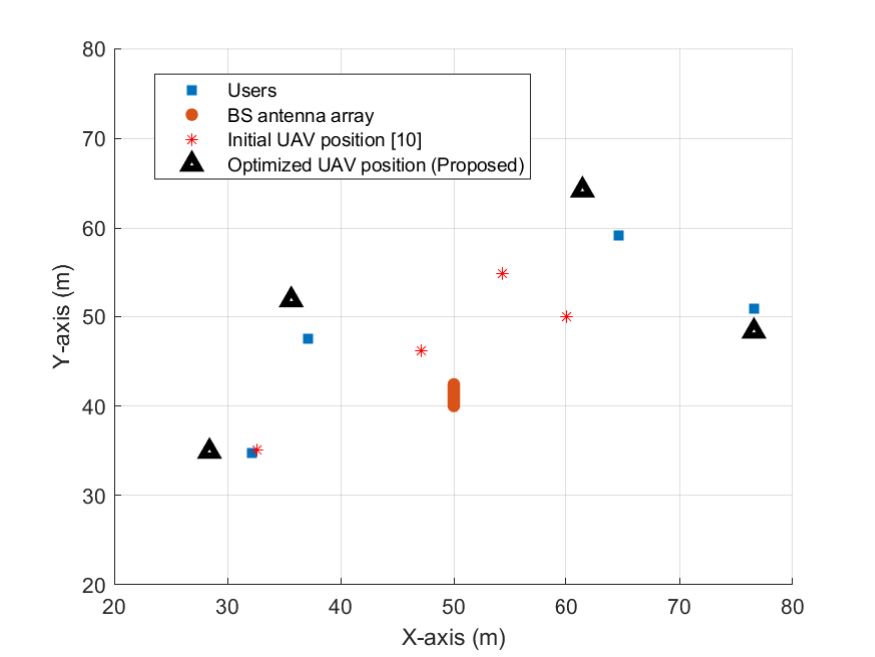}
		\caption{
			The considered scenario illustrates the positions of users, BS, initial UAV positions, and optimized UAV positions.
            % \nm{For all figures, please check the reference number (to equations and/or papers) in the legends, when you have the final version ready. they might have changed}
            % \nm{please use different marker for initial positions, like asterisk. in my mind, circles are fixed positions like users.}
            }
		\label{fig:uav_trajectory}
		% \vspace{-2mm}
	\end{figure}
    
	\begin{figure}[t]		\centering %\setcounter{figure}{0}
		\includegraphics[scale=0.55]{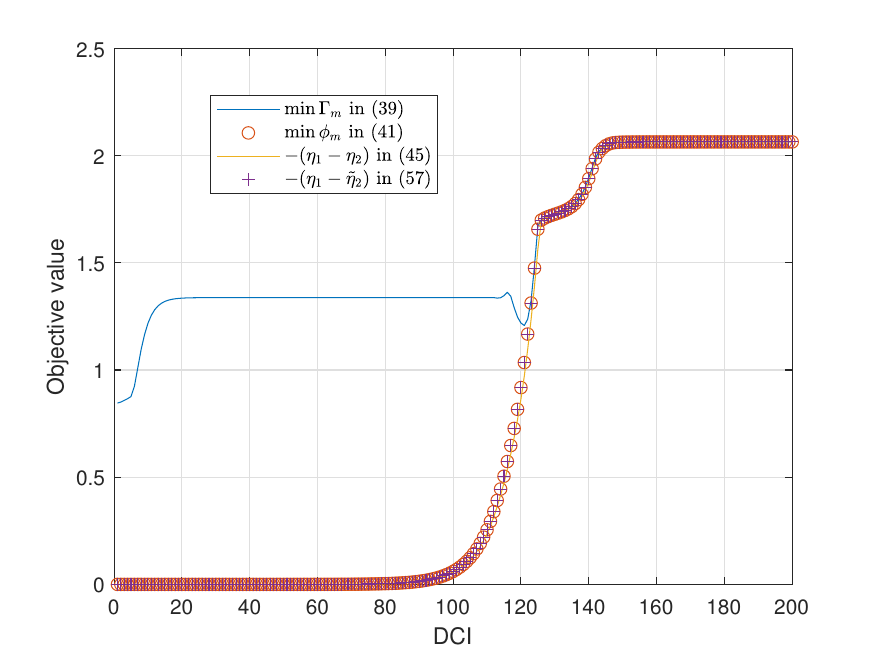}
		\caption{
			The objective values versus the DCI in UAV placement optimization, with $(M,J,N)=(4,4,40)$.	}
		\label{fig:obj_val_vs_DCI}
		% \vspace{-2mm}
	\end{figure}

	% In Fig. \ref{fig:uav_trajectory}, we showcase the 3D placement of the users, BS, and UAV relays. 
	%    We optimize the positions of UAV relays by our proposed UAV placement algorithm, given the initial positions derived by COMP \cite{liu2019comp}. 
	%    The trajectory (dashed line) represents the DCI iteration of our proposed UAV placement algorithm.
	%    The birds eye view is also provided in \ref{fig:uav_trajectory}(b). 
	%    We observe the trend that the optimized UAV positions are distributed close to each user, but separated from the other relays.

	% we compare the proposed UAV placement with SAA and COMP.
	
	% In addition, for the fair comparison, 
	
	%	In Fig. \ref{fig:uav_trajectory}, we showcase the UAV trajectory  along the DC iterations in the optimization procedure of our UAV placement algorithm. The initial UAV positions are the one derived by the algorithm in \cite{liu2019comp}.
	%	To show the advantage of our proposed UAV placement algorithm, we initiate the UAV positions as the one optimized by the algorithm in \cite{liu2019comp}.
	%	Note that the work \cite{liu2019comp} optimize the UAV relay positions to maximize the minimum transmission rate among users, while consider only the first-hop channel with zero-forcing receiver, and ignore the second-hop channel.
	%	Instead, our proposed UAV placement algorithm optimize the UAV relay positions for the system with joint relay beamforming and receive combining, which considers both the first-hop channel and the second-hop channel.
	
	% \vspace{-1em}
	\subsection{Numerical Evaluation of UAV Placement Algorithm}
	% In Fig. \ref{fig:uav_trajectory} (a), we show the UAV position optimized by our UAV placement algorithm. 
	% The figure with a bird's eye view is in Fig. \ref{fig:uav_trajectory} (b). 
        \tcadd{In Fig. \ref{fig:uav_trajectory}, we show the UAV position with a bird's eye view optimized by our UAV placement algorithm.}
	To show the advantage of our UAV placement algorithm over the state-of-the-art, we initiate the UAVs on the positions optimized by the state-of-the-art \cite{liu2019comp}, denoted as the Initial UAV position.
	Note that the work \cite{liu2019comp} optimizes the UAV positions to maximize the minimum transmission rate among users, but it considers only the first-hop channel.
	In contrast, our proposed UAV placement algorithm optimizes the UAV positions for the system with joint relay beamforming and receive combining, which considers both the first-hop and second-hop channels. 	
	% We observe that the optimized UAV positions are near the users and separated from each other.	
    % We observe that each optimized UAV position is close to one user while being farther from the other users. 
    % In our scenario, we observe that the four UAVs are near the four users, respectively, leading to a larger desired signal power. Also, for the UAV close to one user, this UAV is farther from the remaining users, leading to a smaller interference power from other users.
    \tcadd{In our optimized scenario, each of the four UAVs is positioned close to a respective user, while maintaining a greater distance from the other users.
    Note that, in the second-hop channel, the signals from the UAVs are more spatially-separated due to the narrow beam property of a massive MIMO BS, which reduces inter-UAV interference.
    However, the first-hop channels between the users and UAVs lack this spatial separation for multiple users, so each UAV is positioned closer to a certain user and farther from other users to reduce inter-user interference.}
	
    % \nm{I am confused: if paper 8 does not consider the second hop, wouldn't it have a similar incentive in placing UAVs close to one ground node and far from the others?}
    % \tc{[8] applies "the ZF beamforming across UAVs" (processed at a central unit) to eliminate the inter-user interference. In other words, they assume non-orthogonal transmission, but have the ZF receiver. Therefore, "the UAV being far from other users" is not manifested with their algorithm.}

	% \begin{figure}[t]
	% 	\centering
	% 	\begin{minipage}[t]{0.5\textwidth} 
	% 		\centering                
	% 		\includegraphics[width=1\textwidth]{Journal_Fig5_expectedSINRdB_UE_vs_DistanceDiff_V6.eps}\\
	% 		(a)  
	% 	\end{minipage}
	% 	\hfill
	% 	\begin{minipage}[t]{0.5\textwidth}
	% 		\centering                
	% 		\includegraphics[width=1\textwidth]
 %            {Journal_Fig5_GammadB_UE_vs_DistanceDiff_V3.eps}\\
	% 		(b) 
	% 	\end{minipage}	
	% 	\caption[UAV trajectory]{
	% 		\tcadd{We evaluate the performance metrics by considering one UAV deviated from its optimized positions along the x-axis, with the remaining $3$ UAVs at the optimized position, with $(M,J,N)=(4,4,40)$.}
	% 		(a) The expected SINR versus the distance difference. The dash line is the expected value of ${\text{SINR}}_m$ in \eqref{SINRm_V2} with $(\mathbf{w},\mathbf{B})$ optimized by the JRBC algorithm. 
	% 		The circle mark is the expected value of $\hat{\text{SINR}}_m$ in \eqref{eq_SINR_maximized_by_w}. 
 %            (b) $\Gamma_m$ versus the distance difference.
	% 	}		
	% 	\label{fig:Ex_SINR_vs_distance_diff}
	% 	\vspace{-2mm}
	% \end{figure}

        \begin{figure}[t]		\centering %\setcounter{figure}{0}
		\includegraphics[scale=0.55]{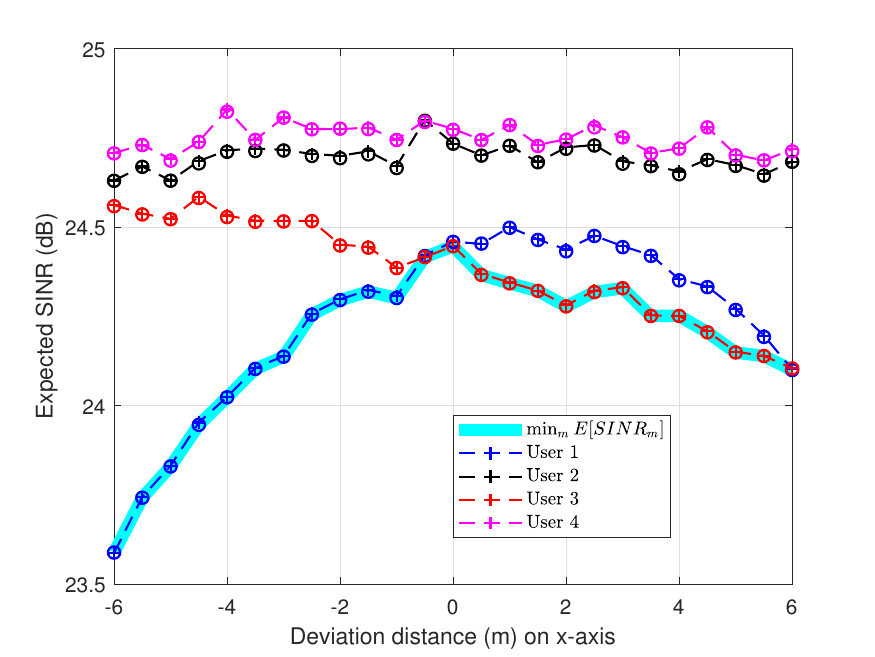}
		\caption{ The expected beamforming SINR of each user versus the deviation of one UAV from its optimized position along the x-axis, with the remaining $3$ UAVs at the optimized position. The dash line is the expected value of ${\text{SINR}}_m$ in \eqref{SINRm_V2} with $(\mathbf{w},\mathbf{B})$ optimized by the JRBC algorithm. The circle mark is the expected value of $\widehat{\text{SINR}}_m$ in \eqref{eq_SINR_maximized_by_w}.
            }
		\label{fig:Ex_SINR_vs_distance_diff}
		% \vspace{-2mm}
	\end{figure}
	
	In Fig. \ref{fig:obj_val_vs_DCI}, for UAV placement optimization, we compare the objective value of $\min_m\Gamma_m$ in \eqref{opt_prob_to_be_solved}  with its approximate functions, $\min_m \phi_m$ in \eqref{sub_obj_func1}, $-(\eta_1 - \eta_2)$ in \eqref{optimization_UAV_movement_with_large_scale_DCP}, and $-(\eta_1 - \tilde{\eta}_2)$ in \eqref{optimization_UAV_movement_DCP_covex}, to evaluate the validity of the approximate functions in the DC framework.
	We observe that $\min_m \phi_m$,  $-(\eta_1 - \eta_2)$, and $-(\eta_1 - \tilde{\eta}_2)$ have almost the same values in the DCIs.
	These approximate functions catch up with the value of $\min_m\Gamma_m$ after $\text{DCI}=120$, and they converge at the maximum of $2.1$ after $\text{DCI}=140$.
	Fig. \ref{fig:obj_val_vs_DCI} shows that $-(\eta_1 - \tilde{\eta}_2)$ is a good approximation for $\min_m\Gamma_m$ in DC framework, which optimizes the UAV positions to have the maximum of $\min_m\Gamma_m$.
    \tcadd{Note that the objective value remains almost constant between $\text{DCI}=30$ and $\text{DCI}=120$, while the decision variables continue to change within this range. This observation validates the stop criterion based on the convergence of decision variables in \textbf{Algorithm \ref{alg_uav_placement}}, rather than the convergence of the objective value.}
	
	%	we evaluate the objective function $\min_m\Gamma_m$ in our formulated UAV placement optimization problem \eqref{opt_prob_to_be_solved} versus the DC iteration (DCI).
	%	We address the problem \eqref{opt_prob_to_be_solved} using DC framework
	%%	we evaluate the objective values $\min_m\Gamma_m$ (as in \eqref{eq_lower_bound}) of the UAV placement optimization problem  versus the DC iterations.
	%	The function $\min_m\Gamma_m$ is the approximation of the minimum expected beamforming SINR among users, i.e., $\min_m \phi_m$ in \eqref{sub_obj_func1}.
	%	To show the tightness between the 
	%	 accompanied with its substitute functions in our derivations, i.e., $\min_m \phi_m$ in \eqref{sub_obj_func1}, $-(\eta_1 - \eta_2)$ in \eqref{optimization_UAV_movement_with_large_scale_DCP}, and $-(\eta_1 - \tilde{\eta}_2)$ in \eqref{optimization_UAV_movement_DCP_covex}.
	%	We observe that $\min_m \phi_m$,  $-(\eta_1 - \eta_2)$, and $-(\eta_1 - \tilde{\eta}_2)$ have almost the same values in the DC iterations.
	%	These substitute functions catch up with the value of $\min_m\Gamma_m$ after $DCI=120$, and they converge at the objective value of $2.1$ after $DCI=140$.
	
	\tcadd{In Fig. \ref{fig:Ex_SINR_vs_distance_diff}, we evaluate the expected SINRs of $M=4$ users as one UAV deviates from its optimized position along the x-axis, while the remaining $3$ UAVs stay at their optimized positions.
    % \nm{grammar of next sentence! compares.. are evaluated..}
    Fig. \ref{fig:Ex_SINR_vs_distance_diff} compares two different expected SINRs,  ${\text{SINR}}_m$ (as defined in \eqref{SINRm_V2}) and $\widehat{\text{SINR}}_m$ (as defined in \eqref{eq_SINR_maximized_by_w}).  
	${\text{SINR}}_m$ represents the beamforming SINR using $(\mathbf{w},\mathbf{B})$ optimized by the JRBC algorithm.
	$\widehat{\text{SINR}}_m$ is an approximated beamforming SINR that leverages the narrow beam property of the BS.
	Fig. \ref{fig:Ex_SINR_vs_distance_diff} shows that $\mathbb{E}[\widehat{\text{SINR}}_m]$ closely matches $\mathbb{E}[{\text{SINR}}_m]$, indicating that $\widehat{\text{SINR}}_m$ is a reliable approximation of the beamforming SINR.
    Additionally, we observe that the local maximum of $\min_m \mathbb{E}[\text{SINR}_m]$ occurs at a deviation distance of $0\text{ m}$, validating the use of the approximation $\Gamma_m$ (as defined in \eqref{eq_lower_bound}) in the DC-based framework for UAV placement optimization and demonstrating the effectiveness of our proposed UAV placement algorithm.}

	\begin{figure}[t]		\centering %\setcounter{figure}{0}
		\includegraphics[scale=0.55]{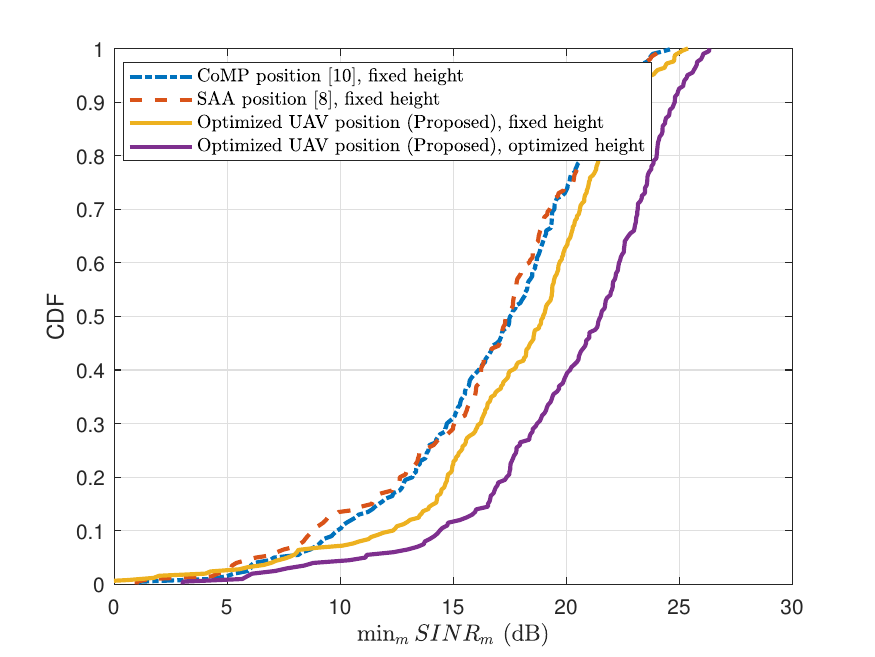}
		\caption{
			The CDF of $\min_m \text{SINR}_m$, with $(M,J,N)\!=\!(4,4,40)$.
			%			 The JRBC algorithm is applied for each UAV placement scheme for fair comparison.
			%			The JRBC is applied as the relay beamforming and receiver design for each scheme.
			%			 layout of users, BS antennas, and UAV relays, $(M,J,N)=(3,3,20)$.
		}
		\label{fig:min_sinr_CDF}
		% \vspace{-2mm}
	\end{figure}
	
	For the proposed UAV placement algorithm, we compare it with the existing schemes listed as follows:
	\begin{itemize}
		\item \textbf{SAA position} \cite{kalogerias2018spatially}: This method uses a direct Monte Carlo approach with sampling to identify the optimal UAV position within its neighboring area. 
		Originally designed for a single-user relay network, we adapt it for multi-user scenarios by iteratively optimizing the UAV position for the worst user, i.e., the user with the smallest beamforming SNR.
        The optimization of the UAV position for the worst user is carried out under the assumption of orthogonal transmission.
        % Note that orthogonal transmission is assumed during the optimization of the UAV position for the worst user.
  %       To optimize the UAV position for the worst user, the orthogonal transmission is assumed.
  %       Orthogonal transmission is assumed
		% \hl{Note that the orthogonal transmission is assumed during the optimization for each user.}\nm{grammar: rephrase.}
		%		\nm{are you assuming orthogonal allocation?} \tc{Do you mean the collision avoidance?}
		%		This approach is designed for the single-user relay network. To enable it in multi-user cases, we apply a sub-gradient descent-like method to iteratively optimize the UAV relay positions for the worst user, i.e., the one with the smallest beamforming SNR. 
		\item \textbf{CoMP position} \cite{liu2019comp}:
		This method is designed for an uplink multi-UAV enabled multi-user system. The UAVs receive signals from users and then forward it to a central processor for joint decoding using a ZF receiver. Note that this design is based solely on the first-hop channel, assuming the second-hop channel is ideal.
		%		 A UAV relay placement algorithm considers an uplink multi-UAV enabled multi-user system, where each UAV forwards the received signals from all users to a central processor for joint decoding with a ZF receiver. However, the design is based on the first-hop channel, while assuming the second-hop channel is perfect.
	\end{itemize}

    % \nm{in this paragraph, please stress more clearly the fact that, even if you keep the uav height at a fixed level, you are outperforming state of art. If you allow height to be optimized, you gain even further. This is needed in response to one reviewer comment.}
	In Fig. \ref{fig:min_sinr_CDF}, we evaluate the cumulative distribution function (CDF) of the minimum SINR among users. 
    To ensure a fair comparison across UAV placement algorithms, we adopt the same transceiver design, i.e., JRBC algorithm.
	Since the works \cite{kalogerias2018spatially, liu2019comp} optimize the UAV placement on a surface with fixed height, we first evaluate the performance of UAV placement algorithms assuming that UAVs operate at a fixed height of $20\text{ m}$.
    At $\text{CDF} =0.5$, the UAV position optimized by our algorithm achieves $\min_m \text{SINR}_m=19\text{ dB}$, which outperforms SAA position and CoMP position by $1.6\text{ dB}$.      
    % Note that, in our scenario, lowering the UAV height shortens the distance to users (i.e., enhancing the first-hop channel gain) but increases the distance to the BS (i.e., degrading the second-hop channel gain). Therefore, a lower UAV height does not necessarily lead to an improved SINR performance.
    % does not better SINR performance.  
    % Also, in our considered scenario, a lower UAV height decreases the distance between the UAVs and the users (i.e., the first-hop channel), but it increases the distance between the UAVs and the BS (i.e., the second-hop channel). 
    % Thus, a lower UAV height does not guarantee better SINR performance. 
	Our UAV placement algorithm can optimize the UAV position in 3-D coordinates. 
    Note that, in our scenario, lowering the UAV height shortens the distance to users (i.e., enhancing the first-hop channel gain) but increases the distance to the BS (i.e., degrading the second-hop channel gain). Therefore, a lower UAV height does not necessarily lead to an improved SINR performance.
    By allowing the UAV height to be optimized within the range of $15$ m to $25$ m, our optimized UAV position achieves $\min_m \text{SINR}_m=22$ dB at $\text{CDF} =0.5$, which is $4.6\text{ dB}$ better than the SAA position and CoMP position.
    This result underscores the importance of considering both the first-hop and second-hop channels in the UAV placement optimization.
    % highlights the importance of considering both the first-hop and second-hop channels in the UAV placement optimization.
    \tcminor{In summary, with a fixed UAV height, the position optimized by our UAV placement algorithm outperforms those optimized by the state-of-the-art \cite{liu2019comp}. 
    When UAV height is also optimized, the performance improvement becomes even more significant.
    Although optimizing UAV height can further enhance SINR performance, it serves as an optional enhancement rather than a requirement for demonstrating the effectiveness of our UAV placement algorithm.}
    % our UAV placement algorithm to demonstrate its effectiveness.}
    % even if we keep the UAV height at a fixed level, our proposed UAV placement algorithm outperforms the state-of-the-arts.
    % If it allows the UAV height to be optimized, the improvement by our UAV placement algorithm is even further.}

	% \vspace{-1mm}
	\subsection{Numerical Evaluation of JRBC Algorithm}
	For the proposed JRBC algorithm, we compare it with the existing relay beamforming schemes as below:
	\begin{itemize}		
		\item \textbf{Multi-user relay beamforming} \cite{phan2013iterative,che2014joint}: 
		This approach designs the relay beamforming to maximize the minimum SINR among users, without the receive combining. It corresponds to a special case of our transceiver design with $\mathbf{B}=\mathbf{I}$  and $N=M$. %\nm{It corresponds to a special case of our transceiver design with ...}
		% but the receive combining is not employed.
		\item \textbf{Transceiver design with ZF constraint} \cite{behbahani2008optimizations}: This approach optimizes the total SINR of multiple data streams.
		% , considering the received signal power constraint. 
		We assume that each data stream corresponds to a user.
		This approach considers the received signal power constraint, so the output power of each relay might exceed the individual relay power constraint.
		The output power of each relay may need to be clipped or scaled down \cite{behbahani2008optimizations}.
		In our simulation, we clip the output power of the relay when it exceeds the individual relay power constraint.
		
		%				\nm{are you doing the clipping in your simulations? You should to ensure the constraint.}
		%		Note that this work considers the received signal power constraint. The output power of each relay has the chance to violate the individual relay power constraint, so the relay may require to clip or scale down the output power (as discussed in \cite{behbahani2008optimizations}).
		% , considering the power constraint on the received signal. 
		\item \textbf{Single-user relay beamforming} \cite{havary2008distributed, zheng2009collaborative,kalogerias2018spatially}: This approach designs the relay beamforming to optimize the SNR for single-user communications, considering a total relay power constraint. TDMA is applied for this method to operate in our considered multi-user scenario. % \nm{+ TDMA?}
		% \vspace{-3mm}
	\end{itemize}
	% \vspace{-1mm}
	The existing multi-user schemes are designed for the case that the number of receive antennas is equal to the number of users, i.e., $N=M$, thus making the works \cite{phan2013iterative,che2014joint} applicable. Thus, the following experiments are evaluated with $(M, J, N) = (4,4,4)$, unless they are otherwise specified.
	
	\begin{figure}[t]		\centering %\setcounter{figure}{0}
		\includegraphics[scale=0.55]{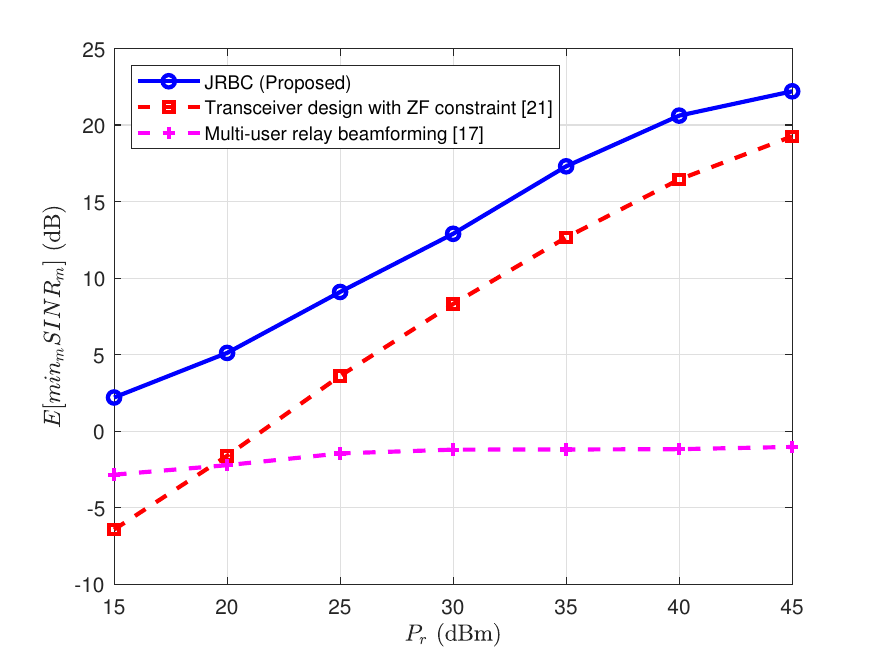}
		\caption{
			The average beamforming SINR versus power budget $P_r$, with $(M,J,N)=(4,4,4)$ and $P_t=30$ dBm.
			%			\nt{to be updated}
		}
		\label{fig:expected_sinr_vs_Pr}
		% \vspace{-2mm}
	\end{figure}

	In Fig. \ref{fig:expected_sinr_vs_Pr}, we evaluate the $\mathbb{E}[\min_m \text{SINR}_m]$ of the relay beamforming schemes versus the relay power budget $P_r$. 
	%	Note that the work \cite{behbahani2008optimizations} (i.e., \text{Transceiver design with ZF constraint}) considers the receive power constraint at the BS. There is the chance that the output power of each relay violates the individual relay power constraint, so the relay require to clip or scale down the output power (as discussed in \cite{behbahani2008optimizations}).	
    \text{JRBC} has the largest $\mathbb{E}[\min_m \text{SINR}_m]$	for $P_r\in[15,45]$ dBm.
    With $P_r=30$ dBm, JRBC achieves $\mathbb{E}[\min_m \text{SINR}_m]=13\text{ dB}$.
    In contrast, the transceiver design with ZF constraint and \text{multi-user relay beamforming} 
    achieve $8$ dB and $-1$ dB in the same configuration, respectively.    
    % \nm{Grammar!: . In contrast, the transceiver design with ZF constraint and \text{multi-user relay beamforming} 
    % achieve $8$ dB and $-1$ dB in the same configuration, respectively.}, as opposed to transceiver design with ZF constraint and \text{multi-user relay beamforming} achieve $8$ dB and $-1$ dB in the same configuration, respectively.
	Multi-user relay beamforming \cite{phan2013iterative} is upper-bounded at $\mathbb{E}[\min_m \text{SINR}_m]=-1$ dB for $P_r>30$ dBm, due to the limited number of relays.
	For JRBC and transceiver design with ZF constraint, $\mathbb{E}[\min_m \text{SINR}_m]$ increases with a larger $P_r$ thanks to the receive combining.

\begin{figure}[t]		\centering %\setcounter{figure}{0}
		\includegraphics[scale=0.55]{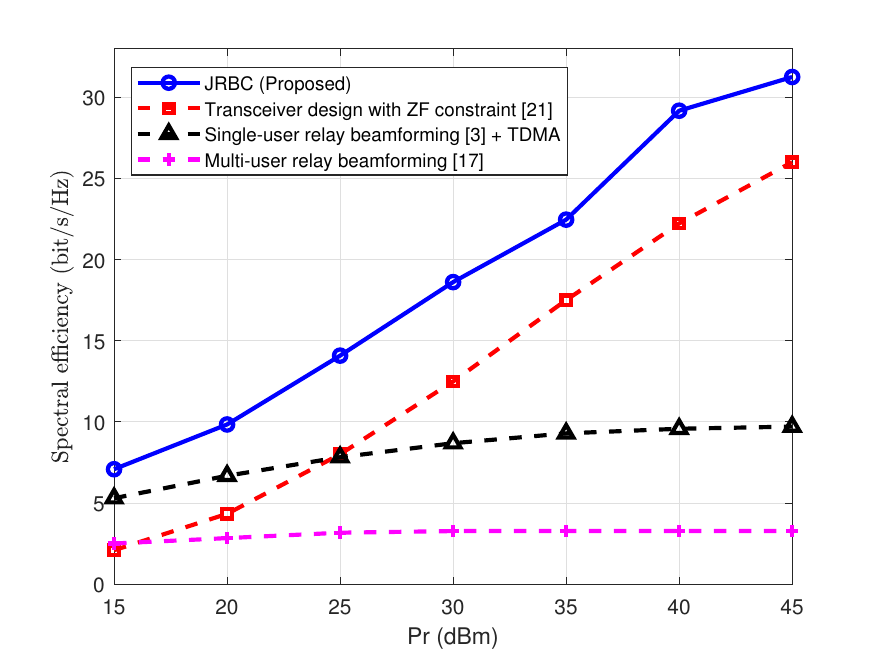}
		\caption{
			The effective rate versus the relay power budget $P_r$, with $(M,J,N)=(4,4,4)$ and $P_t=30$ dBm.
			%			transmit power of the user $P_t$, with $(M,J,N)=(4,4,4)$ and $(M,J,N)=(4,10,4)$. \nt{to be updated}
		}
		\label{fig:expected_SE_vs_Pr}
		% \vspace{-2mm}
	\end{figure}

	In Fig. 8, we evaluate the spectral efficiency (SE) as follows
	%	. The rate expression maximized by Gaussian signaling over the channel can be expressed as 
	\begin{align}
	% \vspace{-1mm}
	\text{SE} = \frac{1}{T_{tot}}\sum_{m\in\mathcal{M}}T_m\log_2(1+\text{SINR}_m),
	% \vspace{-1mm}
	\end{align}
	where $T_{tot}$ is the total transmission time of all users, and $T_{m}$ is the transmission time of user $m$.
	For TDMA, we assume that $T_m = T_{tot}/M$ , so $\text{SE} = \frac{1}{M}\sum_{m\in\mathcal{M}}\log_2(1+\text{SINR}_m)$.
	For the cases with non-orthogonal transmission that all users transmit at the same time and frequency, i.e., $T_m = T_{tot}$, we have $\text{SE} =\sum_{m\in\mathcal{M}}\log_2(1+\text{SINR}_m)$.
    We evaluate the SE with the same transmit power of the users $P_t=30\text{ dBm}$.
    % \hl{We consider the same transmit power of the user $P_t$ since it is the user's capability.}\nm{? unclear}
	%	The non-orthogonal transmission utilizes the time and frequency resource more efficiently, but the interference between users could degrade the SE, which requires the transceiver design to improve the SINR performance.
	%	\nm{what about power constraint? is the average power computed over the entire interval Ttot, or just over the transmission time? For example in the case of TDMA, each user could transmit with power $\times M$ while satisfying the same avg power across Ttot.}
	%	\tc{No, I did not assume the same average power across Ttot, but only the same on the transmission of each user.}
	Fig. \ref{fig:expected_SE_vs_Pr} evaluates the SE versus the relay power budget $P_r$.
	The SE increases with a larger $P_r$. 
	\text{JRBC} achieves the largest SE for $P_r\in[15,45]$ dBm.
	% the $P_r$ from $15$ to $45$ dBm.
	At $P_r =30\text{ dBm}$, \text{JRBC} has $19$ bit/s/Hz, as opposed to $13$ bit/s/Hz for  transceiver design with ZF constraint, $8.5$ bit/s/Hz for single-user relay beamforming with TDMA, and $3$ bit/s/Hz for multi-user relay beamforming.
	\text{Single-user relay beamforming with TDMA} is upper-bounded at $\text{SE}=10$ bit/s/Hz due to the inefficiency of time resource usage in TDMA.

    % \nm{can you include the case of 3 UAVs as well in the figure? And comment on the fact that when you have less uav than users than we are not able to separate the data streams in the spatial doaminb leading to significant interference. you can then point out to future work to incorporate scheduling to handle these situations. This addressed the reviewer comment}
    
\begin{figure}[t]		\centering %\setcounter{figure}{0}
		\includegraphics[scale=0.55]{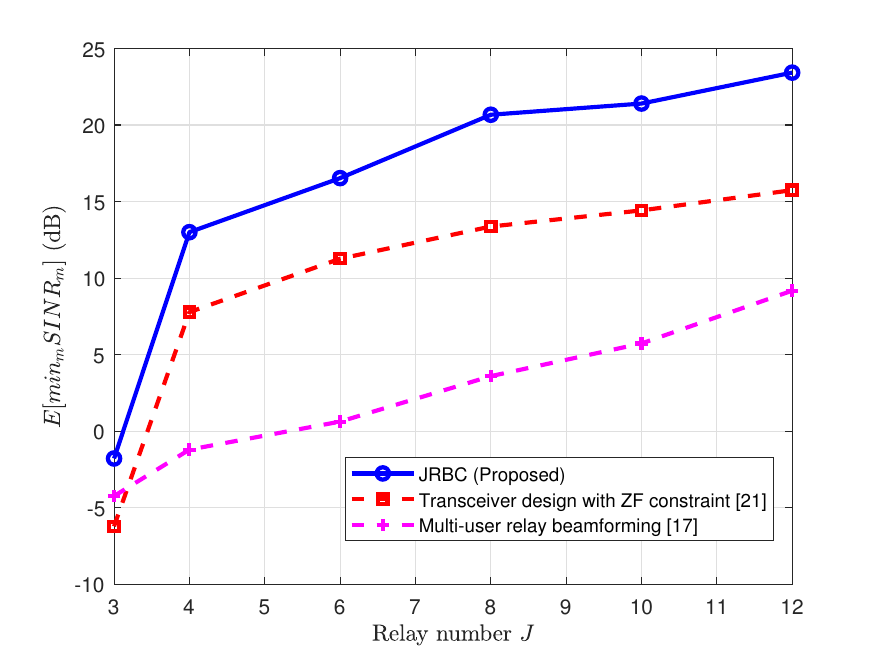}
		\caption{
			The average beamforming SINR versus the number of UAV relays $J$, with $(M,N) = (4,4)$ and $P_t=P_r=30$ dBm.
			%			transmit power of the user $P_t$, with $(M,J,N)=(4,4,4)$ and $(M,J,N)=(4,10,4)$. \nt{to be updated}
		}
		\label{fig:expected_sinr_vs_J}
		% \vspace{-3mm}
	\end{figure}

	In Fig. \ref{fig:expected_sinr_vs_J}, we evaluate $\mathbb{E}[\min_m \text{SINR}_m]$ of the relay beamforming schemes versus the number of UAV relays $J$. 
	For each $J$, the UAV positions are optimized by the UAV placement algorithm.
    The \text{multi-user relay beamforming} \cite{phan2013iterative} improves its $\mathbb{E}[\min_m \text{SINR}_m]$ with a larger $J$, which verifies our argument in Fig. \ref{fig:expected_sinr_vs_Pr} that its $\mathbb{E}[\min_m \text{SINR}_m]$ can only achieve $-1\text{ dB}$ because of the limited number of UAV relays. 
	For $J=12$, \text{JRBC} achieves $\mathbb{E}[\min_m \text{SINR}_m]=23$ dB, which outperforms the state-of-the-art \cite{behbahani2008optimizations} by $7$ dB.	It shows that JRBC is scalable to the number of UAV relays and outperforms the state-of-the-art.
    \tcadd{Note that, for $J=3$ UAV relays, all relay beamforming schemes can only achieve $\mathbb{E}[\min_m \text{SINR}_m]<0\text{ dB}$ due to the overloaded MU scenario $(J<M)$.
    When we increase to $J=4$ (i.e., underloaded MU scenario), we observe a significant improvement in $\mathbb{E}[\min_m \text{SINR}_m]$ for all relay beamforming schemes.
    This highlights the necessity of the underloaded MU scenario assumption for the relay beamforming in non-orthogonal wireless relay networks.
    In overloaded MU scenarios ($M>J$), significant inter-user interference results in poor SINR performance due to non-orthogonal transmission in our system model, requiring user scheduling.
    % \sst{To address this issue, the user scheduling is necessary to divide users into multiple groups, serving each group independently.} 
    Once user scheduling is applied to create an underloaded scenario, our UAV placement and JRBC algorithms become applicable (see \textbf{Remark \ref{overloaded}}).}

	% \vspace{-3mm}
	\section{Conclusion}\label{sec_conclusion}
	We developed a design to enhance the minimum SINR among users in multi-user relay networks by jointly optimizing the UAV relay placement, relay beamforming, and receive combining.
	For beamforming-aware UAV placement optimization, we proposed a UAV placement algorithm that provided UAV positions with a better minimum expected beamforming SINR among users based on statistical CSI, using the DC framework.
	For transceiver design, we proposed a JRBC algorithm to refine the relay beamforming and receive combining based on estimated CSI, using the BCD approach.
	Numerical results showed the effectiveness of the UAV placement algorithm in deploying UAVs to positions that led to an improved minimum expected beamforming SINR among users. 
    The JRBC algorithm yielded enhanced SINR compared to the state-of-the-art relay beamforming schemes.

	\bibliographystyle{IEEEtran}
	\bibliography{IEEEabrv,reference} 

\begin{thebibliography}{10}
\providecommand{\url}[1]{#1}
\csname url@samestyle\endcsname
\providecommand{\newblock}{\relax}
\providecommand{\bibinfo}[2]{#2}
\providecommand{\BIBentrySTDinterwordspacing}{\spaceskip=0pt\relax}
\providecommand{\BIBentryALTinterwordstretchfactor}{4}
\providecommand{\BIBentryALTinterwordspacing}{\spaceskip=\fontdimen2\font plus
\BIBentryALTinterwordstretchfactor\fontdimen3\font minus
  \fontdimen4\font\relax}
\providecommand{\BIBforeignlanguage}[2]{{%
\expandafter\ifx\csname l@#1\endcsname\relax
\typeout{** WARNING: IEEEtran.bst: No hyphenation pattern has been}%
\typeout{** loaded for the language `#1'. Using the pattern for}%
\typeout{** the default language instead.}%
\else
\language=\csname l@#1\endcsname
\fi
#2}}
\providecommand{\BIBdecl}{\relax}
\BIBdecl

\bibitem{havary2008distributed}
V.~Havary-Nassab, S.~Shahbazpanahi, A.~Grami, and Z.-Q. Luo, ``{Distributed
  beamforming for relay networks based on second-order statistics of the
  channel state information},'' \emph{IEEE Trans. Signal Process.}, vol.~56,
  no.~9, pp. 4306--4316, 2008.

\bibitem{havary2009optimal}
V.~Havary-Nassab, S.~Shahbazpanahi, and A.~Grami, ``{Optimal distributed
  beamforming for two-way relay networks},'' \emph{IEEE Trans. Signal
  Process.}, vol.~58, no.~3, pp. 1238--1250, 2009.

\bibitem{zheng2009collaborative}
G.~Zheng, K.-K. Wong, A.~Paulraj, and B.~Ottersten, ``{Collaborative-relay
  beamforming with perfect CSI: Optimum and distributed implementation},''
  \emph{IEEE Signal Process. Lett.}, vol.~16, no.~4, pp. 257--260, 2009.

\bibitem{li2010cooperative}
J.~Li, A.~P. Petropulu, and H.~V. Poor, ``Cooperative transmission for relay
  networks based on second-order statistics of channel state information,''
  \emph{IEEE Trans. Signal Process.}, vol.~59, no.~3, pp. 1280--1291, 2010.

\bibitem{brinton2025key}
{Brinton, Christopher G and Chiang, Mung and Kim, Kwang Taik and Love, David J
  and Beesley, Michael and Repeta, Morris and Roese, John and Beming, Per and
  Ekudden, Erik and Li, Clara and others}, ``{Key focus areas and enabling
  technologies for 6G},'' \emph{{IEEE Communications Magazine}}, vol.~63,
  no.~3, pp. 84--91, 2025.

\bibitem{10615846}
B.~Keshavamurthy and N.~Michelusi, ``{Orchestrating UAVs for Prioritized Data
  Harvesting: A Cross-Layer Optimization Perspective},'' in \emph{IEEE
  International Conference on Communications Workshops}, 2024, pp. 1268--1273.

\bibitem{zeng2019accessing}
Y.~Zeng, Q.~Wu, and R.~Zhang, ``{Accessing from the sky: A tutorial on UAV
  communications for 5G and beyond},'' \emph{Proceedings of the IEEE}, vol.
  107, no.~12, pp. 2327--2375, 2019.

\bibitem{kalogerias2018spatially}
D.~S. Kalogerias and A.~P. Petropulu, ``Spatially controlled relay
  beamforming,'' \emph{IEEE Trans. Signal Process.}, vol.~66, no.~24, pp.
  6418--6433, 2018.

\bibitem{evmorfos2022reinforcement}
S.~Evmorfos, K.~I. Diamantaras, and A.~P. Petropulu, ``{Reinforcement learning
  for motion policies in mobile relaying networks},'' \emph{IEEE Trans. Signal
  Process.}, vol.~70, pp. 850--861, 2022.

\bibitem{liu2019comp}
L.~Liu, S.~Zhang, and R.~Zhang, ``{CoMP in the sky: UAV placement and movement
  optimization for multi-user communications},'' \emph{IEEE Trans. Commun.},
  vol.~67, no.~8, pp. 5645--5658, 2019.

\bibitem{xie2020common}
L.~Xie, J.~Xu, and Y.~Zeng, ``{Common Throughput Maximization for UAV-Enabled
  Interference Channel With Wireless Powered Communications},'' \emph{IEEE
  Trans. Commun.}, vol.~68, no.~5, pp. 3197--3212, 2020.

\bibitem{ma2011pilot}
J.~Ma, P.~Orlik, J.~Zhang, and G.~Y. Li, ``{Pilot matrix design for estimating
  cascaded channels in two-hop MIMO amplify-and-forward relay systems},''
  \emph{IEEE Trans. Wireless Commun.}, vol.~10, no.~6, pp. 1956--1965, 2011.

\bibitem{sanguinetti2012tutorial}
L.~Sanguinetti, A.~A. D'Amico, and Y.~Rong, ``{A tutorial on the optimization
  of amplify-and-forward MIMO relay systems},'' \emph{IEEE J. Sel. Areas
  Commun.}, vol.~30, no.~8, pp. 1331--1346, 2012.

\bibitem{phan2012beamforming}
A.~H. Phan, H.~D. Tuan, H.~H. Kha, and H.~H. Nguyen, ``{Beamforming
  optimization in multi-user amplify-and-forward wireless relay networks},''
  \emph{IEEE Trans. Wireless Commun.}, vol.~11, no.~4, pp. 1510--1520, 2012.

\bibitem{cheng2012joint}
Y.~Cheng and M.~Pesavento, ``{Joint optimization of source power allocation and
  distributed relay beamforming in multiuser peer-to-peer relay networks},''
  \emph{IEEE Trans. Signal Process.}, vol.~60, no.~6, pp. 2962--2973, 2012.

\bibitem{rashid2013relay}
U.~Rashid, H.~D. Tuan, and H.~H. Nguyen, ``{Relay beamforming designs in
  multi-user wireless relay networks based on throughput maximin
  optimization},'' \emph{IEEE Trans. Commun.}, vol.~61, no.~5, pp. 1739--1749,
  2013.

\bibitem{phan2013iterative}
A.~H. Phan, H.~D. Tuan, H.~H. Kha, and H.~H. Nguyen, ``{Iterative DC
  optimization of precoding in wireless MIMO relaying},'' \emph{IEEE Trans.
  Wireless Commun.}, vol.~12, no.~4, pp. 1617--1627, 2013.

\bibitem{ruan2019distributed}
H.~Ruan and R.~C. de~Lamare, ``{Distributed robust beamforming based on
  low-rank and cross-correlation techniques: Design and analysis},'' \emph{IEEE
  Trans. Signal Process.}, vol.~67, no.~24, pp. 6411--6423, 2019.

\bibitem{che2014joint}
E.~Che, H.~D. Tuan, and H.~H. Nguyen, ``{Joint optimization of cooperative
  beamforming and relay assignment in multi-user wireless relay networks},''
  \emph{IEEE Trans. Wireless Commun.}, vol.~13, no.~10, pp. 5481--5495, 2014.

\bibitem{dimas2019cooperative}
A.~Dimas, D.~S. Kalogerias, and A.~P. Petropulu, ``Cooperative beamforming with
  predictive relay selection for urban mmwave communications,'' \emph{IEEE
  Access}, vol.~7, pp. 157\,057--157\,071, 2019.

\bibitem{behbahani2008optimizations}
A.~S. Behbahani, R.~Merched, and A.~M. Eltawil, ``{Optimizations of a MIMO
  relay network},'' \emph{IEEE Trans. Signal Process.}, vol.~56, no.~10, pp.
  5062--5073, 2008.

\bibitem{evmorfos2022deep}
S.~Evmorfos and A.~P. Petropulu, ``{Deep actor-critic for continuous 3D motion
  control in mobile relay beamforming networks},'' in \emph{Proc. IEEE Int.
  Conf. on Acoust., Speech Signal Process.}, 2022, pp. 5353--5357.

\bibitem{hanna2019uav}
S.~Hanna, E.~Krijestorac, H.~Yan, and D.~Cabric, ``{UAV swarms as
  amplify-and-forward MIMO relays},'' in \emph{Proc. IEEE 20th Int. Workshop
  Signal Process. Adv. Wireless Commun. (SPAWC)}, 2019, pp. 1--5.

\bibitem{hanna2021uav}
S.~Hanna, E.~Krijestorac, and D.~Cabric, ``{UAV swarm position optimization for
  high capacity MIMO backhaul},'' \emph{IEEE J. Sel. Areas Commun.}, vol.~39,
  no.~10, pp. 3006--3021, 2021.

\bibitem{kang20203d}
Z.~Kang, C.~You, and R.~Zhang, ``{3D placement for multi-UAV relaying: An
  iterative gibbs-sampling and block coordinate descent optimization
  approach},'' \emph{IEEE Trans. Commun.}, vol.~69, no.~3, pp. 2047--2062,
  2020.

\bibitem{zhang20213d}
C.~Zhang, L.~Zhang, L.~Zhu, T.~Zhang, Z.~Xiao, and X.-G. Xia, ``{3D Deployment
  of Multiple UAV-Mounted Base Stations for UAV Communications},'' \emph{IEEE
  Trans. Commun.}, vol.~69, no.~4, pp. 2473--2488, 2021.

\bibitem{ding20203d}
R.~Ding, F.~Gao, and X.~S. Shen, ``{3D UAV Trajectory Design and Frequency Band
  Allocation for Energy-Efficient and Fair Communication: A Deep Reinforcement
  Learning Approach},'' \emph{IEEE Trans. Wireless Commun.}, vol.~19, no.~12,
  pp. 7796--7809, 2020.

\bibitem{huang2021joint}
Y.~Huang and A.~Ikhlef, ``{Joint design of fronthaul and access links in
  massive MIMO multi-UAV-enabled CRANs},'' \emph{IEEE Wireless Commun. Lett.},
  vol.~10, no.~11, pp. 2355--2359, 2021.

\bibitem{mahmood2023joint}
A.~Mahmood, T.~X. Vu, S.~Chatzinotas, and B.~Ottersten, ``{Joint Optimization
  of 3D Placement and Radio Resource Allocation for Per-UAV Sum Rate
  Maximization},'' \emph{IEEE Trans. Veh. Technol.}, vol.~72, no.~10, pp.
  13\,094--13\,105, 2023.

\bibitem{dinh2019joint}
P.~Dinh, T.~M. Nguyen, S.~Sharafeddine, and C.~Assi, ``{Joint location and
  beamforming design for cooperative UAVs with limited storage capacity},''
  \emph{IEEE Trans. Commun.}, vol.~67, no.~11, pp. 8112--8123, 2019.

\bibitem{gao20203d}
N.~Gao, S.~Jin, and X.~Li, ``{3-D deployment of UAV swarm for massive MIMO
  communications},'' in \emph{Proc. ACM MobiArch 15th Workshop Mobility
  Evolving Internet Architect.}, 2020, pp. 24--29.

\bibitem{khuwaja2019optimum}
A.~A. Khuwaja, G.~Zheng, Y.~Chen, and W.~Feng, ``{Optimum Deployment of
  Multiple UAVs for Coverage Area Maximization in the Presence of Co-Channel
  Interference},'' \emph{IEEE Access}, vol.~7, pp. 85\,203--85\,212, 2019.

\bibitem{wang2022resource}
C.~Wang, D.~Deng, L.~Xu, and W.~Wang, ``{Resource Scheduling Based on Deep
  Reinforcement Learning in UAV Assisted Emergency Communication Networks},''
  \emph{IEEE Trans. Commun.}, vol.~70, no.~6, pp. 3834--3848, 2022.

\bibitem{wu2018joint}
Q.~Wu, Y.~Zeng, and R.~Zhang, ``{Joint trajectory and communication design for
  multi-UAV enabled wireless networks},'' \emph{IEEE Trans. Wireless Commun.},
  vol.~17, no.~3, pp. 2109--2121, 2018.

\bibitem{gao2024irs}
{Gao, Ying and Wu, Qingqing and Chen, Wen and Liu, Yang and Li, Ming and da
  Costa, Daniel Benevides}, ``{IRS-aided overloaded multi-antenna systems:
  Joint user grouping and resource allocation},'' \emph{IEEE Trans. Wireless
  Commun.}, vol.~23, no.~8, pp. 8297--8313, 2024.

\bibitem{you20193d}
C.~You and R.~Zhang, ``{3D trajectory optimization in Rician fading for
  UAV-enabled data harvesting},'' \emph{IEEE Trans. Wireless Commun.}, vol.~18,
  no.~6, pp. 3192--3207, 2019.

\bibitem{keshavamurthy2023maestro}
B.~Keshavamurthy, M.~A. Bliss, and N.~Michelusi, ``{MAESTRO-X: Distributed
  Orchestration of Rotary-Wing UAV-Relay Swarms},'' \emph{IEEE Trans. Cogn.
  Commun. Netw.}, vol.~9, no.~3, pp. 794--810, 2023.

\bibitem{yan2019comprehensive}
C.~Yan, L.~Fu, J.~Zhang, and J.~Wang, ``{A Comprehensive Survey on UAV
  Communication Channel Modeling},'' \emph{IEEE Access}, vol.~7, pp.
  107\,769--107\,792, 2019.

\bibitem{chou2023compressed}
T.-H. Chou, N.~Michelusi, D.~J. Love, and J.~V. Krogmeier, ``{Compressed
  training for dual-wideband time-varying sub-terahertz massive MIMO},''
  \emph{IEEE Trans. Commun.}, vol.~71, no.~6, pp. 3559--3575, 2023.

\bibitem{golub2013matrix}
G.~H. Golub and C.~F. Van~Loan, \emph{Matrix computations}, 4th~ed.\hskip 1em
  plus 0.5em minus 0.4em\relax Baltimore, MD, USA: The Johns Hopkins Univ.
  Press, 2012.

\bibitem{horn2012matrix}
R.~A. Horn and C.~R. Johnson, \emph{Matrix analysis}, 2nd~ed.\hskip 1em plus
  0.5em minus 0.4em\relax Cambridge, United Kingdom: Cambridge Univ. Press,
  2012.

\bibitem{boyd2004convex}
S.~P. Boyd and L.~Vandenberghe, \emph{Convex optimization}.\hskip 1em plus
  0.5em minus 0.4em\relax Cambridge university press, 2004.

\bibitem{cvx}
M.~Grant and S.~Boyd, ``{CVX}: Matlab software for disciplined convex
  programming, version 2.1,'' \url{http://cvxr.com/cvx}, Mar. 2014.

\bibitem{ben2001lectures}
A.~Ben-Tal and A.~Nemirovski, \emph{Lectures on modern convex optimization:
  analysis, algorithms, and engineering applications}.\hskip 1em plus 0.5em
  minus 0.4em\relax SIAM, 2001.

\bibitem{3GPP38901}
{3rd Generation Partnership Project (3GPP)}, ``{Study on channel model for
  frequencies from 0.5 to 100 GHz},'' 3rd Generation Partnership Project
  (3GPP), Technical Report (TR) 38.901, 2024, version 17.1.0.

\bibitem{3gpp_ts_22125_v1920}
------, ``{Technical Specification Group Services and System Aspects; Unmanned
  Aerial System (UAS) support in 3GPP systems (Release 19)},'' {3rd Generation
  Partnership Project (3GPP)}, Technical Specification (TS) 22.125, 2024,
  version 19.2.0.

\end{thebibliography}

    \begin{IEEEbiography}
        [{\includegraphics[width=1.1in,height=1.3in,clip,keepaspectratio]{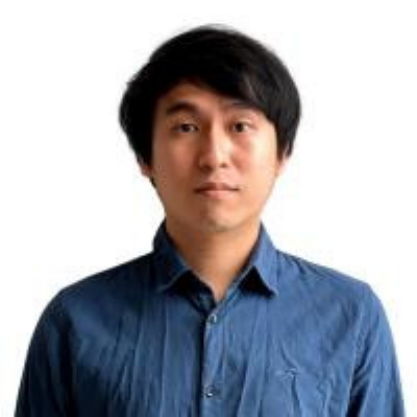}}]{Tzu-Hsuan Chou}(Member, IEEE) received the B.S. degree in electrical engineering and the M.S. degree in electrical and control engineering from National Chiao Tung University, Hsinchu, Taiwan, in 2011 and 2013, respectively, and the Ph.D. degree in electrical and computer engineering from Purdue University, West Lafayette, IN, USA, in 2022. He is currently a Wireless Systems Engineer with Qualcomm Inc., San Diego, CA, USA. From 2014 to 2017, he was a Software Engineer with MediaTek, Hsinchu, Taiwan. His research interests include wireless communications, massive MIMO, compressed sensing, transceiver design, and UAV communications.
    \end{IEEEbiography}

    \begin{IEEEbiography}[{\includegraphics[width=1.1in,height=1.3in,clip,keepaspectratio]{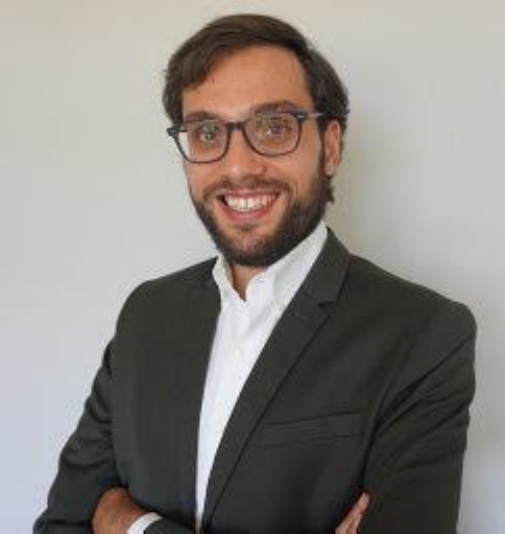}}]{Nicol\`{o} Michelusi}(Senior Member, IEEE) received the B.Sc. (with honors), M.Sc. (with honors), and Ph.D. degrees from the University of Padova, Italy, in 2006, 2009, and 2013, respectively, and an M.Sc. degree in telecommunications engineering from the Technical University of Denmark, Denmark, in 2009. 
    From 2013 to 2015, he was a Postdoctoral Research Fellow with the Ming Hsieh Department of Electrical Engineering, University of Southern California, Los Angeles, CA, USA. From 2016 to 2020, he was an Assistant Professor with the School of Electrical and Computer Engineering, Purdue University, West Lafayette, IN, USA. He is currently an Associate Professor with the School of Electrical, Computer and Energy Engineering, Arizona State University, Tempe, AZ, USA.     
    His research interests include 5G wireless networks, millimeter-wave communications, stochastic optimization, and decentralized and federated learning over wireless systems.     
    He served as an Associate Editor for the \emph{IEEE Transactions on Wireless Communications} (2016–2021) and is currently an Editor for the \emph{IEEE Transactions on Communications}. He is also a member of the IEEE Signal Processing for Communications and Networking Technical Committee.     
    He co-chaired the Distributed Machine Learning and Fog Networking Workshop at IEEE INFOCOM in 2021, 2023, and 2024; the Wireless Communications Symposium at IEEE GLOBECOM 2020; the IoT, M2M, Sensor Networks, and Ad-Hoc Networking Track at IEEE VTC 2020; and the Cognitive Computing and Networking Symposium at ICNC 2018. He served as Technical Area Chair for the Communication Systems track at Asilomar 2023.     
    He is the recipient of several awards, including the NSF CAREER Award in 2021, the IEEE Communication Theory Technical Committee (CTTC) Early Achievement Award in 2022, the IEEE Communications Society William R. Bennett Prize in 2024, and the IEEE ICC Best Paper Award for the Communication Theory Symposium in 2025.
    \end{IEEEbiography}

    \begin{IEEEbiography}
        [{\includegraphics[width=1.1in,height=1.3in,clip,keepaspectratio]{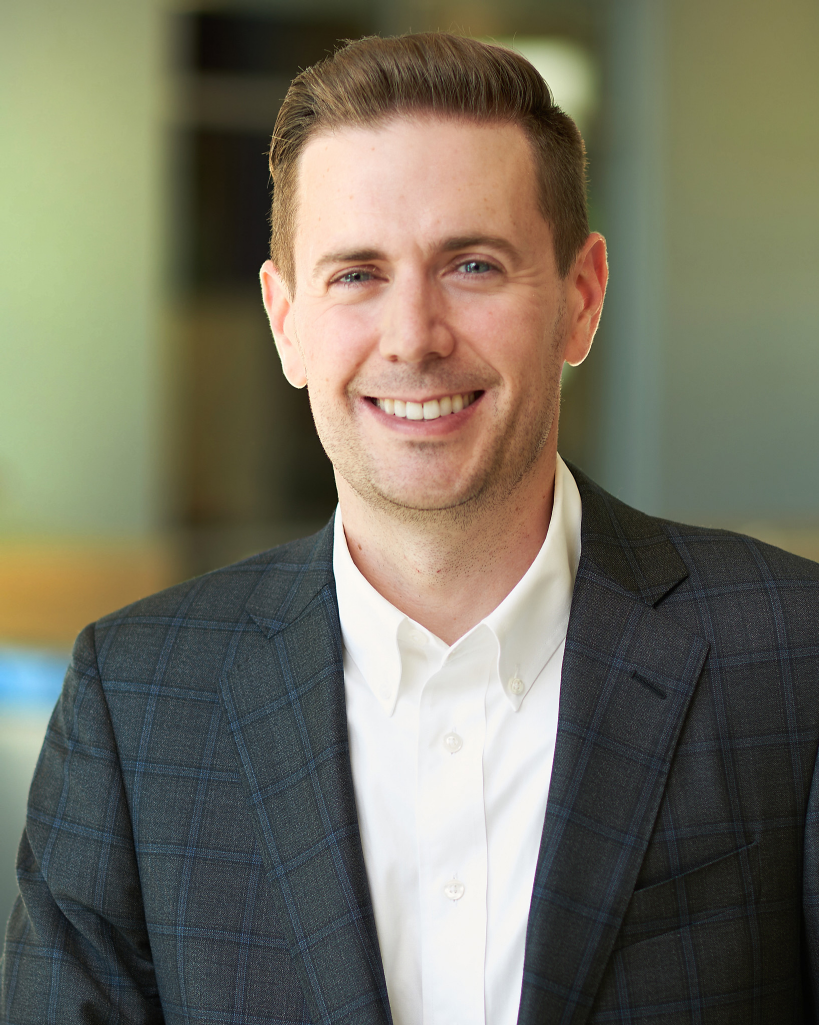}}]{David J. Love}(Fellow, IEEE) is the Nick Trbovich Professor of Electrical and Computer Engineering at Purdue University.  He received the B.S., M.S.E., and Ph.D. degrees from UT-Austin. He is currently a Senior Editor for IEEE Journal on Selected Areas in Communications and held editorial positions for IEEE Signal Processing Magazine, IEEE Trans. Communications, and IEEE Trans. Signal Processing. He holds 32 U.S. patents. His research interests include 6G and beyond wireless, MIMO communications, millimeter-wave wireless, software-defined radios, and coding theory.  He is a Fellow of the IEEE, American Association for the Advancement of Science (AAAS), and National Academy of Inventors (NAI).  His research has been recognized by the IEEE Communications Society (2016 Stephen O. Rice Prize, 2020 Fred W. Ellersick Prize, and 2024 William R. Bennett Prize), IEEE Signal Processing Society (2015 SPS Best Paper Award), and IEEE VT Society (2010 Jack Neubauer Memorial Award).
    \end{IEEEbiography}

    \begin{IEEEbiography}
        [{\includegraphics[width=1.1in,height=1.3in,clip,keepaspectratio]{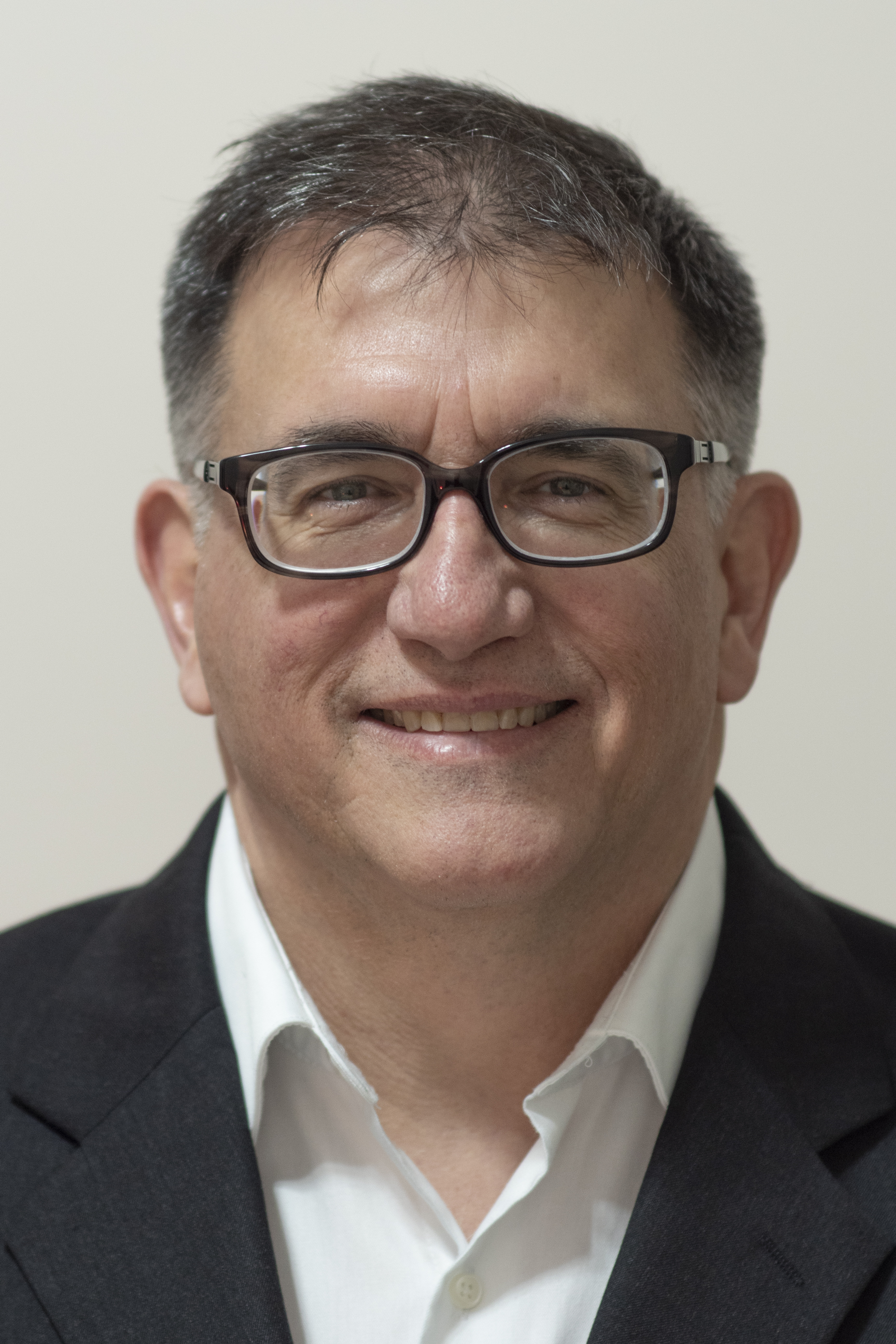}}]{James V. Krogmeier}(Senior Member, IEEE) received the B.S.E.E. degree from the University of Colorado Boulder, Boulder, CO, USA, and the M.S. and Ph.D. degrees from the University of Illinois at Urbana-Champaign, Champaign, IL, USA. He has industry experience in telecommunications and is a Founding Member of two software startup companies. He is currently a Professor of electrical and computer engineering with Purdue University, West Lafayette, IN, USA. He has authored or coauthored many technical papers in refereed journals and conference proceedings of the IEEE, the ASABE, and the Transportation Research Board, and is a Co-Inventor of five U.S. patents. His research interests include the applications of statistical signal and image processing in agriculture, intelligent transportation systems, sensor networking, and wireless communications. His research has been funded by the USDA-NIFA, the NSF, the DARPA, the Indiana Department of Transportation, the Federal Highway Administration, and industry. He was on a number of IEEE technical program committees and an Associate Editor for several IEEE journals.

    \end{IEEEbiography}

\end{document}